\documentclass[superscriptaddress,preprintnumbers,amsmath,amssymb,nofootinbib,eqsecnum,a4paper,secnumarabic,11pt]{revtex4}
\usepackage{graphicx}
\usepackage{color}
\usepackage{dcolumn}      
\usepackage{bm,bbm}       
\usepackage{cancel}       
\usepackage{latexsym}     
\usepackage{mathrsfs}     
\usepackage{multirow}     
\usepackage{hyperref}     
\usepackage{subfigure}    
\usepackage{titlesec}     
\usepackage{ulem}         
\usepackage{comment}         
\def\slashchar#1{\setbox0=\hbox{$#1$} 
\dimen0=\wd0 
\setbox1=\hbox{/} \dimen1=\wd1 
\ifdim\dimen0>\dimen1 
\rlap{\hbox to \dimen0{\hfil/\hfil}} 
#1 
\else 
\rlap{\hbox to \dimen1{\hfil$#1$\hfil}} 
/ 
\fi}

\makeatletter
\let\start@align@nopar\start@align
\let\start@gather@nopar\start@gather
\let\start@multline@nopar\start@multline
\long\def\start@align{\par\start@align@nopar}
\long\def\start@gather{\par\start@gather@nopar}
\long\def\start@multline{\par\start@multline@nopar}
\makeatother

\newcommand*{\justifyheading}{\raggedright}
\titleformat{\chapter}[display]
  {\normalfont\huge\bfseries\justifyheading}{\chaptertitlename\ \thechapter}
  {20pt}{\Huge}
\titleformat{\section}
  {\normalfont\Large\bfseries\justifyheading}{\thesection}{1em}{}
\titleformat{\subsection}
  {\normalfont\large\bfseries\justifyheading}{\thesubsection}{1em}{}
\titleformat{\subsubsection}
  {\normalfont\bfseries\justifyheading}{\thesubsubsection}{1em}{}
\usepackage[english]{babel}

\def\a{\alpha}
\def\b{\beta}

\def\d{\delta}

\def\g{\gamma}

\def\l{\lambda}
\def\m{\mu}
\def\n{\nu}

\def\q{\partial}
\def\s{\sigma}

\def\D{\Delta}
\def\G{\Gamma}

\def\sla#1{\rlap/#1}

\begin{document}



\preprint{KEK-TH-1787}

\title{
  Quantum corrections to the spin-independent cross section\\
  of the inert doublet dark matter
}
\author{Tomohiro Abe}
\affiliation{Theory Center, Institute of Particle and Nuclear Studies,
High Energy Accelerator Research Organization (KEK), Tsukuba, 305-0801, Japan}
\author{Ryosuke Sato}
\affiliation{Theory Center, Institute of Particle and Nuclear Studies,
High Energy Accelerator Research Organization (KEK), Tsukuba, 305-0801, Japan}
%
%
%
\begin{abstract}
The inert Higgs doublet model contains a stable neutral boson as a candidate of dark matter.
We calculate cross section for spin-independent scattering of the dark matter on nucleon.
We take into account electroweak and scalar quartic interactions,
and evaluate effects of scattering with quarks at one-loop level and with gluon at two-loop level.
These contributions give an important effect for the dark matter mass to be around $m_h/2$,
because a coupling with the standard model Higgs boson which gives the
 leading order contribution should be suppressed
to reproduce the correct amount of the thermal relic abundance
 in this mass region. 
In particular, we show that the dark matter self coupling changes the value of the spin-independent cross section significantly.
\end{abstract}

~
\vskip 3cm 
\maketitle
\break


\section{Introduction}
The discovery of the Higgs boson at the Large hadron collider (LHC) in 2012~\cite{1207.7214, 1207.7235}
is one of the biggest achievements of the standard model (SM).
In spite of its success, the SM does not include a candidate of the dark matter
which has many evidences for existing in our universe \cite{Bertone:2004pz}.
Hence, we need some extension of the SM to explain the dark matter as an elementary particle.

The inert two-Higgs doublet model~\cite{PHRVA.D18.2574, hep-ph/0603188}
is a simple extension of the
SM with a dark matter candidate. 
It was originally discussed in an analysis of 
electroweak symmetry breaking
in the two Higgs doublet model by Deshpande and Ma \cite{PHRVA.D18.2574},
and recently, it draws attention as a model of dark matter \cite{hep-ph/0603188}.
In this model, an additional $SU(2)_L$ doublet scalar field with $Y=1/2$, which is called \textit{inert} doublet,
and a $Z_2$ parity
are introduced.
Under this parity,
all of the SM fields are even and the inert doublet is odd.
Then the lightest neutral boson with the $Z_2$ odd charge becomes the dark matter candidate.
The $Z_2$ odd particles have electroweak interaction and scalar quartic interactions with the SM Higgs boson.
Thus, they are thermalized in 
the
early universe,
and the amount of the dark matter in the present universe is generated as a thermal relic \cite{PRLTA.39.165, NUPHA.B310.693, NUPHA.B360.145}.

The Higgs sector in the inert doublet model sometimes appears in a part of beyond the standard models,
\textit{e.g.,} left-right Twin Higgs model \cite{Chacko:2005un, Goh:2007dh, 0712.1234},
a composite Higgs model \cite{1105.5403},
a radiative seesaw model \cite{Ma:2006km, 1302.3936, 1303.7356}
and models of neutrino flavor with non-Abelian discrete symmetry \cite{0808.1729,1007.0871,1205.3442,1304.1603}.
Also, the inert doublet model is analyzed in contexts of 
strong first order electroweak phase transition \cite{1110.5334, 1204.4722, 1207.0084, 1302.2614, 1304.2055},
Coleman-Weinberg mechanism driven by the inert doublet \cite{0707.0633}, and inflation \cite{1202.0288}.
In spite of its simplicity, the inert doublet model has rich phenomenology.
In addition to the dark matter candidate, the model has a heavier neutral scalar and a charged scalar boson.
These $Z_2$ odd particles can be probed directly
at the LHC Run II \cite{PHRVA.D76.095011, PHRVA.D81.035003, PHRVA.D82.035009, 1206.6316, 1303.7102}
and the ILC \cite{1303.6191, 1401.6698}.
The measurements of the branching fraction of the Higgs decay
{\it e.g.}, diphoton signal and invisible decay
will be a probe of the $Z_2$ odd sector \cite{PHRVA.D85.095021, 1212.4100,1305.6266, 1401.6698}.
Also, there is a possibility of the inert doublet dark matter to be probed by indirect search \cite{astro-ph/0703512, 0811.1798, 0901.1750, 1306.4681}.
Thus, the inert doublet model is well motivated dark matter model in both theoretical and phenomenological points of view.

The direct detection experiments give an important constraint
on the inert doublet dark matter \cite{hep-ph/0603188, hep-ph/0607067, JCAP.0702.028}. 
At the leading order, the inert doublet dark matter scatters with the quarks at the tree level,
and with the gluon at the one-loop level by exchanging the SM Higgs boson.
These contributions to the cross section for scattering of the dark matter on nucleon
can be calculated in the same manner 
as the singlet scalar dark matter model \cite{Silveira:1985rk, McDonald:1993ex, Burgess:2000yq}.
It is proportional to $\l_A^2$, where $\l_A$ is the effective Higgs-dark matter coupling which is defined in Sec.~\ref{sec:model}.
If $\l_A$ is not so small,
they give dominant contribution to the cross section.
However, if the dark matter mass $m_A$ is around a half of the SM Higgs boson mass,
$\l_A$ should be suppressed because the SM Higgs boson $s$-channel exchange diagrams significantly contribute to
the annihilation cross section which determines the relic amount of the dark matter.
In this case, contributions which does not depends on $\l_A$ become
important for the spin-independent cross section.
For example, as shown in Ref.~\cite{PHRVA.D87.075025},
one-loop electroweak correction for the scattering with the light quarks gives an important correction.

In this paper, we revisit the radiative correction on the spin-independent
cross section in the inert two-Higgs doublet model for
the dark matter mass to be around a half of the Higgs boson mass.
In particular, Ref.~\cite{PHRVA.D87.075025} does not take into account for the effect of various scalar quartic couplings.
We take into account for the non-zero values of the inert doublet couplings,
which are equivalent to the mass difference between the dark matter and other $Z_2$ odd particles.
They cannot be neglected in a viable parameter region
in the light of the LEP II collider constraint \cite{hep-ph/0703056, PHRVA.D79.035013}.
In addition to them, there is an interesting coupling, namely
the self-coupling of the $Z_2$ odd particles, $\lambda_2$.
This coupling is irrelevant for the phenomenology at the tree level, but
we find it also plays the significant role here. 
Furthermore, we also evaluate contributions from twist-2 quark operators and two-loop diagrams of dark matter-gluon scattering.
These contributions give the same order
corrections as the scattering with quark at the one-loop level.

This paper is organized as follows.
We briefly review the inert two-Higgs doublet model in
Sec.~\ref{sec:model}. 
In Sec.~\ref{sec:sigmaSI}, we review the calculation of the spin-independent cross
section at the tree level, and introduce our strategy to incorporate the
loop corrections to it.
In Sec.~\ref{sec:result}, we show our result.
We conclude in Sec.~\ref{sec:conclusion}.
The details of the loop calculations are in the Appendices.

\section{Model} \label{sec:model}
In this section, we briefly review the inert doublet model.
In addition to the SM Higgs field $H$, we introduced a new $SU(2)_L$ doublet scalar field $\Phi$ with $Y=1/2$.
We impose $Z_2$ parity, under which the scalar fields behave as,
\begin{align}
   H \to H, \quad
\Phi \to -\Phi.
\end{align}
Other quark and lepton fields are also invariant under the $Z_2$ parity as the SM Higgs field.
Hence, $\Phi$ cannot have Yukawa interactions with the SM fermions.
The generic potential of $H$ and $\Phi$ under the $Z_2$ parity is,
\begin{align}
 - V(H, \Phi)
=&
- m_1^2 H^{\dagger} H
- m_2^2 \Phi^{\dagger} \Phi
- \lambda_1 (H^{\dagger} H)^2
- \lambda_2 (\Phi^{\dagger} \Phi)^2
\nonumber\\
&
- \lambda_3 (\Phi^{\dagger} \Phi) (H^{\dagger} H)
- \lambda_4 (\Phi^{\dagger} H) (H^{\dagger} \Phi)
- \left(
     \frac{\lambda_5}{2} (\Phi^{\dagger} H)^2 + h.c.
\right).
\end{align}
We assume that $\Phi$ does not get any vacuum expectation value (VEV),
then, the $Z_2$ parity which we have imposed is unbroken in the vacuum, and
$m_1^2$ is related to the Higgs VEV and the coupling $\l_1$ as,
\begin{align}
  m_1^2 =  - 2 \lambda_1 v^2,
\end{align}
where $v$ is the Higgs VEV, $v^2 = (\sqrt{2} G_F)^{-1} \simeq
(246\textrm{~GeV})^2$. $G_F$ is the Fermi constant.
Compared to the SM, we have additional five free parameters, $m_2^2$, $\l_2$, $\l_3$, $\l_4$ and $\l_5$.
For the stability of this potential, the following relations are required \cite{PHRVA.D18.2574}:
\begin{align}
\lambda_1>0, \qquad
\lambda_2>0, \qquad
\lambda_3> -2 \sqrt{ \lambda_1 \lambda_2}, \qquad
\lambda_3 + \lambda_4 - |\lambda_5| > -2 \sqrt{ \lambda_1 \lambda_2}.
\label{eq:potential_stability}
\end{align}
We can always take $\lambda_5$ as a real positive by a redefinition of the phase of $\Phi$ field.
For example, when $\arg \lambda_5 = \theta \neq 0$, we redefine $\Phi$ as $e^{i\theta/2}\Phi$.  
Therefore, the inert doublet Higgs does not contribute to $CP$ violation.
Hereafter we take a basis in which $\lambda_5$ is a real positive.
In this basis, we parametrize the component fields of $H$ and $\Phi$ as follows,
\begin{align}
 H=
\left(
\begin{matrix}
 -i \pi_W^{+} \\
\frac{v + h + i \pi_Z}{\sqrt{2}}
\end{matrix}
\right)
,
\quad
\Phi=
\left(
\begin{matrix}
 -i H^{+} \\
 \frac{S + i A}{\sqrt{2}}
\end{matrix}
\right),
\end{align}
where each component fields correspond to mass eigenstates.
We can find mass eigenvalues of each particles and interaction terms.
The mass eigenvalues are,
\begin{align}
m_{h}^2
=&
 2 v^2 \lambda_1
, \\
 m_{H^{\pm}}^2
=&
 m_2^2 + \frac{1}{2}\lambda_3 v^2
\label{eq:charged_mass}
, \\
m_{S}^2
=&
 m_2^2 + \frac{1}{2}(\lambda_3 + \lambda_4 + \lambda_5) v^2
\label{eq:nonDM_mass}
, \\
m_{A}^2
=&
 m_2^2 + \frac{1}{2}(\lambda_3 + \lambda_4 - \lambda_5) v^2
.
\end{align}
As we mentioned in the above, we take $\lambda_5 > 0$ in this paper,
hence $A$ is the lightest neutral $Z_2$ odd particle, and it is the dark matter candidate \footnote{
Some references assume $S$ is the lightest $Z_2$ odd particle.
However, this is just a difference of the basis of $\Phi$.
For example, if we define $\Phi' \equiv i\Phi$, we can see $S' = -A$ and $A'=S$.
Hence, there is no physical difference.
}. 

The three-point interaction terms for the Higgs boson and the $Z_2$ odd particles are,
\begin{align}
{\cal L} \ni &
-\frac{1}{2}(\lambda_3 + \lambda_4 - \lambda_5) v h A^2
- \lambda_3 v h H^{+} H^{-} 
- \frac{1}{2}(\lambda_3 + \lambda_4 + \lambda_5) v h S^2
.
\end{align}
The Higgs coupling to the dark matter is important to study dark matter
phenomenology, and it is proportional to $\lambda_3 + \lambda_4 -
\lambda_5$. So we denote it as
\begin{align}
 \lambda_A 
\equiv&
 \lambda_3 + \lambda_4 - \lambda_5
.
\end{align}
We also introduce other short-handed notations,
\begin{align}
\Delta m_{H^{\pm}}
\equiv&
 m_{H^{\pm}} - m_{A}
,
\\
\Delta m_{S}
\equiv&
 m_{S} - m_{A}
.
\end{align}
We treat ($m_A$, $\Delta m_{H^{\pm}}$, $\Delta m_{S}$,
$\lambda_A$) as input parameters and determined ($m_2^2$, $\lambda_3$,
$\lambda_4$, $\lambda_5$) from these input parameters.
Note that $\lambda_2$ is not related with these input parameters, and
irrelevant for the analysis at tree level. However, $\lambda_2$ plays an
important role at the loop level as we will see later.
The loop correction to the dark matter mass is small for the light dark
matter mass regime~\cite{1303.3010}, so we keep using the above tree
level relations among the mass and couplings in this paper.

In the following of this paper, we assume almost all of the energy density of the dark matter is comprised of
the inert doublet dark matter which is generated as a thermal relic.
The amount of thermal relic is controlled by the annihilation cross section of the dark matter \cite{PRLTA.39.165, NUPHA.B310.693, NUPHA.B360.145}.
There are some comprehensive studies on
viable parameter regions \cite{JCAP.0702.028, PHRVA.D80.055012, 1107.1991, 1303.3010, JCAP.1406.030, Abe:2014gua}.
Because of its $SU(2)_L$ charge, $AA\to WW^{(*)}$ channel gives a significant contribution to
the annihilation cross section for the case of $m_A \gtrsim m_W$ \cite{1003.3125},
and it tends to be too large to obtain the correct abundance $\Omega_{\rm DM} h^2 = 0.1196 \pm 0.0031$ \cite{Ade:2013zuv}.
It is known that there are two parameter regions to obtain the correct relic abundance \cite{JCAP.1406.030, Abe:2014gua}.
One region is the light mass region with $m_A \lesssim 72~{\rm GeV}$,
in which $AA\to WW^*$ 
becomes less significant
because it is well below energy threshold of two body $WW$ mode.
The other region is the heavy mass region with $m_A \gtrsim 600~{\rm GeV}$,
in which the annihilation cross section is suppressed by its mass\footnote{
Ref.~\cite{JCAP.1101.002} pointed out another parameter region in which some of diagrams of $AA\to WW$ cancel out.
However, this parameter region is severely constrained by the LUX experiment. See, Ref.~\cite{JCAP.1406.030}.}.

Since the inert doublet dark matter couples with
the SM Higgs field via the coupling $\l_A$,
 the dark matter can scatter with nucleus and the direct detection experiment gives
an important constraint on the coupling $\l_A$ \cite{hep-ph/0603188, hep-ph/0607067, JCAP.0702.028}.
Especially, this constraint gives a large impact on the light mass region.
This is because the amount of the relic abundance is also controlled by the
same coupling. 
%
%
As a result, the region with $m_A \lesssim 53~{\rm GeV}$ is already excluded by the LUX experiment,
and viable region in the light mass range is $53~{\rm GeV} \lesssim m_A \lesssim 72~{\rm GeV}$ \cite{JCAP.1406.030, Abe:2014gua}.
In this viable range, although the coupling $\l_A$ is small,
the annihilation cross section is enhanced because of the propagator of
the SM Higgs boson in $s$-channel.
However, the scattering of a nucleon and a dark matter
does not hit the SM Higgs pole, and thus the
spin-independent cross section is just suppressed by the coupling $\l_A$.
Therefore
the contributions which is independent of $\l_A$,
\textit{i.e.}, the radiative corrections on the spin-independent cross
section becomes important in this mass range.

\section{Spin-independent cross section}\label{sec:sigmaSI}
In this section, we formulate how to include 
radiative
 corrections to the spin-independent cross section. 
To calculate the cross section of elastic scattering of dark matter and nucleon,
first, we construct the effective interaction of the dark matter and quark/gluon.
The relevant terms for our calculation are written as,
\begin{align}
 {\cal L}_{\rm eff.}
=&
\frac{1}{2} \sum_{q=u,d,s} \Gamma^{q} A^2 (m_q \bar{q}q)
- \frac{1}{2} \frac{\alpha_s}{4 \pi} \Gamma^{G} A^2 G^{a}_{\mu \nu} G^{a \mu \nu}
\nonumber\\
&
+ \frac{1}{2 m_A^2} \sum_{q = u,d,s,c,b} \left[
      (\partial^{\mu}A) (\partial^{\nu}A) \Gamma^q_{\text{t2}} {\cal O}^{q}_{\mu \nu}
      - A  (\partial^{\mu} \partial^{\nu}A) \Gamma'^q_{\text{t2}} {\cal O}^{q}_{\mu \nu}
\right],
\label{eq:eff_int}
\end{align}
where ${\cal O}^q_{\m\n}$ is the quark twist-2 operator which is defined as, 
\begin{align}
{\cal O}^{q}_{\mu \nu}
\equiv
 \frac{i}{2} \bar q \left( \q_\m \g_\n + \q_\n \g_\m - \frac{1}{2}g_{\m\n} \sla{\q} \right) q.
\end{align}
In the effective Lagrangian given in Eq.~(\ref{eq:eff_int}), we neglect higher twist gluon operators
because their contributions are suppressed by $\a_s$ compared to the twist-0 gluon operator \cite{Hisano:2010ct}. 
The coefficients $\G$ are determined by matching with UV Lagrangian, which will be explained later.
To calculate the scattering amplitude of nucleon, we also need matrix elements of quark/gluon operators,
which are given as,
\begin{align}
\langle N | m_q \bar q q |N \rangle &= m_N f_q, \\
-\frac{9\a_s}{8\pi} \langle N | G_{\m\n}^a G^{a\m\n} |N \rangle &= m_N f_g, \\
\langle N | {\cal O}_{\m\n}^q |N \rangle &= \frac{1}{m_N} \left( p_\m p_\n - \frac{1}{4} m_N^2 g_{\m\n}\right)(q(2) + \bar q(2) ).
\end{align}
$f_g$ is related to $f_q$ as,
\begin{align}
f_g = 1-\sum_{q=u,d,s} f_q.
\end{align}
This relation is derived by using the relation
obtained from the trace anomaly~\cite{Shifman:1978zn}, 
\begin{align}
m_N
=
\langle N | T^{\mu}_{\mu} | N \rangle
=&
-
\frac{9 \alpha_s}{8 \pi}
\langle N | G_{\mu \nu}^a G^{ a \mu \nu} | N \rangle
+ \sum_{q=u,d,s} \langle N | m_q \bar{q} q | N \rangle.
\label{eq:trace-anomaly}
\end{align}
From this discussion, we can see $\langle N | m_q \bar q q | N \rangle$ and 
$(\a_s / 4\pi) \langle N | G^a_{\m\n}G^{a\m\n} | N \rangle$ are same order.
Thus, 
the calculation at 
the $n$-loop order requires 
the ($n+1$)-loop order
calculation for diagrams with $G_{\mu \nu}^a G^{a \mu \nu}$. 
For $q(2)$ and $\bar{q}(2)$,
we can see that they are the second moments of the quark and anti-quark parton distribution functions
by using a discussion of operator product expansion as\footnote{
For example, see section 18.5 in Peskin-Schroeder's textbook \cite{Peskin:1995ev}.
},
\begin{align}
 q(2) + \bar{q}(2) =& \int_0^1 dx ( q(x) + \bar{q}(x)).
\end{align}
We use the CTEQ parton distribution functions \cite{Pumplin:2002vw} to evaluate them, and use the same value used in \cite{Hisano:2010fy}.

We have checked that the spin-independent cross section of
a dark matter and a proton is the almost same as of the
a dark matter and a neutron. Their difference is smaller than 
a few percent in almost all of the parameter region.
In the following of this paper, we calculate the scattering cross section of a dark matter and a neutron.
The matrix elements which are used are summarized in Tab.~\ref{tab:matrixelements}.
\begin{table}
\centering
\begin{tabular}{|c|c|}
\hline
$f_u$ & 0.0110 \\ 
$f_d$ & 0.0273 \\ 
$f_s$ & 0.0447 \\ 
\hline
\end{tabular}
    \qquad
\begin{tabular}{|c|c||c|c|}
\hline
$u(2)$ & 0.11 &
$\bar u(2)$ & 0.036 \\
$d(2)$ & 0.22 &
$\bar d(2)$ & 0.034 \\
$s(2)$ & 0.026 &
$\bar s(2)$ & 0.026 \\
$c(2)$ & 0.019 &
$\bar c(2)$ & 0.019 \\
$b(2)$ & 0.012 &
$\bar b(2)$ & 0.012 \\
\hline
\end{tabular}
\caption{
Matrix elements for neutron.
Left panel shows the matrix elements for quark twist-0 operators, which are taken from the default values of \texttt{micrOMEGAs} \cite{Belanger:2013oya}.
Right panel shows the second moments for quark distribution function,
which are evaluated at the scale of $\m=m_Z$ by using the CTEQ parton distribution functions \cite{Pumplin:2002vw}.
}\label{tab:matrixelements}
\end{table}
By using the above matrix elements and the coefficients $\G$'s
in the effective interaction given in Eq.~(\ref{eq:eff_int}),
the scattering amplitude of the nucleon and the dark matter is given as,
\begin{align}
 i {\cal M}
=&
i m_N \left[
     \sum_q \Gamma^q f_q
     + \frac{2}{9} \Gamma^G f_g
     + \frac{3}{4} \sum_q (\Gamma^q_{\text{t2}} + {\Gamma'}^q_{\text{t2}} )(q(2)+\bar q(2) )
\right],\\
\label{eq:M-for-xsecSI}
\sigma_{\text{SI}} =& \frac{\m^2}{4\pi m_A^2} |{\cal M}|^2,
\end{align}
where $\m$ is the reduced mass, which is defined as $\m \equiv m_N m_A / (m_N + m_A)$.
Hence, 
what we have to calculate is the effective coupling $\G$'s.

\subsection{At the leading order} \label{sec:3.1}
\begin{figure}[tb]
\subfigure[]{
 \includegraphics[width=0.23\hsize]{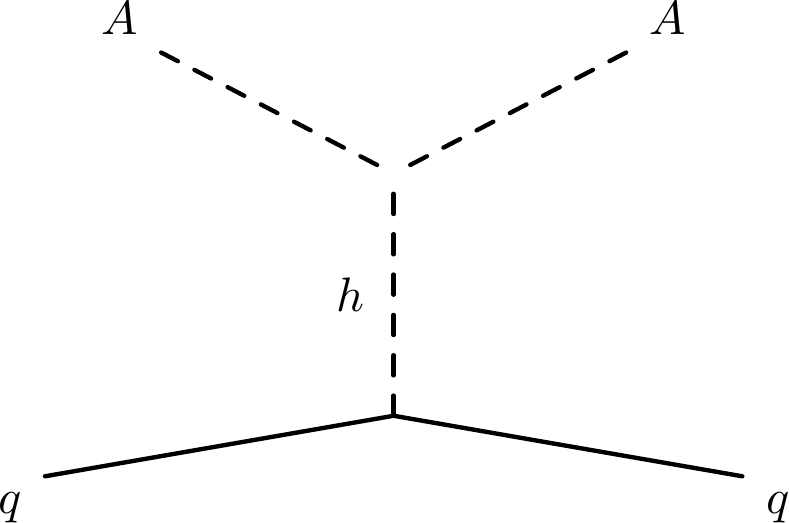} 
\label{fig:LO_qq}
}  
\subfigure[]{
\qquad
 \includegraphics[width=0.23\hsize]{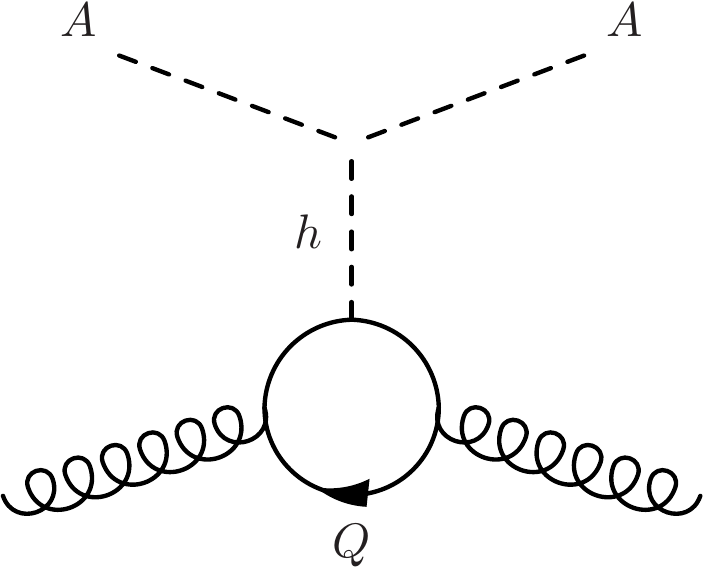} 
\label{fig:LO_GG}
}
\caption{
The diagrams which contribute to the spin-independent cross section at the leading order.
}
\label{fig:LO-diagrams}
\end{figure}

We start to give a brief review on the calculation at the 
leading order.
We need to calculate the elastic scattering cross section for the dark
matter and nucleon system, $\sigma(\textrm{DM} \ N \to \textrm{DM} \ N)$, where $N$ stands for the nucleon.
As described before, we construct the effective Lagrangian with the gluon and the light quarks $q=u,d,s$
by integrating out the heavy quarks $Q=c,b,t$ and the SM Higgs boson.
We should take into account the one-loop diagrams for the scattering with gluon,
because their contributions are same order as the tree-level scattering with the light quarks.
The dark matter scatters with the SM quarks at the tree level
and the gluon at the one-loop level as shown in Fig.~\ref{fig:LO_qq} and \ref{fig:LO_GG}, respectively.
Their amplitudes are proportional to the effective Higgs-dark matter coupling $\l_A$.
From these processes, the following relevant operators for the spin-independent
cross section are generated,
\begin{align}
 A^2 \bar{q}q, \quad 
 A^2 G^a_{\m\n} G^{a\m\n}.
\end{align}
%
The coefficients of the effective Lagrangian given at the leading order is determined as,
\begin{align}
\G^q = \G^G = \frac{\l_A}{m_h^2},\qquad
\G^q_{\text{t2}} = {\G'}^q_{\text{t2}} = 0.
\end{align}
Using these coefficients and Eq.~(\ref{eq:M-for-xsecSI}),
we can calculate the amplitude of the process and the spin-independent cross section as,
\begin{align}
\sigma_{\text{SI}}
=&
\frac{1}{4\pi} \frac{\l_A^2 \m^2 m_N^2 f_N^2}{m_A^2 m_h^4}, \label{eq:xsec_tree}
\end{align}
where,
\begin{align}
 f_N \equiv {2\over 9} + {7\over 9}\sum_q f_q.
\end{align}

\subsection{At the next leading order}
We move to calculate the loop corrections to the spin-independent cross section.
We need to consider the loop corrections to the four relevant operators for the spin-independent cross section, 
\begin{align}
A^2\bar{q}q,\quad
A^2G_{\mu \nu}^a G^{a \mu \nu},\quad
(\q^\m A) (\q^\n A) {\cal O}_{\m\n}^q,\quad
A (\q^\m \q^\n A) {\cal O}_{\m\n}^q.
\end{align}
%
There are some remarks on this calculation.
First, trace anomaly relation Eq.~(\ref{eq:trace-anomaly}) is suffered from QCD correction at the next-leading order.
However, we consider $\l_A$ is not so large, and assume corrections of the order of $\l_A \a_s/4\pi$ can be neglected.
Also, for the contribution which is independent of $\l_A$, we only take into account the leading order of $\a_s$.
Thus, for the scattering with the gluon,
we can still use Eq.~(\ref{eq:trace-anomaly}) even in the loop level calculation.
Second, we evaluate the effect of twist-2 operator ${\cal O}^q_{\m\n}$ at the scale $\m=m_Z$.
Thus, we take into account $q=u,d,s,c$ and $b$ and evaluate the matrix element of ${\cal O}^q_{\m\n}$ by using 
the parton distribution functions at $\m=m_Z$.

\begin{figure}[tb]
\subfigure[]{
 \includegraphics[width=0.23\hsize]{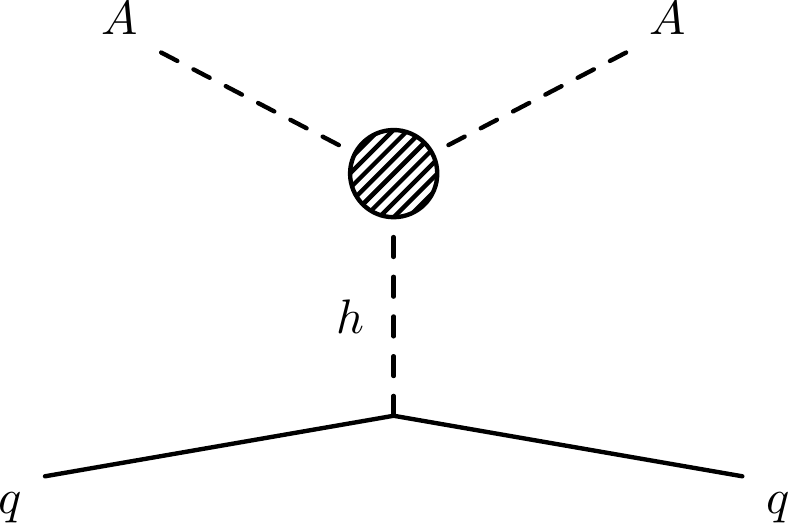} 
\label{fig:2loop1}
}  
\qquad
\subfigure[]{
 \includegraphics[width=0.23\hsize]{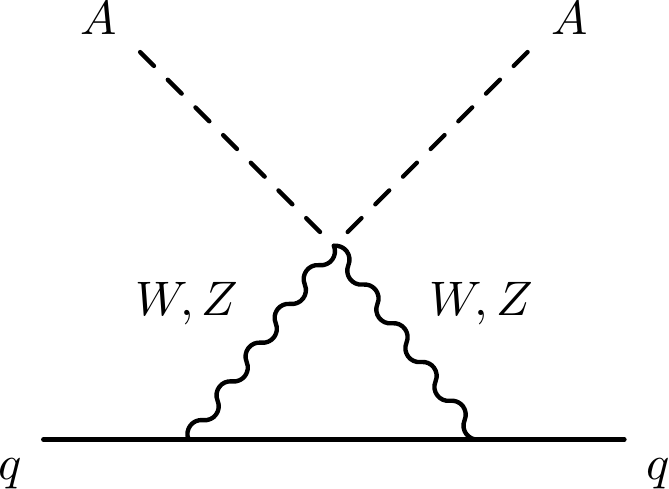} 
\label{fig:2loop2}
}  
\qquad
\subfigure[]{
 \includegraphics[width=0.23\hsize]{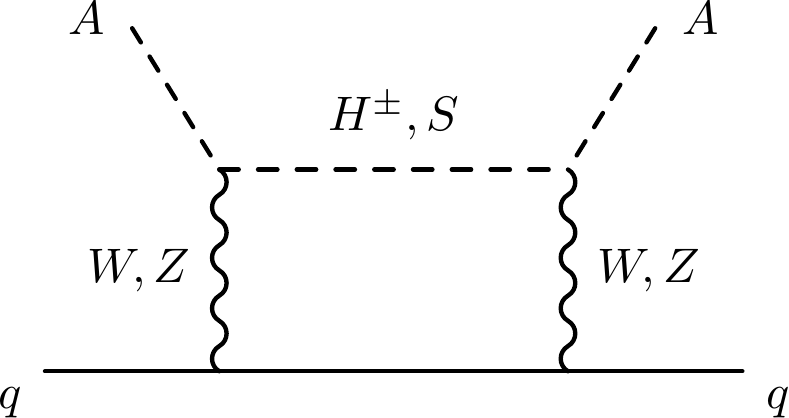} 
\label{fig:2loop3}
}  
\qquad
\subfigure[]{
 \includegraphics[width=0.23\hsize]{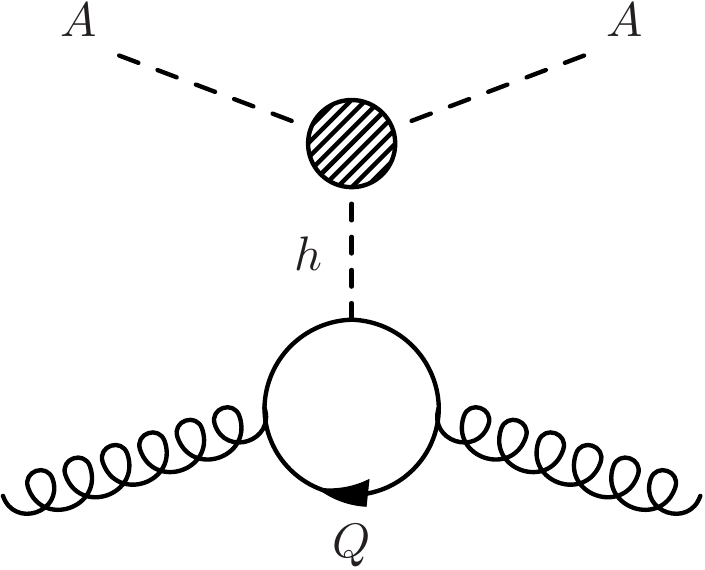} 
\label{fig:2loop4}
}  
\qquad
\subfigure[]{
 \includegraphics[width=0.23\hsize]{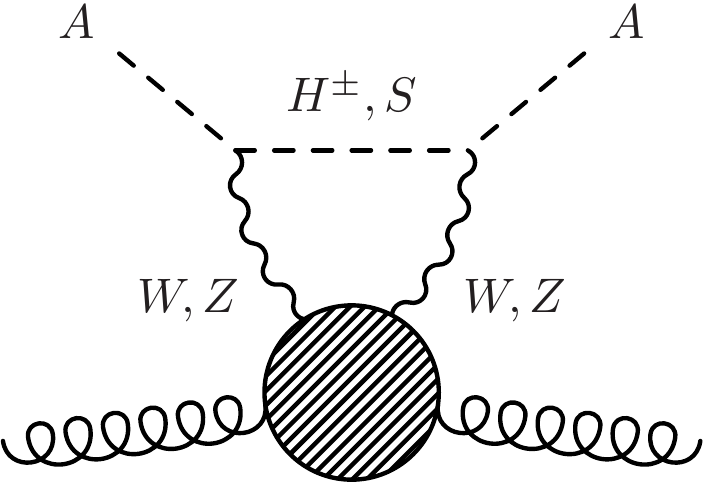} 
\label{fig:2loop5}
}  
\qquad
\subfigure[]{
 \includegraphics[width=0.23\hsize]{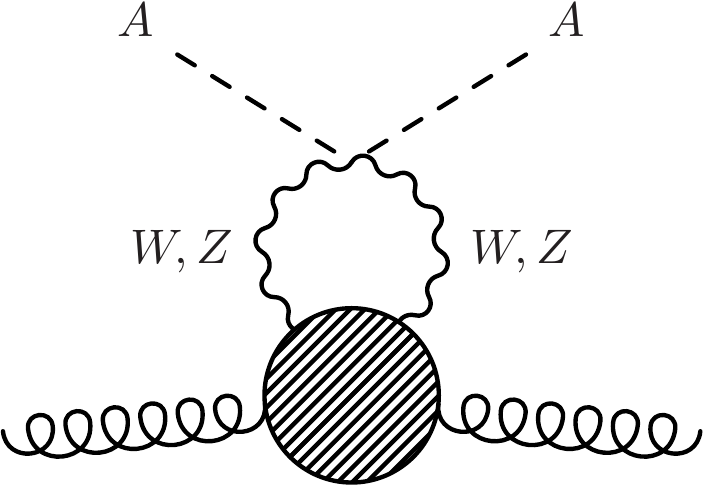} 
\label{fig:2loop6}
}  
\caption{
The diagrams we calculate. The shaded region is one-loop correction.
}
\label{fig:NLO-diagrams}
\end{figure}

The diagrams we need to calculate are shown in
Fig.~\ref{fig:NLO-diagrams}. The diagrams with gluons are 
two-loop
diagrams but contribute to the spin-independent cross section as 
the one-loop
order correction as we mentioned in Sec.~\ref{sec:3.1}. There are
some diagrams which are the same order but not shown
in Fig.~\ref{fig:NLO-diagrams}. They are proportional to the
Higgs coupling to the dark matter, $\lambda_A$. We are interested in the
case that this coupling is 
very small. Thus the diagrams with this coupling 
give much smaller
contributions than the diagrams shown in Fig.~\ref{fig:NLO-diagrams},
and do not need to be calculated.
Here we parametrized the loop corrections to the $\lambda_A$ as $\delta \Gamma_h(q_h^2)$,
and denote the correction from the box 
and triangle 
diagrams as $\Gamma^q_{\textrm{Box}}$.
Here $q_h^2$ 
is the momentum squared of the Higgs boson.
What we need is the scattering amplitude in the non-relativistic limit.
In 
the 
limit of zero momentum transfer,
the amplitudes of the diagrams given in Fig.~\ref{fig:NLO-diagrams} are written as,
\begin{align}
 \text{Fig.~}\ref{fig:2loop1} &= \frac{i\d\G_h(0)}{m_h^2} m_q \bar u u, \label{eq:fig2a}\\
 \text{Fig.~}\ref{fig:2loop2}
+\text{Fig.~}\ref{fig:2loop3} &= i\G_{\text{Box}}^q m_q \bar u u
                                 + \frac{i}{m_A^2}(\G^q_{\text{t2}} + {\G'}^q_{\text{t2}} ) \bar u \left( (pq) \sla{p} - \frac{1}{4} p^2 \sla{q} \right) u, \label{eq:fig2bc}\\
 \text{Fig.~}\ref{fig:2loop4} &= \frac{i\d\G_h(0)}{m_h^2} \times \frac{2}{9}\left( -\frac{9\a_s}{8\pi} G_{\m\n}^a G^{a\m\n} \right), \label{eq:fig2d}\\
 \text{Fig.~}\ref{fig:2loop5}
+\text{Fig.~}\ref{fig:2loop6} &= i\G^G_{\text{Box}} \times \frac{2}{9}\left( -\frac{9\a_s}{8\pi} G_{\m\n}^a G^{a\m\n} \right) \label{eq:fig2ef}.
\end{align}
In Eq.~(\ref{eq:fig2bc}), $p^\m$ and $q^\m$ is momentum of the dark matter and quark, respectively.
We have used equation of motion of quark, $\sla{q} u = m_q u$.
$\G^q_{\text{t2}} + {\G'}^q_{\text{t2}}$ can be read from the above amplitudes,
and $\G^q$ and $\G^G$ is determined as,
\begin{align}
\G^q = \frac{\d \G_h(0)}{m_h^2} + \G_{\text{Box}}^q,\qquad
\G^G = \frac{\d \G_h(0)}{m_h^2} + \G_{\text{Box}}^g.
\end{align}
Here we treat the gluon field as the background field and
neglect its higher twist operators.
For the detail of the calculation of $\G$'s, see the Appendices.

\begin{figure}[tb]
 \includegraphics[width=0.24\hsize]{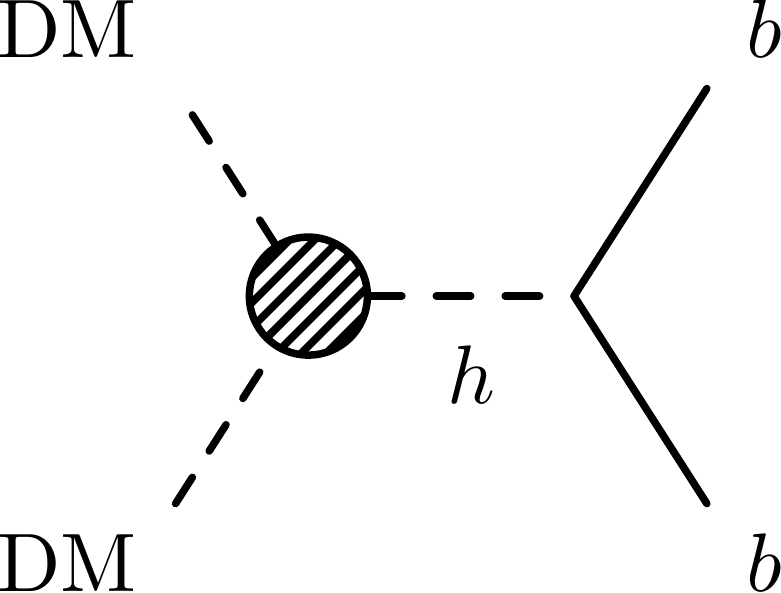} 
\caption{
The diagram giving the dominant contribution in the relic abundance
 calculation for $m_{DM} \simeq m_h/2$. The shaded region contains tree
 and loop corrections. Other diagrams, such as box diagrams, give a
 small correction to this diagrams for $m_{DM} \simeq m_h/2$. 
}
\label{fig:xsec_abundance}
\end{figure}

We need to discuss how to calculate the value of $\lambda_A$ and renormalization condition. 
In the tree level calculation, we set this coupling to reproduce the
current relic abundance of the dark matter in our universe.
Now we need to take into account the one-loop effect. 
Since our focus is $m_A \simeq m_h/2$ regime, the dominant
contribution for the relic abundance calculation is coming from the
diagram shown in Fig.~\ref{fig:xsec_abundance} because this diagram
picks up the Higgs resonance. Hence
it is only the vertex correction that we should take into account, and we can
ignore other one-loop 
corrections,
such as box diagrams, in the relic
abundance calculation. 
Therefore we can set $\lambda_A$ by the following relation,
\begin{align}
\left| \lambda_A + \delta \Gamma_h(m_h^2) + \delta_{\lambda_A} \right|^2
= \left| \lambda_{\text{relic}} \right|^2,
\label{eq:relic_coup}
\end{align}
where 
$\delta_{\lambda_A}$ is the counter-term.
$\lambda_{\text{relic}}$ is the effective Higgs boson coupling to
the dark matter, and is determined as to reproduce 
the
correct relic abundance.
 Since the annihilation cross 
section determine the relic abundance, the square of the couplings appear
in the relation above. 
Thus, we have two solution for $\lambda_A$,
\begin{align}
\l_A = \pm |\l_{\text{relic}}| - \d \G_h(m_h^2) -\delta_{\lambda_{A}}.
\label{eq:lamrelic}
\end{align}
This is crucial in $\sigma_{\text{SI}}$ calculation at the loop
level because  
there is interference between the tree and the loop
diagrams as we can see in 
Eq.~(\ref{eq:M-for-xsecSI}). Depending on the sign in Eq.~(\ref{eq:lamrelic}), the interference is destructive or constructive,
and we find two solutions for $\sigma_{\text{SI}}$. This point was overlooked in Ref.~\cite{PHRVA.D87.075025}.
Now the value of $\lambda_A$ is set by Eq.~(\ref{eq:lamrelic}).
It is useful to renormalize $\l_A$ to make that 
$\d \G_h(m_h^2) = - \delta_{\lambda_A} $ is satisfied.
By using this condition, we can take $\l_A$ as $\pm |\l_{\text{relic}}|$.

 We would like to mention on the stability condition here. Since $\lambda_A =
 \pm |\lambda_{\text{relic}}|$, there are two parameter sets for
 $(\lambda_3, \lambda_4, \lambda_5)$ for each
 $\lambda_{A}$. These parameter sets have to
 satisfy the stability condition given in Eq.~(\ref{eq:potential_stability}).
For 
53~GeV $< m_{\text{DM}} <$ 71~GeV, 
100~GeV $< m_{S} <$ 250~GeV, 
and 100~GeV $< m_{H^{\pm}} <$ 250~GeV, 
we find the first three conditions in Eq.~(\ref{eq:potential_stability})
 are always satisfied, and the last one is satisfied if $\lambda_2
 \gtrsim 0.001$. This constraint on $\lambda_2$ is very weak and almost
 harmless. 
%
%
%
%

It is useful to define ``effective coupling'' $\l_A^{\rm eff.} \equiv \l_A + \d\l$ which is relevant for $\s_{\rm SI}$,
where $\d \l$ is defined as,
\begin{align}
\delta \lambda
\equiv&
\delta \Gamma_h(0)
 + \delta_{\lambda_A}
+ \frac{m_h^2}{f_N} \left( \sum_q \Gamma^q_{\text{Box}} f_q \right)
+ \frac{2}{9} \frac{m_h^2}{f_N} \Gamma^G_{\text{Box}} f_g
+ \frac{3}{4} \frac{m_h^2}{f_N} \sum_q ( \G^q_{\text{t2}} + {\G'}^q_{\text{t2}} )(q(2)+\bar q(2)). \label{eq:def_delta_lambda}
\end{align}
Note that we determined $\delta_{\lambda_A} = - \d \G_h(m_h^2)$
in the previous paragraph.
By using $ \l_A^\text{eff.} \equiv \l_A + \d \l$, 
the spin-independent cross section at the next-leading order is written
in the similar way as the tree level formula
Eq.~(\ref{eq:xsec_tree}), 
\begin{align}
\s_{\text{SI}} 
= 
 \frac{1}{4\pi} \frac{(\l_A^\text{eff.})^2 \m^2 m_N^2 f_N^2}{m_A^2
 m_h^4} 
 = 
 \frac{1}{4\pi}  \frac{(\pm |\l_\text{relic}| + \delta \lambda)^2 \m^2
  m_N^2 f_N^2}{m_A^2  m_h^4}.
\label{eq:xsec_eff}
\end{align}
In the next section, we show our numerical results by using the relation
we find in this section. The analytic expressions and the details of the
calculation are in the appendix. 

When $m_{\text{DM}} > m_h/2$, it is kinematically forbidden to
hit the pole of the Higgs propagator, and the enhancement of the cross
section due to the Higgs resonance does not happen. The
dominant contribution to the dark matter annihilation cross section does
not come from $\sqrt{s} = m_h^2$ but from $\sqrt{s} \simeq 4
m_{\text{DM}}^2 > m_h^2$ for $m_{\text{DM}} > m_h/2$.
Therefore we replace $\delta \Gamma_h(m_h^2)$ in the
above equations into $\delta \Gamma_h(4 m_{\text{DM}}^2)$ for
$m_{\text{DM}} > m_h/2$. 
\section{Results}\label{sec:result}

\begin{figure}[tb]
 \includegraphics[width=0.6\hsize]{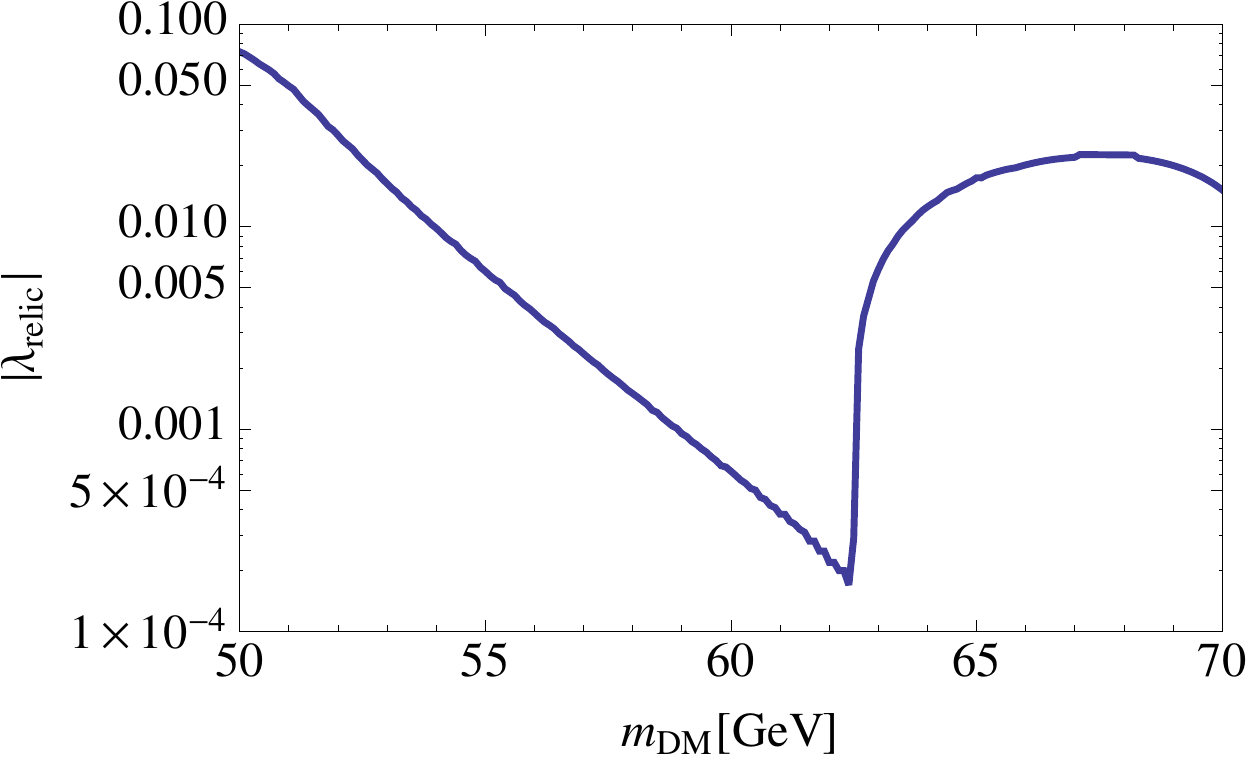} 
\caption{
The absolute value of the effective couplings as a function of the dark
 matter mass for  $m_{H^{\pm}}=m_S=m_{DM}+ 50$~GeV.  
This coupling is determined so as to reproduce the correct relic
 abundance, and is the same as the $\lambda_A$ determined at the tree
 level analysis.
}
\label{fig:LO-coupling}
\end{figure}

We start by showing the tree level result on $\lambda_A$ to find
the mass region in which the loop correction becomes significant.
In Figure~\ref{fig:LO-coupling}, 
we show the absolute value of the Higgs boson coupling to the dark
matter, $\lambda_A$ at the tree level as a function of the dark matter
mass. This coupling is determined by requiring to reproduce the current
relic abundance of the dark matter in our universe, and is the same as
$|\lambda_{\textrm{relic}}|$ defined in Eq.~(\ref{eq:relic_coup}). 
It is calculated by using \texttt{micrOMEGAs} \cite{Belanger:2013oya}. 
Since we are 
interested in the small coupling regime, 
we focus on $53$~GeV $< m_{\textrm{DM}} < 64$~GeV.
In this plot, we take $\Delta m_{H^{\pm}} =
\Delta m_{S} = 
50$~GeV, but these parameter dependence is very week as 
long as the mass difference is large enough to ignore the
co-annihilation process, namely $\Delta m_{S, H^{\pm}} \gtrsim
20$~GeV. 
%
%
%

\begin{figure}[tb]
\includegraphics[width=0.50\hsize]{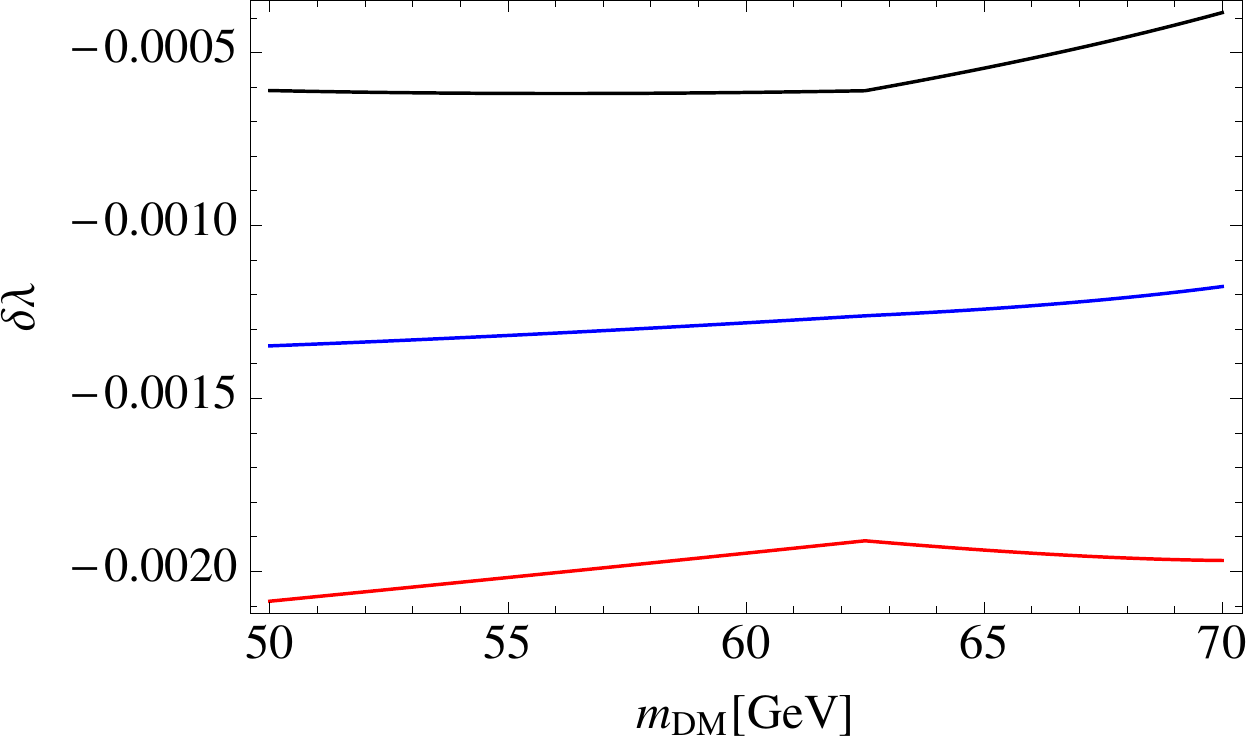} 
\caption{
The value of $\delta \lambda$ defined in
 Eq.~(\ref{eq:def_delta_lambda}).
The red, blue, and black lines are for $\lambda_2=0, 0.5$, and $1.0$,
 respectively. Here we fixed $m_{H^{\pm}}-m_{DM} = m_{S}- m_{DM} = 50$~GeV.
}
\label{fig:delta_lambda}
\end{figure}

We move to discuss on the effect of the loop correction.
We show the value of $\delta \lambda$  for
$\Delta m_{H^{\pm}} = \Delta m_{S} = 50$~GeV in
Fig.~\ref{fig:delta_lambda}.
The three lines correspond to the different $\lambda_2$ choices.
We find $\d\l$ is the order of $10^{-3}$.
Thus, the radiative correction becomes important for $|\lambda_{\text{relic}}| \lesssim {\cal O}(10^{-3})$,
namely $55~\text{GeV} \lesssim m_{\text{DM}} \lesssim 63~\text{GeV}$, where the tree level coupling is
comparable or even smaller than the one-loop level value as we can see
from Fig.~\ref{fig:LO-coupling}.

\begin{figure}[tb]
\includegraphics[width=0.30\hsize]{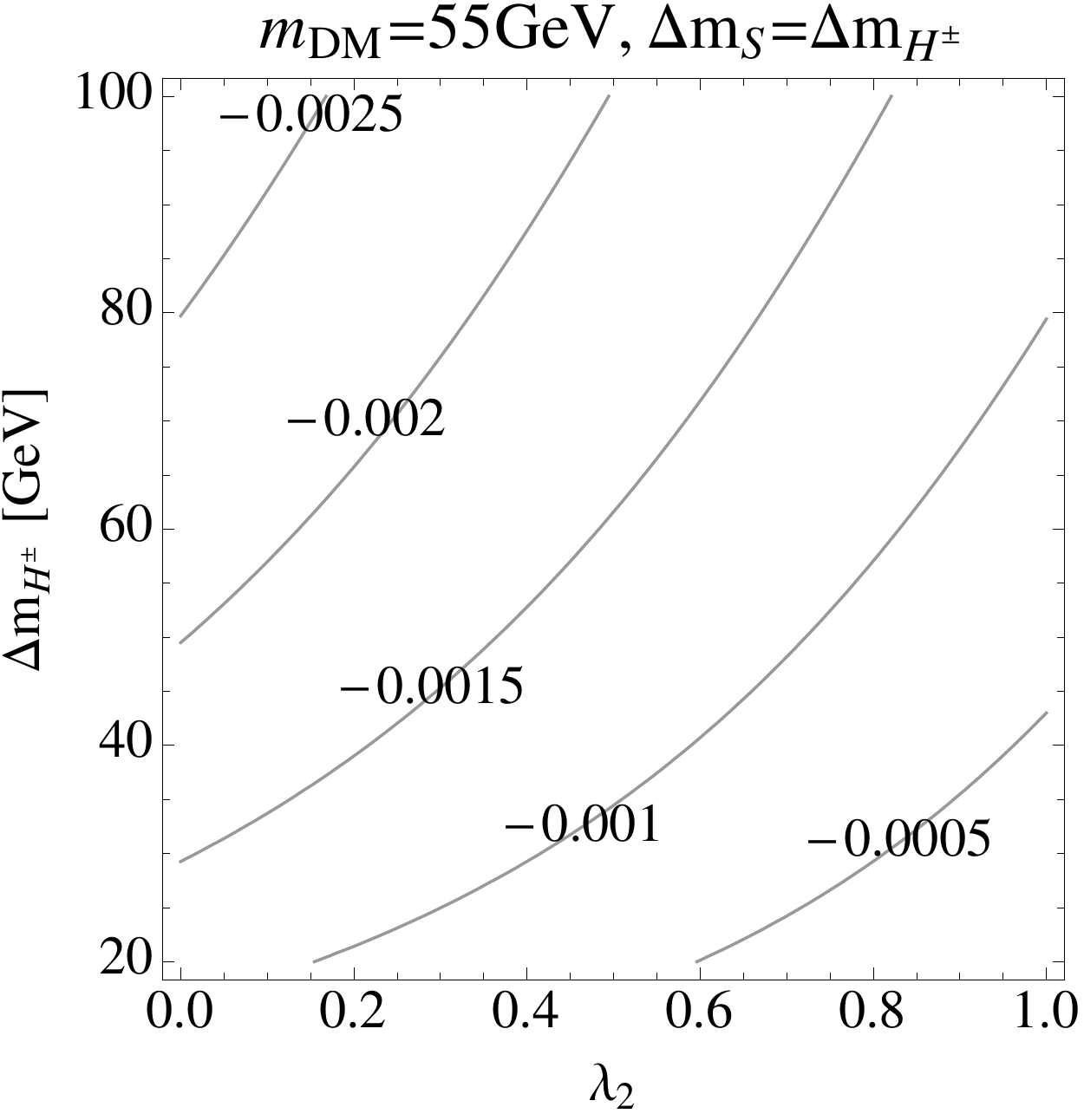} 
\quad
\includegraphics[width=0.30\hsize]{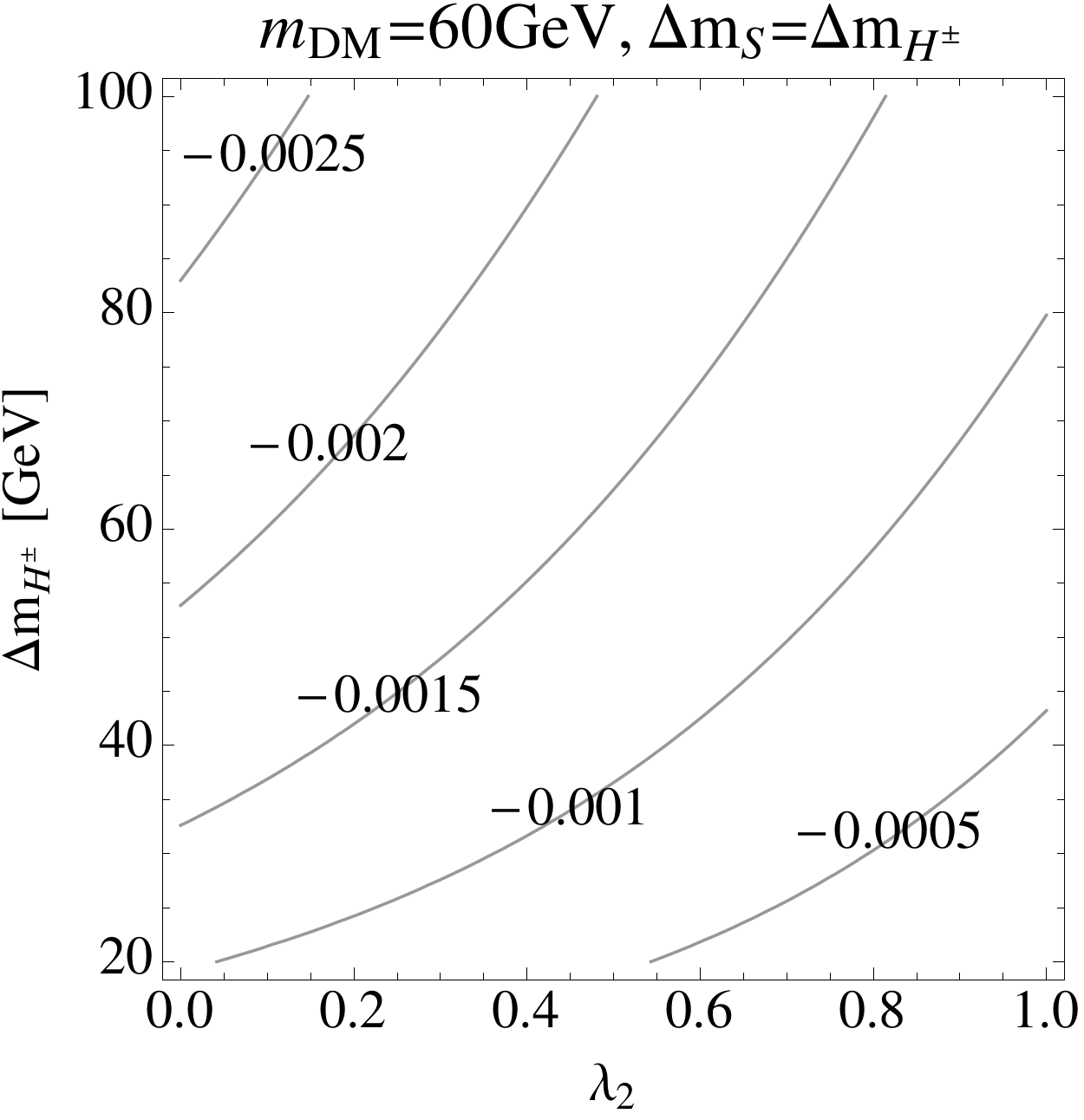} 
\quad
\includegraphics[width=0.30\hsize]{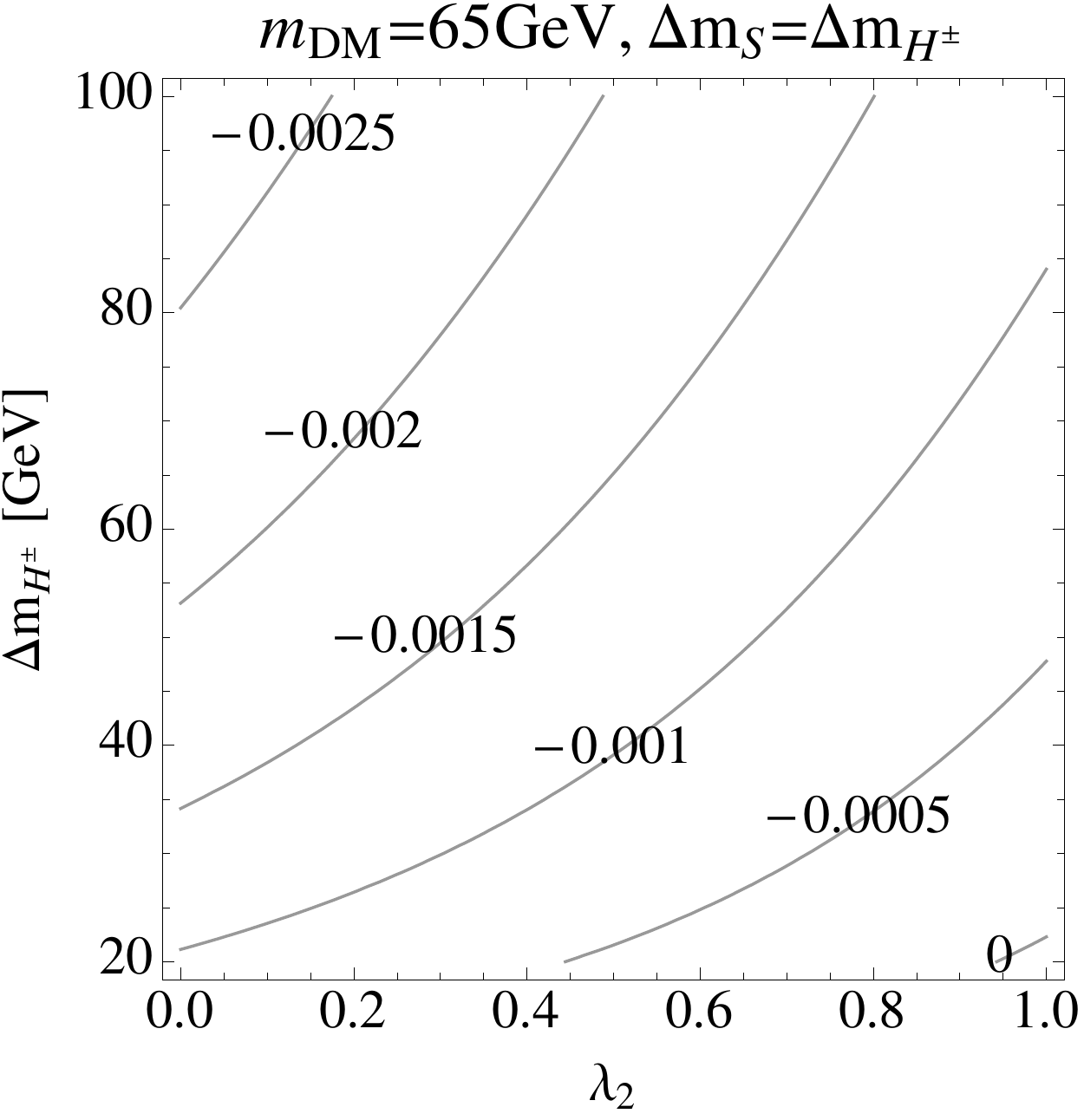} 
\caption{
The value of $\delta \lambda$ as a function of $\lambda_2$ and mass
 difference between the dark matter and other $Z_2$ odd particles.
 Here we take $m_{H^{\pm}} = m_{S}$. The dark matter mass of each panels are
$m_{DM}=$55~GeV (left), 60~GeV (middle), and 65~GeV (right).
}
\label{fig:lam2_delM}
\end{figure}

Now $\delta \lambda$ depends on the four parameters, 
$\lambda_2$, $m_{\text{DM}}$, $\Delta m_{H^{\pm}}$, $\Delta m_{S}$.
We show these parameter dependence of $\delta\lambda$ in
Fig.~\ref{fig:lam2_delM}. Here we take $m_{H^{\pm}} = m_{S}$. This
parameter choice enhances the custodial symmetry in $Z_2$ odd sector and
suppress the contributions to the $T$ parameter from $Z_2$ odd sector.
We find that $\delta\lambda$ weakly depends on $m_{\text{DM}}$, and is sensitive to the value of $\Delta m_{S, H^{\pm}}$ and
$\lambda_2$. The dependence on $\Delta m_{S, H^{\pm}}$ is contrast to the
tree level analysis where $|\lambda_{\text{relic}}|$ is almost independent
from $\Delta m_{S, H^{\pm}}$ as long as $\Delta m_{S, H^{\pm}} \gtrsim 20$~GeV. 
Another feature is the larger $\lambda_2$ makes $\delta \lambda$ to be
zero. This means the terms proportional to $\lambda_2$ cancel the other
loop contributions.
%
%
%

\begin{figure}[tb]
\centering
\includegraphics[width=0.30\hsize]{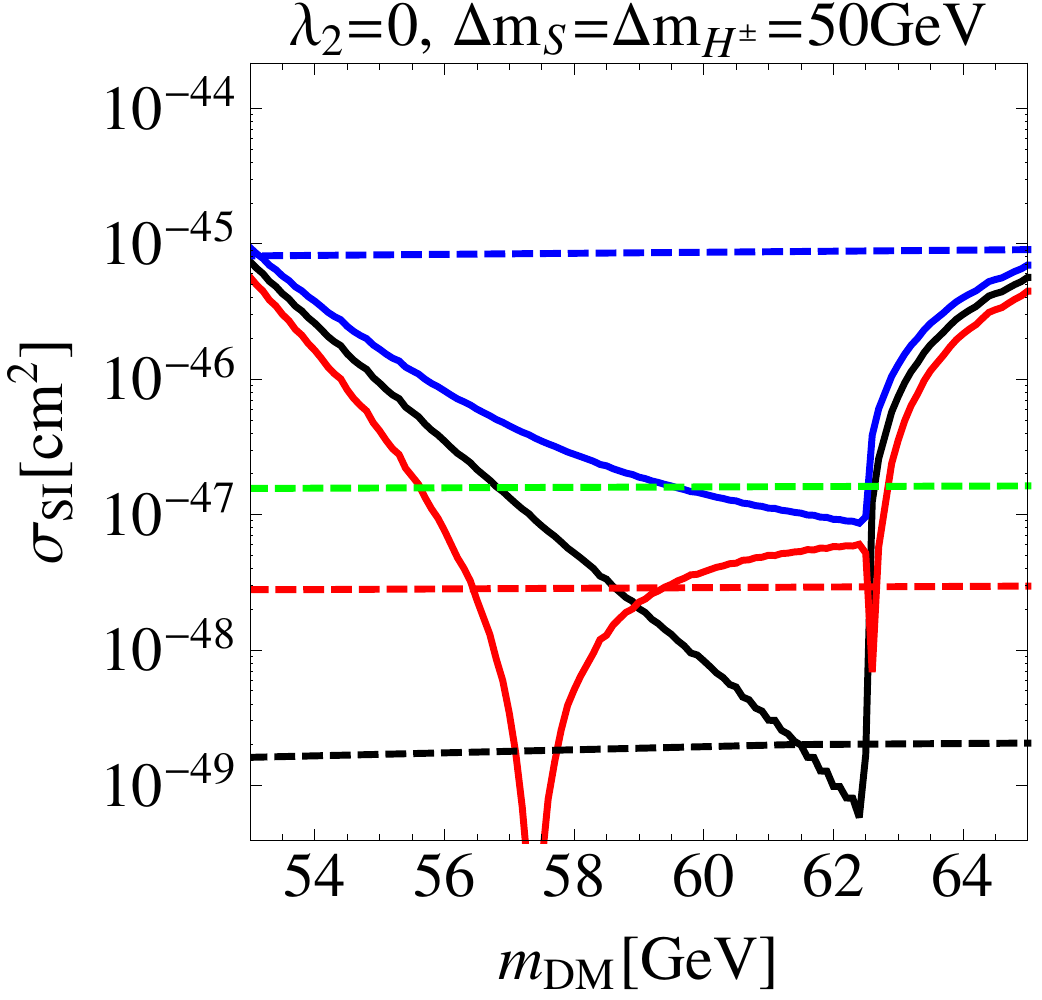} 
\quad
\includegraphics[width=0.30\hsize]{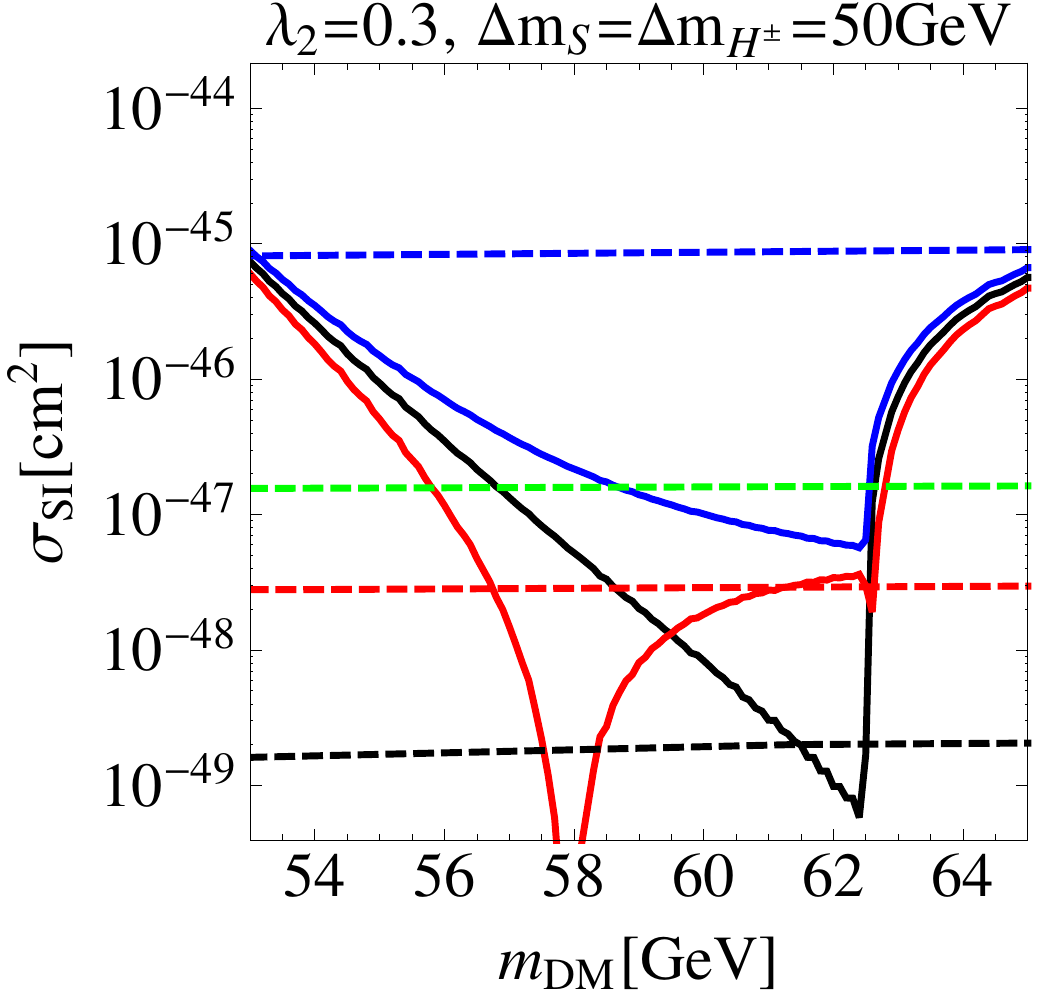} 
\quad
\includegraphics[width=0.30\hsize]{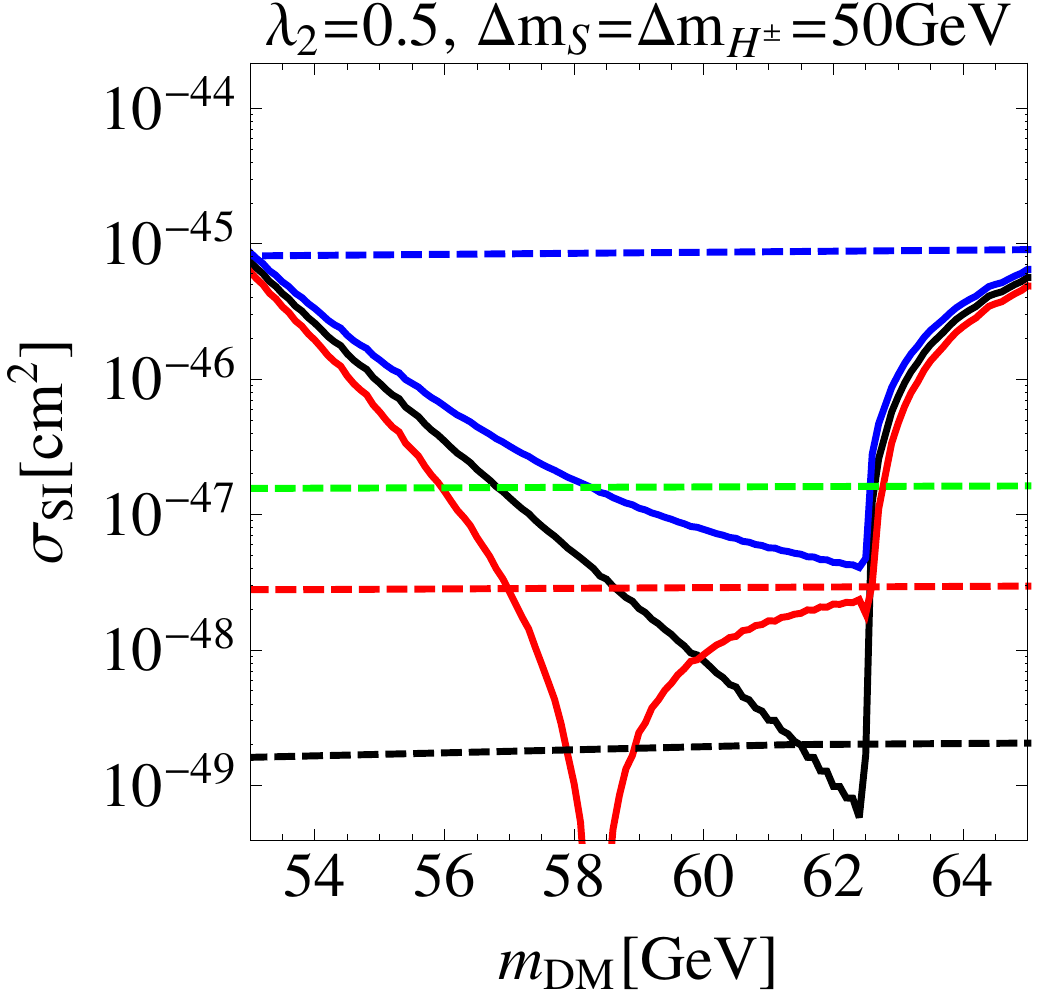} 
\\ \quad \\
\includegraphics[width=0.30\hsize]{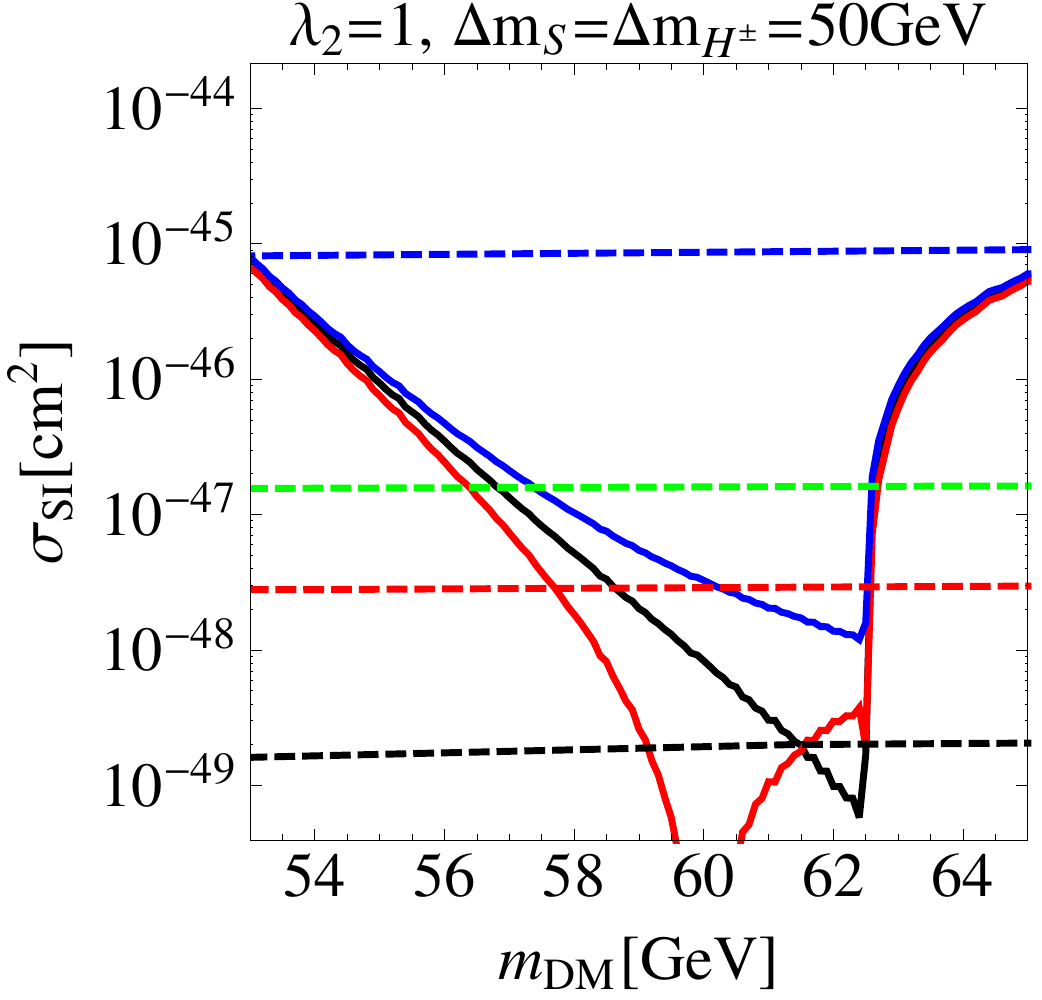} 
\quad
\includegraphics[width=0.30\hsize]{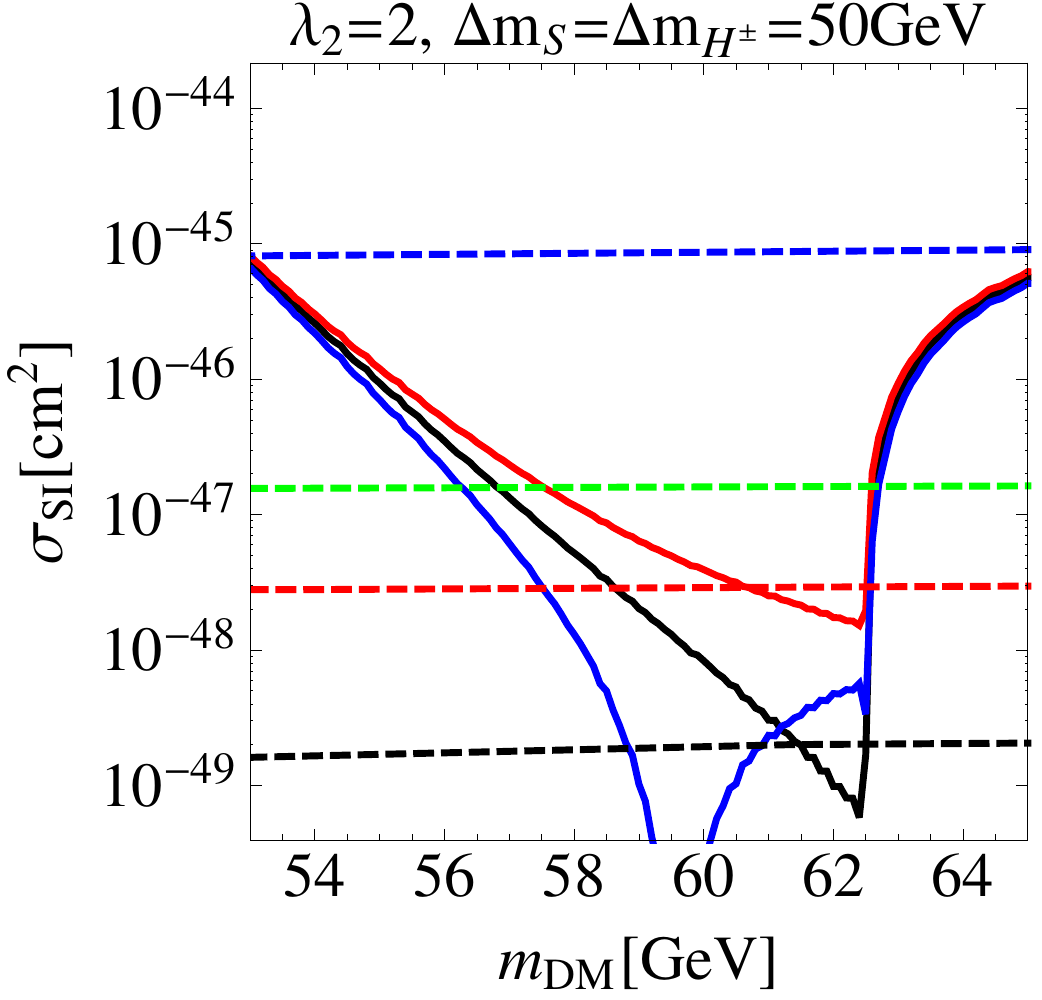} 
\quad
\includegraphics[width=0.30\hsize]{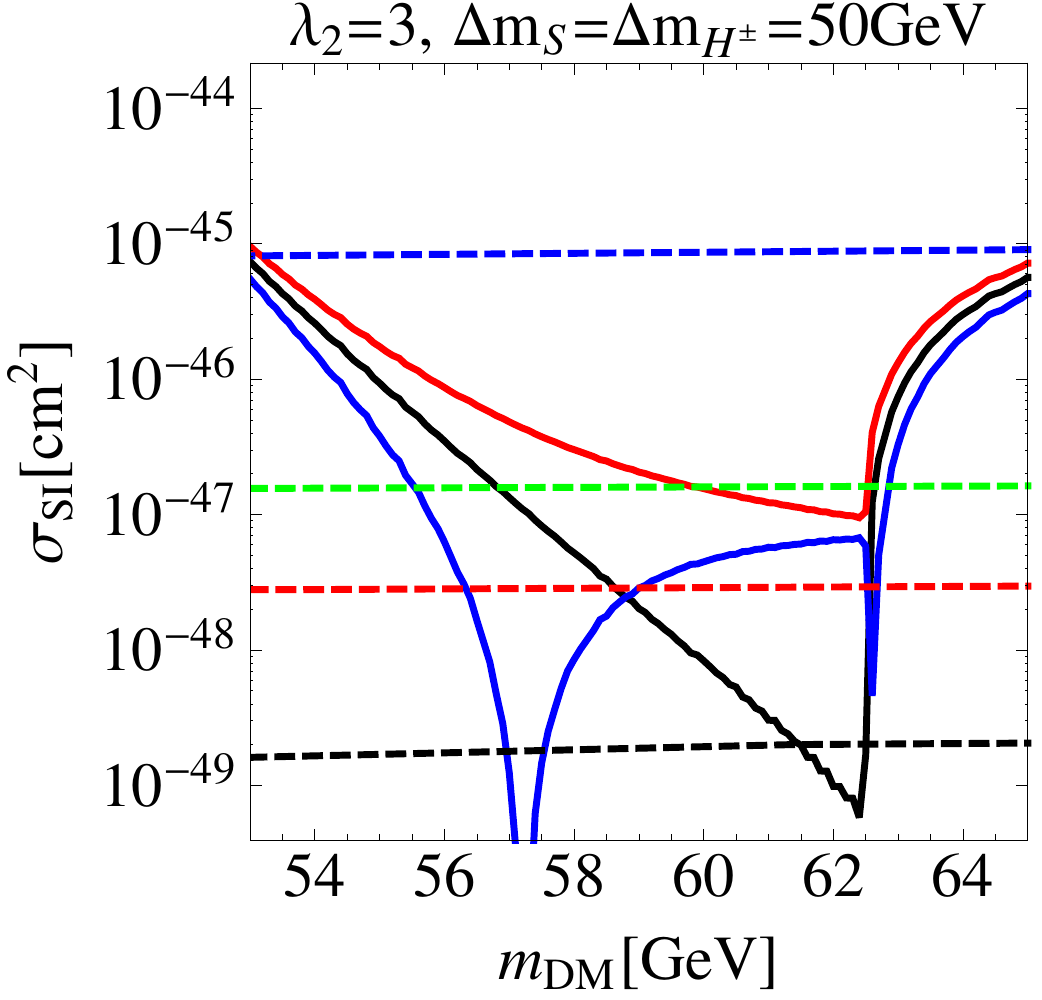} 
\caption{
The spin-independent cross section at tree
 level (black-solid line), and loop level (red-solid and blue-solid
 lines). Since the sign 
 of the tree level coupling, $\lambda_{\textrm{relic}}$, is unknown,
 there are two possibility for the result at loop level. If the
 couplings at tree and loop levels are constructive (destructive), the
 effective coupling is blue (red) line. 
 Here $\lambda_2 = 0$ (upper-left),
 $\lambda_2 = 0.3$ (upper-middle),
 $\lambda_2 = 0.5$ (upper-right),
 $\lambda_2 = 1.0$ (lower-left),
 $\lambda_2 = 2.0$ (lower-middle), and
 $\lambda_2 = 3.0$ (lower-right).
 The current bound and future prospects are also shown. The blue-dashed
 line is the current LUX bound. The green-dashed, red-dashed lines are
 the future prospect by XENON1T and LZ, respectively, and the
 black-dashed line is the discovery limit caused by atmospheric and
 astrophysical neutrinos.
}
\label{fig:mass_vs_xsecSI}
\end{figure}

\begin{figure}[tb]
\centering
\includegraphics[width=0.4\hsize]{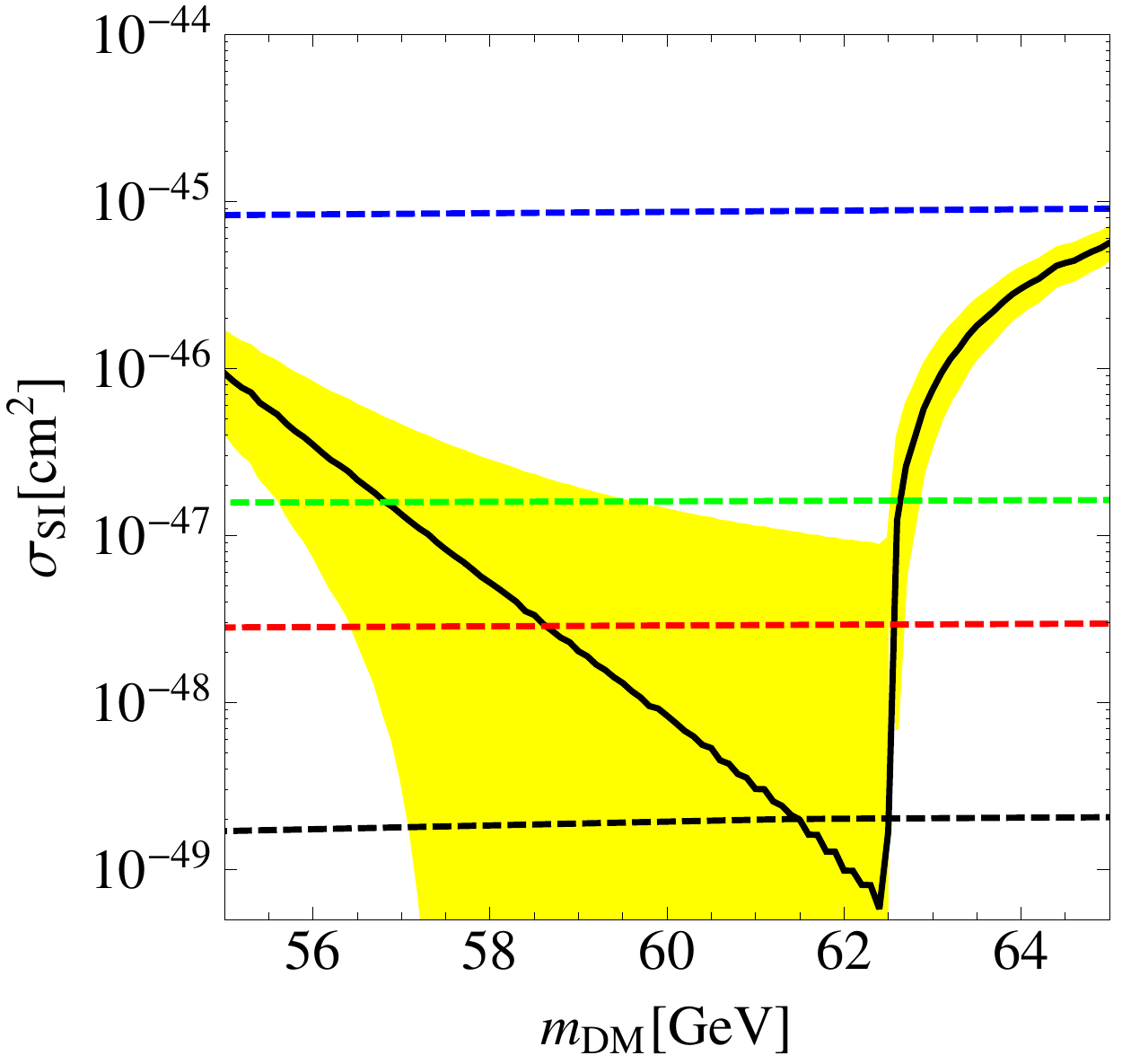} 
\caption{
The spin-independent cross section at tree
 level (black-solid line), and loop level (yellow shaded region).
 Here we vary $\lambda_2$ for $0< \lambda_2 < 1.45$.
 The blue-dashed
 line is the current LUX bound. The green-dashed, red-dashed lines are
 the future prospect by XENON1T and LZ, respectively, and the
 black-dashed line is the discovery limit caused by atmospheric and
 astrophysical neutrinos.
Here we take $\Delta m_{H^{\pm}}=50$~GeV,  $\Delta m_{S}=50$~GeV.
}
\label{fig:mass_vs_xsecSI_filled}
\end{figure}

We show the spin-independent cross section
both at the tree and loop levels as 
a function of the dark matter mass 
in Fig.~\ref{fig:mass_vs_xsecSI}, 
 with the current bound \cite{1310.8214} and future
prospects \cite{Aprile:2012zx, Feng:2014uja, Billard:2013qya}.
The value of $\lambda_2$ is different in each panels.
We take $\Delta m_{H^{\pm}} = \Delta m_{S} = 50$~GeV as a benchmark.
Since the sign of the tree level coupling, 
$\lambda_A$,
is unknown, there are two 
possibilities 
for the result at the loop level. 
The feature is highly depend on the sign of 
$\lambda_A$,
and we see that the spin-independent cross section at the loop level 
is both larger and
smaller than the one at the tree level value. 
For large $\lambda_2$ region, the sign of the loop correction to the
effective coupling is flipped as we can see from the upper-left and
lower-right panels.
In this benchmark, the loop corrections vanish when $\lambda_2
\simeq 1.45$ because the loop corrections depending on $\lambda_2$ cancel the
other loop corrections.
%
%
%
%
%
%
%
From the figure, we
 can see the importance of the loop corrections in this dark
matter mass region. For $\lambda_2 = 0.3$ case, for example, we have a chance to
detect 62~GeV dark matter in the future, although it is impossible according
to the tree level analysis. On the other hand, it might be impossible
for $\sim$58~GeV dark matter to be detected, although it is possible
according to the tree level analysis. Thus the detectable dark matter
mass range is modified due to the loop correction, and it is also depend
on the model parameters, especially the dark
matter self-interacting 
coupling $\lambda_2$.
Since we do not know the value of $\lambda_2$, we can not give a strict
prediction on the spin-independent cross section in this dark matter
mass region. We varied the value of $\lambda_2$ for $0<\lambda_2 <
1.45$, where the perturbative calculation works well,
and make a plot in Fig.~\ref{fig:mass_vs_xsecSI_filled}. The yellow
region is the model prediction for $\Delta m_S = \Delta
m_{H^{\pm}} = 50$~GeV.  

\begin{figure}[tb]
\includegraphics[width=0.35\hsize]{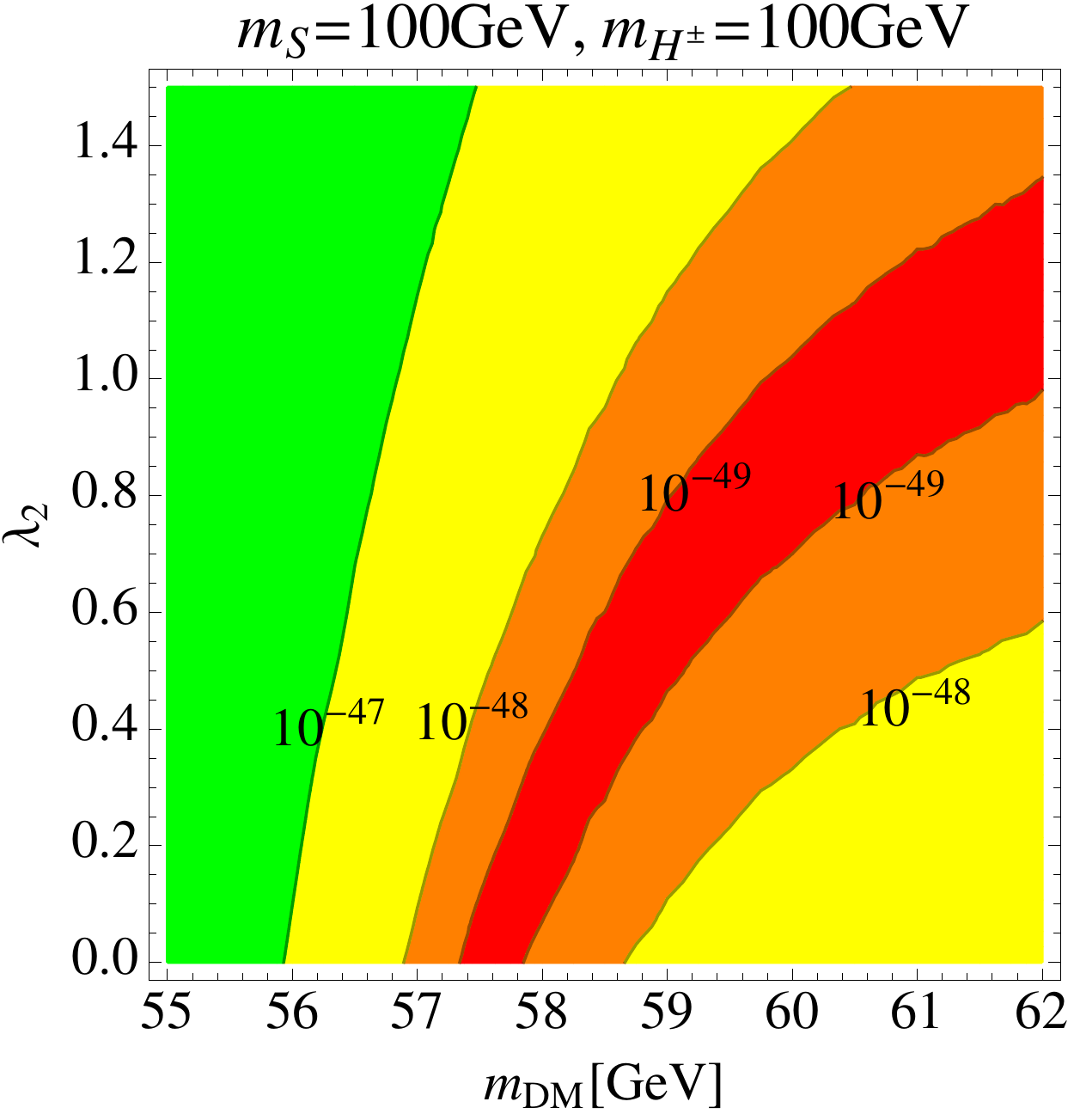} 
\includegraphics[width=0.35\hsize]{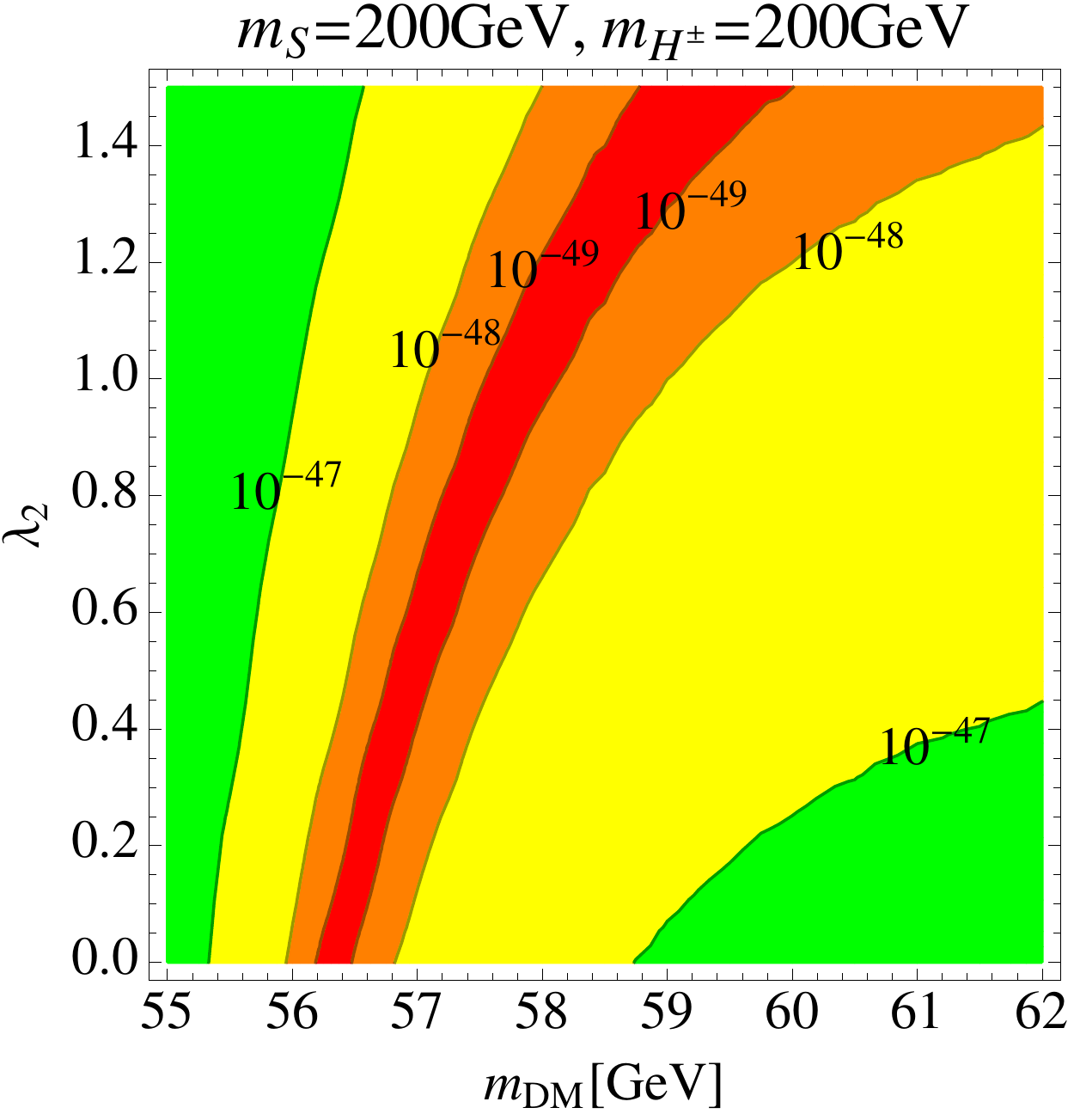} 
\\
\includegraphics[width=0.35\hsize]{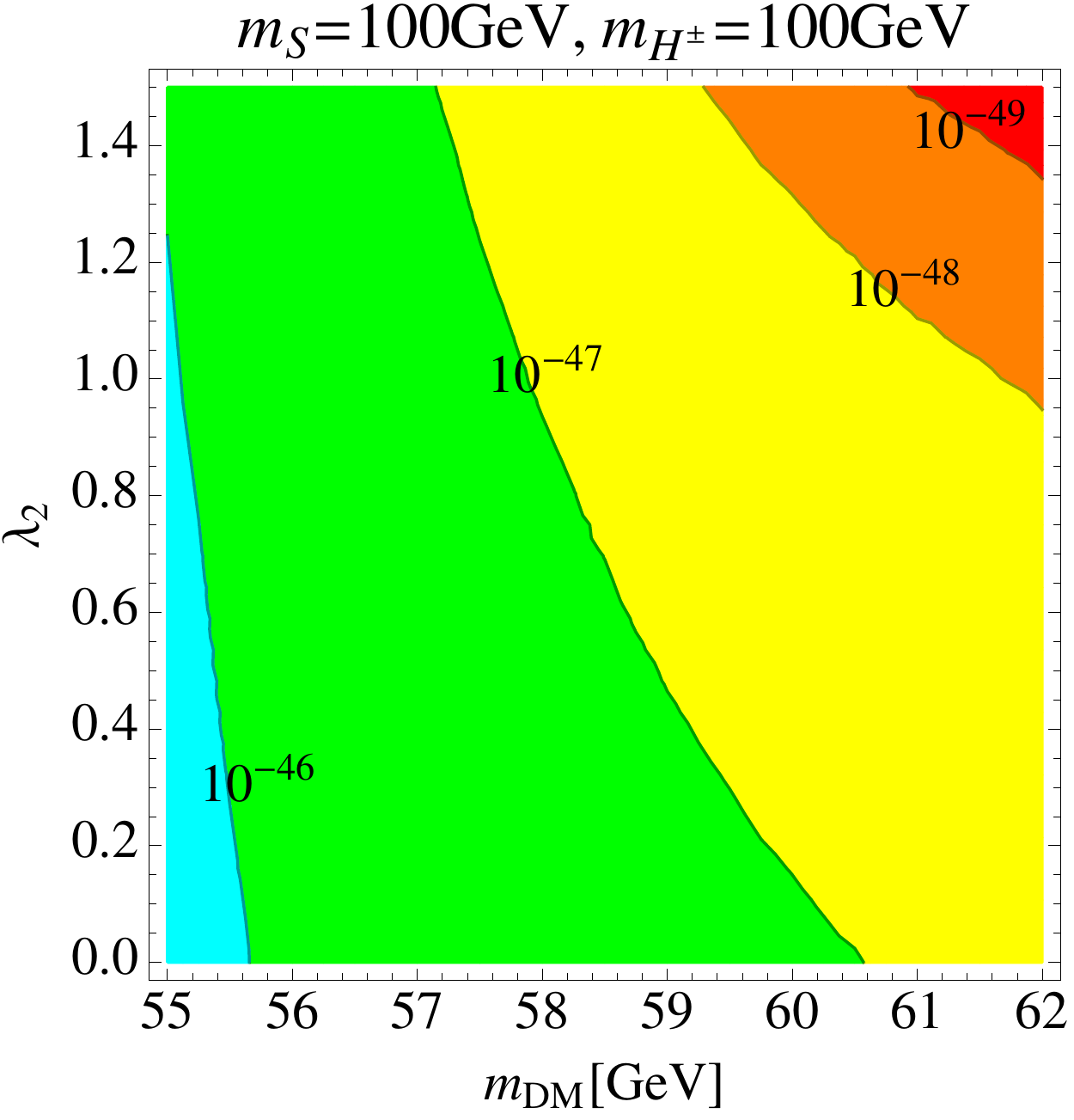}  
\includegraphics[width=0.35\hsize]{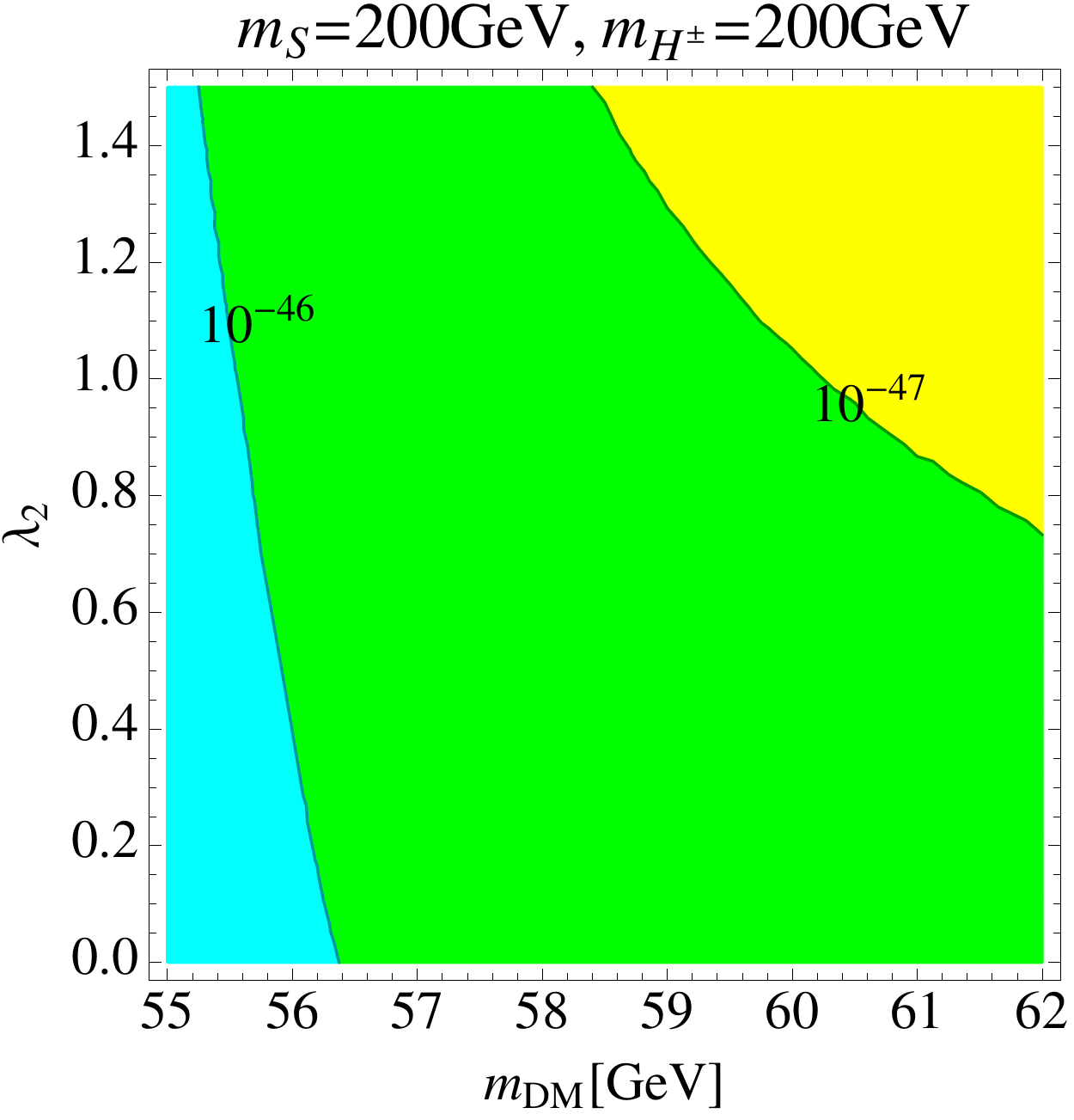}  
\caption{
The $\sigma_{\text{SI}}$ in ($m_{\text{DM}}, \lambda_2$)-plain.
The value of $\sigma_{\text{SI}}$ is
 $\sigma_{\text{SI}} < 10^{-49}$~cm$^{2}$,
 $10^{-49}$~cm$^2 < \sigma_{\text{SI}} < 10^{-48}$~cm$^{2}$,
 $10^{-48}$~cm$^2 < \sigma_{\text{SI}} < 10^{-47}$~cm$^{2}$, 
 $10^{-47}$~cm$^2 < \sigma_{\text{SI}} < 10^{-46}$~cm$^{2}$, and
 $10^{-46}$~cm$^2 < \sigma_{\text{SI}}$
 in the red, orange, yellow, green, and cyan regions, respectively.
 In the left (right) panel, we take 
 $\Delta m_{S} = \Delta m_{H^{\pm}} = 100 ~(200)$~GeV.
 In the upper (lower) panel, the sign of the $|\lambda_{\text{relic}}|$ is
  positive (negative), see Eq.~(\ref{eq:lamrelic}).
}
\label{fig:mDM_vs_lam_xsecSI}
\end{figure}

\begin{figure}[tb]
\includegraphics[width=0.3\hsize]{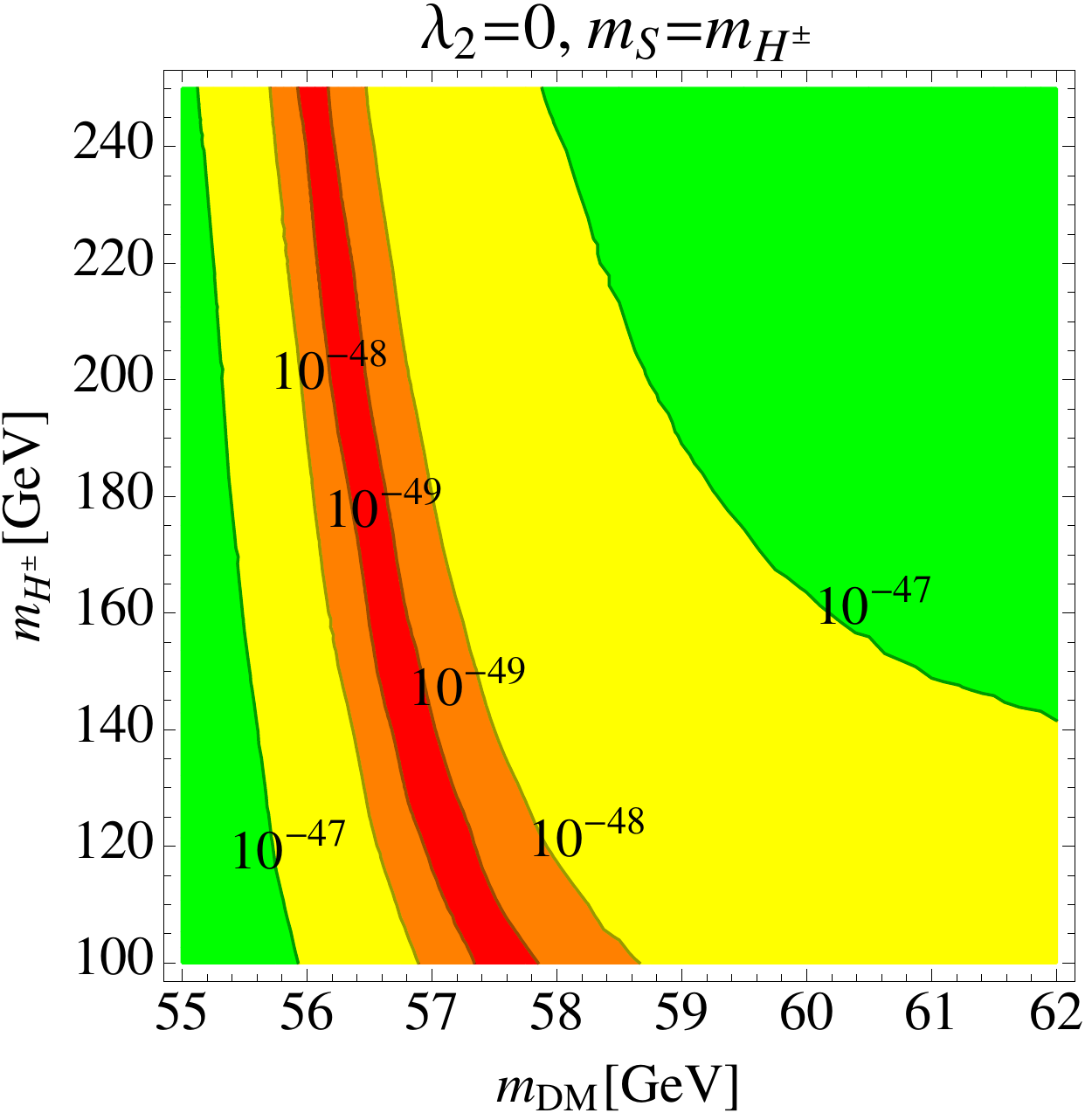} 
\includegraphics[width=0.3\hsize]{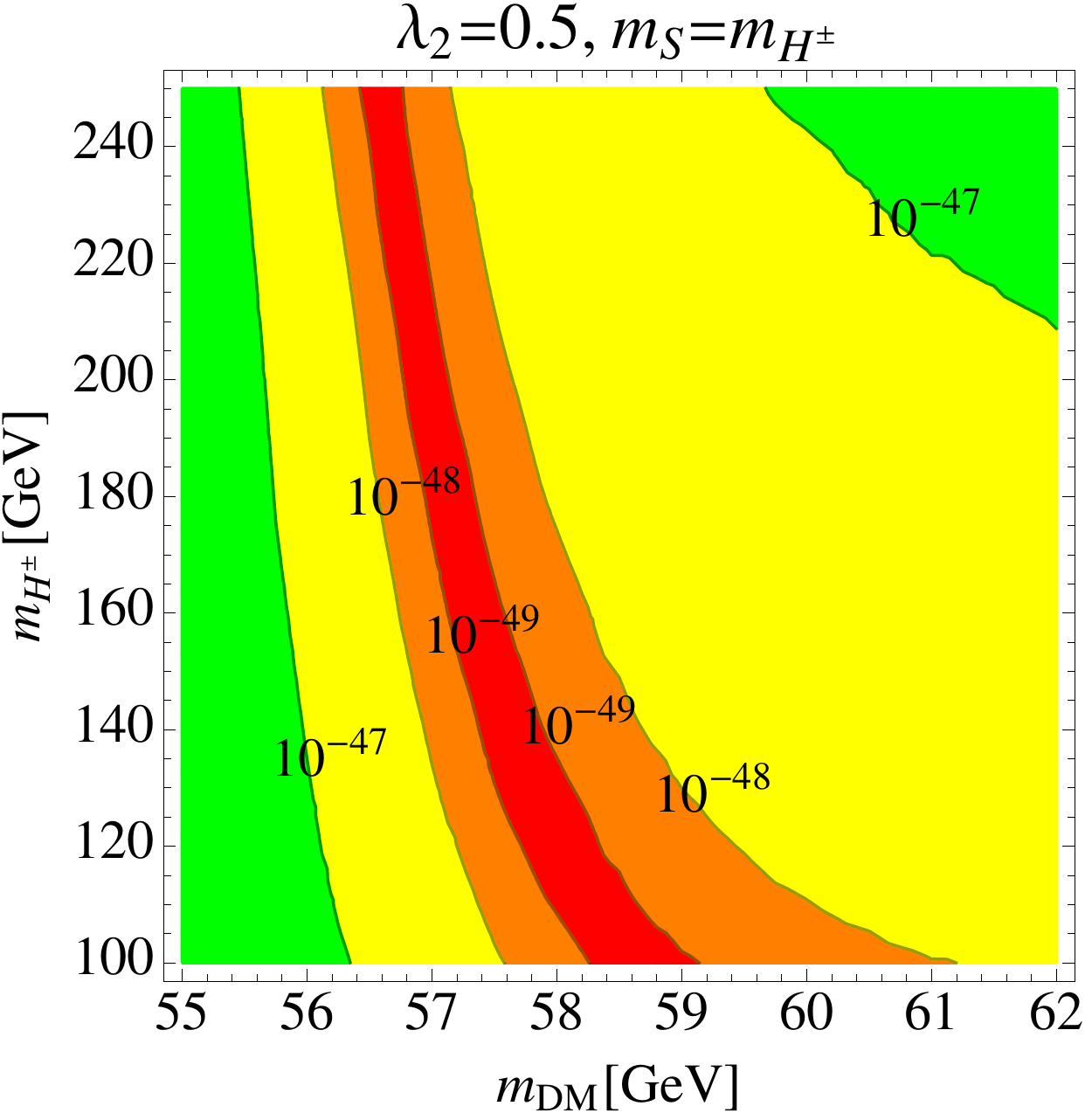} 
\includegraphics[width=0.3\hsize]{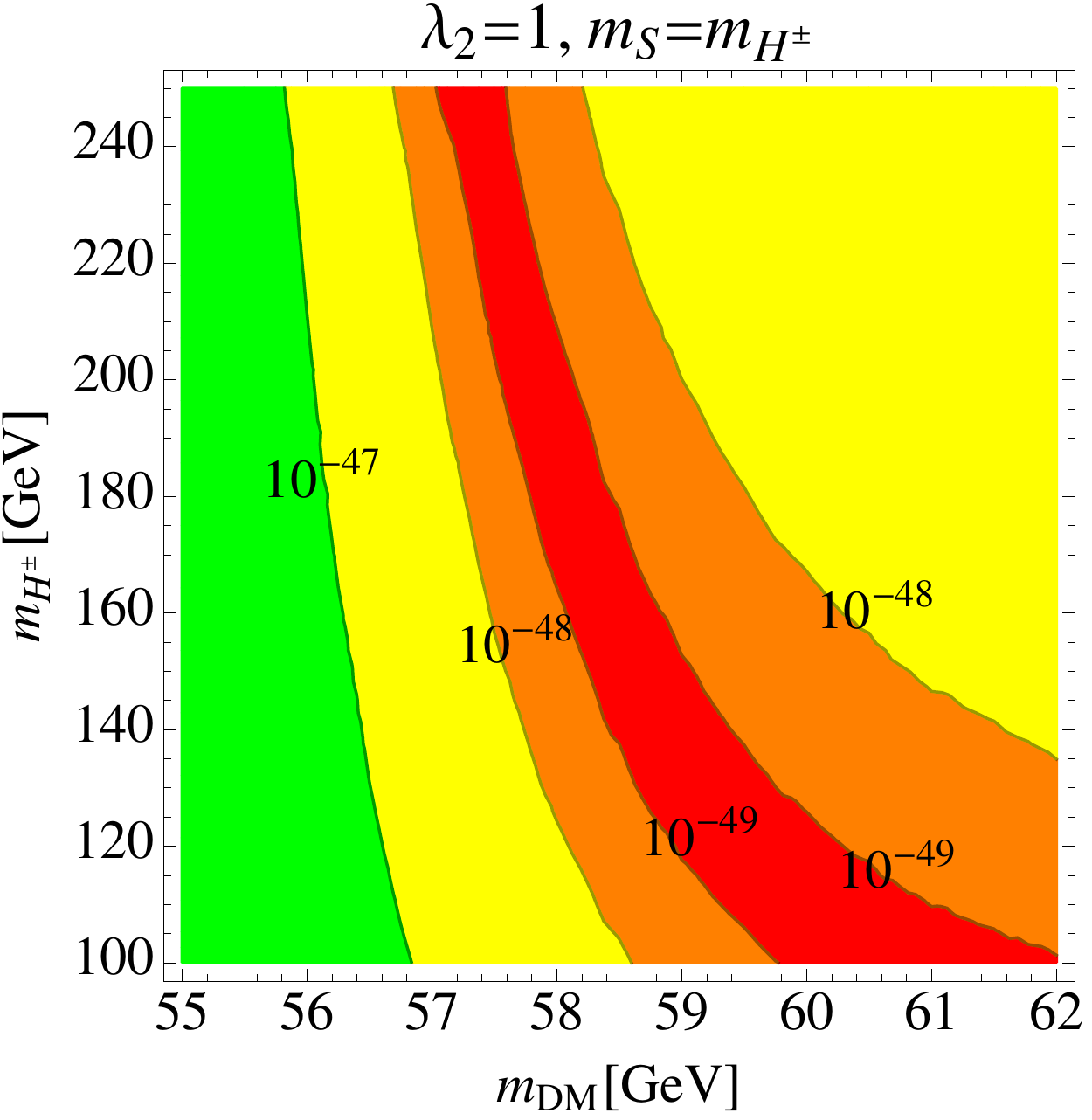} 
\includegraphics[width=0.3\hsize]{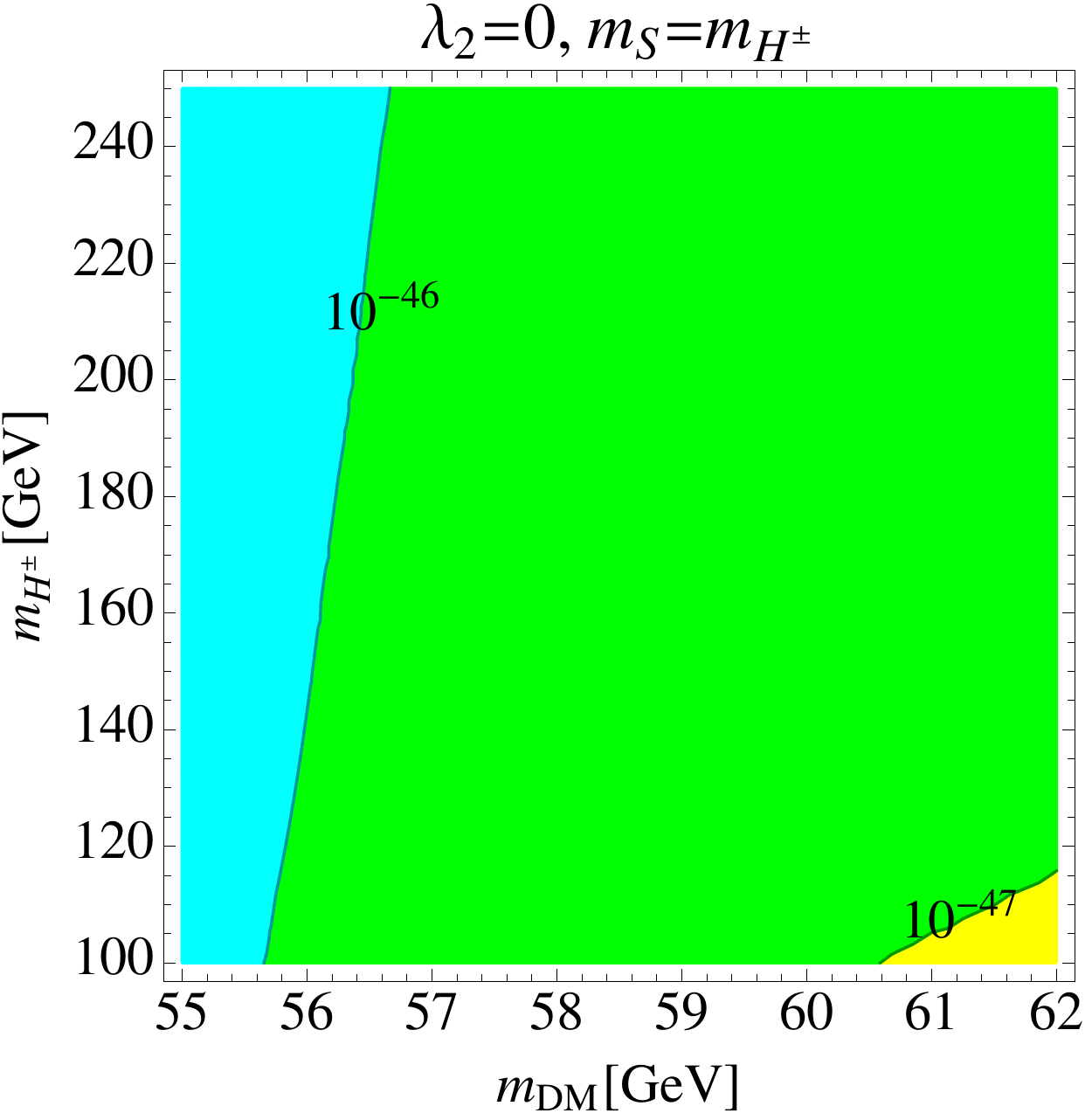}  
\includegraphics[width=0.3\hsize]{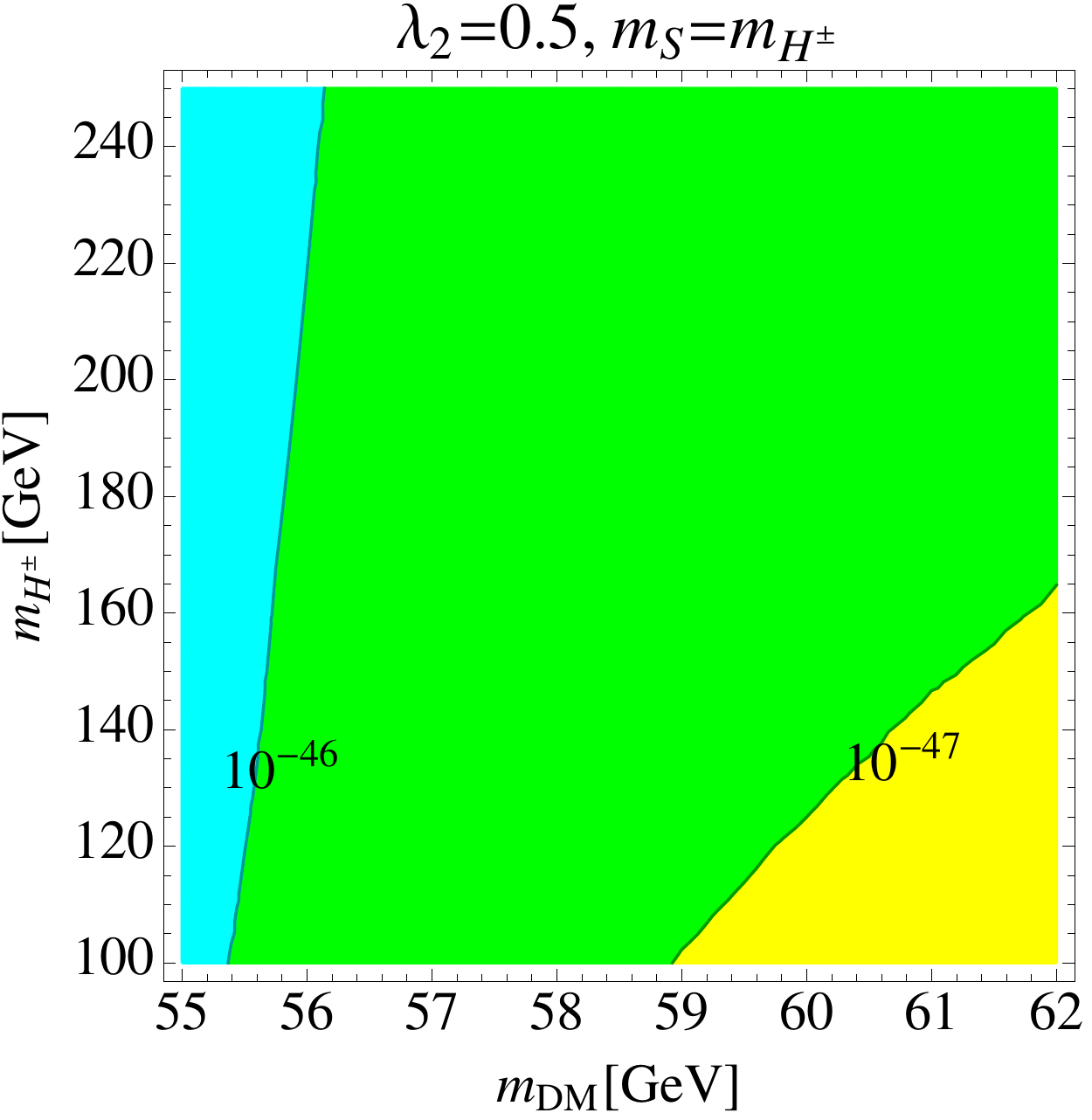}  
\includegraphics[width=0.3\hsize]{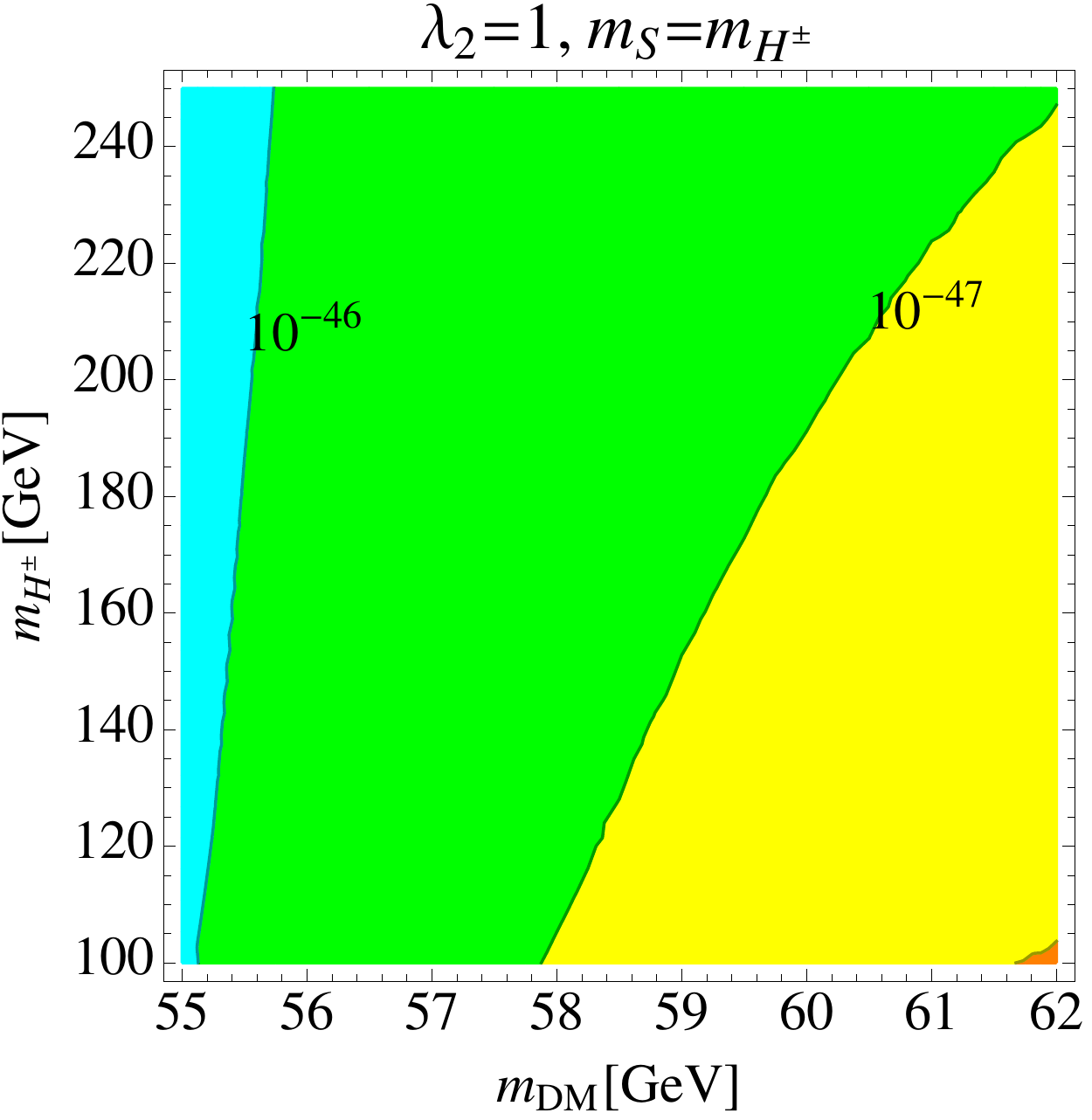}  
\caption{
The $\sigma_{\text{SI}}$ in ($m_{\text{DM}}, m_{H^{\pm}}$)-plain.
We take $m_{S} = m_{H^{\pm}}$.
The value of $\sigma_{\text{SI}}$ is
 $\sigma_{\text{SI}} < 10^{-49}$~cm$^{2}$,
 $10^{-49}$~cm$^2 < \sigma_{\text{SI}} < 10^{-48}$~cm$^{2}$,
 $10^{-48}$~cm$^2 < \sigma_{\text{SI}} < 10^{-47}$~cm$^{2}$, 
 $10^{-47}$~cm$^2 < \sigma_{\text{SI}} < 10^{-46}$~cm$^{2}$, and
 $10^{-46}$~cm$^2 < \sigma_{\text{SI}}$
 in the red, orange, yellow, green, and cyan regions, respectively.
 From the left to the right panel, we take
 $\lambda_2 =$ 0, 0.5, and 1, respectively. 
 In the upper (lower) panel, the sign of the $|\lambda_{\text{relic}}|$ is
  positive (negative), see Eq.~(\ref{eq:lamrelic}).
}
\label{fig:mDM_vs_mH_xsecSI}
\end{figure}

So far we have chosen $\Delta m_{S} = \Delta m_{H^{\pm}} =
$50~GeV. However, the choice of these mass difference also play the
significant role for $\sigma_{\text{SI}}$ as we can see from
Fig.~\ref{fig:lam2_delM}. In this paragraph, we vary these parameter
keeping the custodial symmetric limit, $\Delta m_{S} = \Delta m_{H^{\pm}}$.
We make plots 
the $\sigma_{\text{SI}}$ in ($m_{\text{DM}}, \lambda_2$)-plain
in Fig.~\ref{fig:mDM_vs_lam_xsecSI}, and
in ($m_{\text{DM}}, m_{H^{\pm}}$)-plain
in Fig.~\ref{fig:mDM_vs_mH_xsecSI}.
The red region is basically beyond the discovery limit caused by atmospheric and
 astrophysical neutrinos, and we can see that the dark matter mass range in which
 the dark matter is possible to be detected in the future direct
 detection experiments is highly depending on the model parameter.

Finally, we give an approximate formula for $\d\l$ which is defined in Eq.~(\ref{eq:def_delta_lambda}).
In the case of $m_H^\pm = m_S$,
\begin{align}
\d \l 
=&
-0.00409
 m_{\text{DM}} 
\left( 0.0000144
 - 7.77 \times 10^{-8} m_{H^{\pm}} 
 - 0.00334 \frac{1}{m_{H^{\pm}}} 
\right)
\nonumber\\
&+ 
\l_2 
\left(
0.00183
-  
7.87 \times 10^{-10} m_{H^{\pm}}^2 
+ 
m_{\text{DM}}^2 
\left(-4.13 \times 10^{-8} - \frac{0.00113}{m_{H^{\pm}}^2}
\right)
\right).
\end{align}
By using the above expression and Eq.~(\ref{eq:xsec_eff}), an approximate value of the cross section can be obtained.
We have checked its error is less than 2\% in the range of $50<m_{\rm DM}<62.5~\text{GeV}$ and $100<m_{H^\pm} = m_S < 250~\text{GeV}$.

%

\section{Conclusion and discussion}\label{sec:conclusion}
In this paper, we discussed the
spin-independent cross section $\s_{\text{SI}}$ of 
nucleon and the dark matter in the inert doublet model.
We revisited the radiative corrections to the
spin-independent cross section with taking into
account the effect of the non-zero values of the inert doublet couplings,
namely the mass differences among $Z_2$ odd particles and the dark matter self
coupling $\l_2$.
The effect of these couplings were ignored in the previous
work~\cite{PHRVA.D87.075025}, 
but we find they actually control the main contribution in the radiative
corrections.  

The sign of the tree level coupling is important for
precise prediction of the spin-independent cross section. Depending on its
sign, the spin-independent cross section at the one-loop level becomes
bigger or smaller than the tree level prediction. When it becomes
bigger, the direct detect experiments have chance to detect the dark
matter even if its mass is a half of the Higgs mass. This feature can
not found at the tree level analysis.

The unknown model parameters are the origin of the  uncertainty for the
model prediction to the spin-independent cross section. Once the LHC
experiment find the extra scalars, $S$ and $H^{\pm}$, and determined
their masses, the uncertainty will be reduced.


\section*{Acknowledgments}
The authors thank Natsumi Nagata for useful discussion.
The work is supported by
MEXT Grant-in-Aid for Scientific Research on Innovative Areas (No. 23104006 [TA])
and JSPS Research Fellowships for Young Scientists [RS].

\section*{Appendix}
\appendix
\renewcommand{\thesubsection}{\thesection .\arabic{subsection} }
\renewcommand{\thesubsubsection}{\thesection
.\arabic{subsection}.\arabic{subsubsection} } 
In the appendices, we give explicit formulae for the loop corrections to the spin-independent cross section.
Electroweak gauge couplings are defined as,
\begin{align}
g_W = \frac{e}{s},\qquad
g_Z = \frac{e}{sc},\qquad
g_{f_L} = g_Z(T_{3,f} - s^2 Q_f),\qquad
g_{f_R} = -g_Z s^2 Q_f,
\end{align}
where $f$ runs through $u,d,s,c,b$ and $t$.

\section{One-loop box type diagrams}\label{sec:oneloop_box}
We calculate one-loop box diagrams which contribute to the $q A \to q A$ process.
We consider only the light quarks. We expand the diagrams by the masses of
the light quarks and keep only its leading order. This calculation is
for the spin-independent cross section, and we can assume the momentum
transfer is small, we take it zero. 
The sum of the diagrams we calculate in this section give the contributions to
$\Gamma^{q}_{\text{Box}}$,
$\Gamma^q_{t2}$, and $\Gamma'^q_{t2}$ through,
\begin{align}
 i \Gamma^{q}_{\text{Box}} m_q
+
\frac{i}{m_A^2}(\Gamma^q_{t2} + \Gamma'^q_{\text{t2}})
\left(
   p^{\mu}q_{\mu} \slashchar{p}
   - \frac{1}{4} p^2 \slashchar{q}
\right)
.
\end{align}
The definitions of
$\Gamma^{q}_{\text{Box}}$,
$\Gamma^q_{t2}$, and $\Gamma'^q_{t2}$ 
are given in Eq.~(\ref{eq:eff_int}).

\subsection{$Z$ boson contribution}
\begin{figure}[tb]
\subfigure[]{
 \includegraphics[width=0.23\hsize]{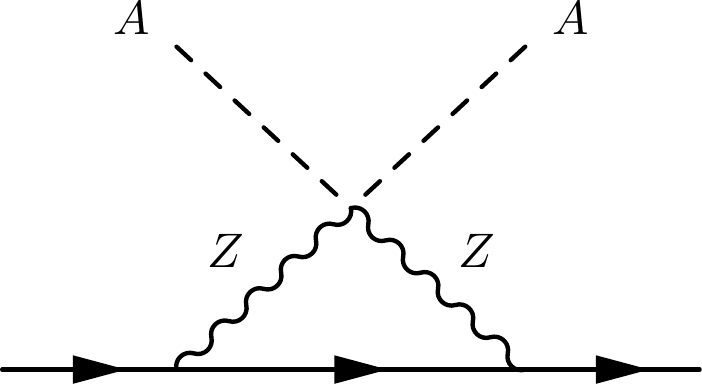} 
\label{fig:Z1}
}  
\subfigure[]{
 \includegraphics[width=0.23\hsize]{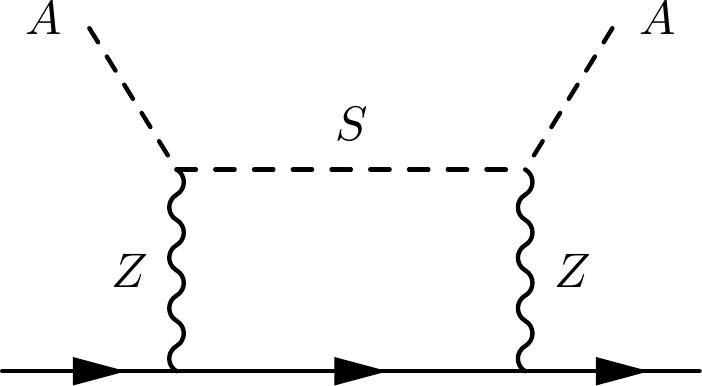} 
\label{fig:Z2}
}  
\subfigure[]{
 \includegraphics[width=0.23\hsize]{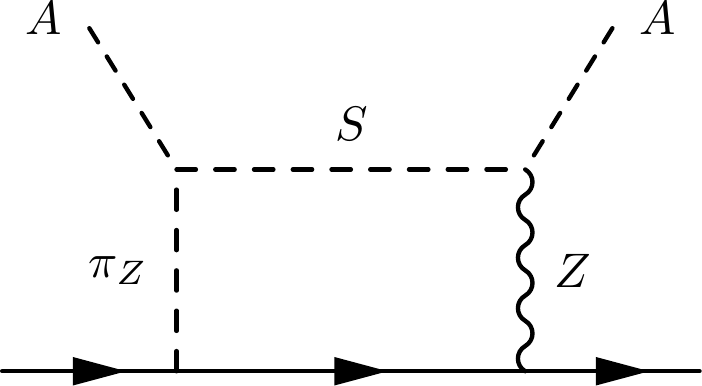} 
\label{fig:Z3}
}  
\subfigure[]{
 \includegraphics[width=0.23\hsize]{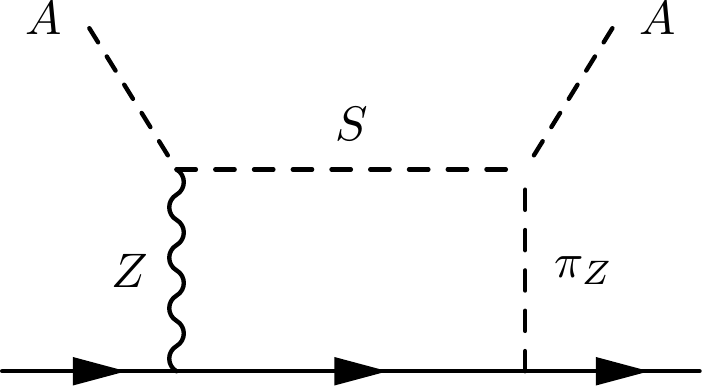} 
\label{fig:Z4}
}  
\caption{
For the box diagrams, we also have ``crossed'' diagrams in which the
 vertices $A$ attached are flipped.
}
\label{fig:box-Z}
\end{figure}

We calculate the contributions from $Z$ boson and its would-be NG boson
depicted by the diagrams in Fig.~\ref{fig:box-Z}.
In the followings, ``crossed'' means diagrams in which the
 vertices which $A$ attached are flipped.
The box-diagrams without would-be NG bosons (Fig.~\ref{fig:Z2}) contribute to twist-2 operator.
%
%
%
\begin{align}
\text{Fig.}~\ref{fig:Z1}
=&
\frac{i}{(4\pi)^2} g_Z^2 \frac{m_f}{m_Z^2}
\left( 2 g_{f_L} g_{f_R} - \frac{1}{4} ( g_{f_L}^2 + g_{f_R}^2 ) \right),
\\
\text{Fig.}~\ref{fig:Z2} + \text{(crossed)}
=&
\frac{i}{(4\pi)^2} \frac{1}{2} g_Z^2 m_f \nonumber\\
& \times \Biggl( \frac{(g_{f_L} - g_{f_R})^2}{2} f_{B1}
+ m_A^2 (g_{f_L}^2 + g_{f_R}^2) \left( f_{B2} -3 f_{B3} \right)
+ 4 m_A^2 g_{f_L} g_{f_R} f_{B2} \Biggr) \nonumber\\
&
+ \frac{i}{(4\pi)^2} g_Z^2 
\left( p^{\mu}q_{\mu} \slashchar{p} - \frac{1}{4} p^2 \slashchar{q} \right)
(g_{f_L}^2 + g_{f_R}^2) 2 (f_{B2} - f_{B3} )
\\
\text{Fig.}~\ref{fig:Z3},\ref{fig:Z4} + 
\text{(crossed)}
=&
-
\frac{i}{(4\pi)^2} \frac{1}{2} g_Z^2 \frac{m_S^2 - m_A^2}{v^2} 
m_f \Biggl( f_{B1} + 2 m_{A}^2 f_{B4} \Biggr) ,
\end{align}
%
%
%
and where $p$ and $q$ are four-momenta of the dark matter and the quark,
respectively. Note that we ignore the momentum transfer between the dark
matter and the quark. 
The definitions of $f_{B1}$, $f_{B2}$, and $f_{B3}$ are
given in Appendix~\ref{sec:loopfunction_box}, and their argument here
is $(m_Z, m_S, m_A)$.

\subsection{$W$ boson contribution}
\begin{figure}[tb]
\subfigure[]{
 \includegraphics[width=0.23\hsize]{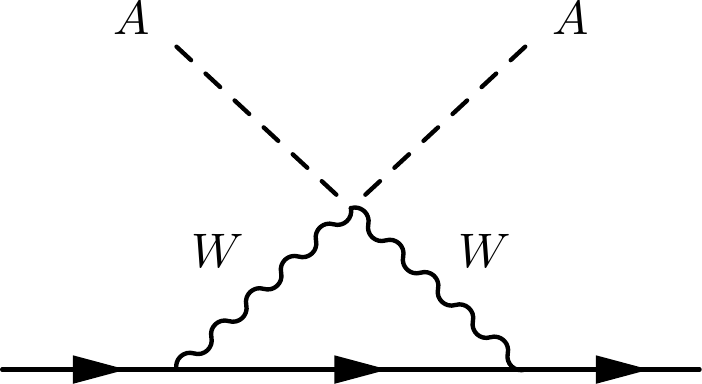} 
\label{fig:W1}
}  
\subfigure[]{
 \includegraphics[width=0.23\hsize]{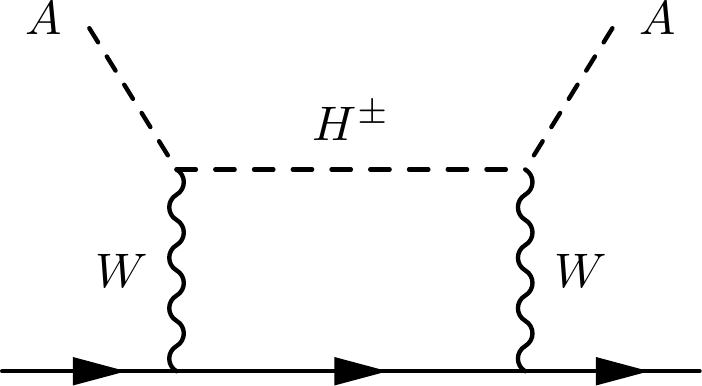} 
\label{fig:W2}
}  
\subfigure[]{
 \includegraphics[width=0.23\hsize]{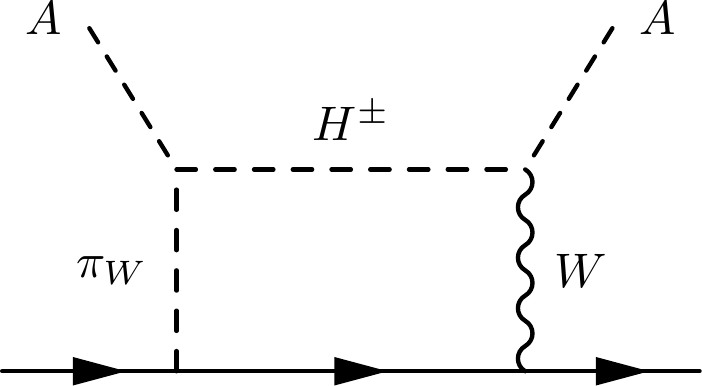} 
\label{fig:W3}
}  
\subfigure[]{
 \includegraphics[width=0.23\hsize]{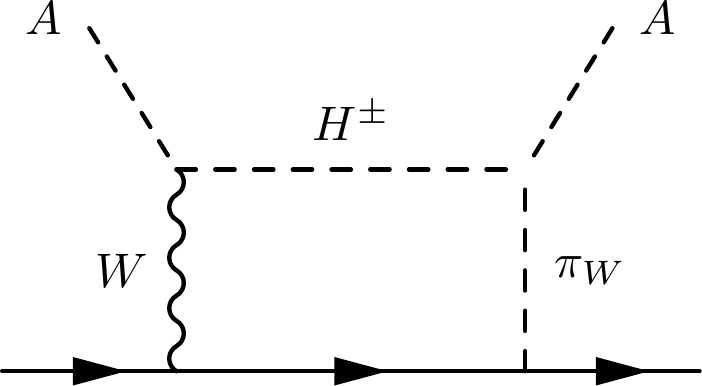} 
\label{fig:W4}
}  
\caption{
For the box diagrams, we also have ``crossed'' diagrams in which the
 vertices $A$ attached are flipped.
}
\label{fig:box-W}
\end{figure}

We calculate the contributions from $W$ boson and its would-be NG boson
depicted by the diagrams in Fig.~\ref{fig:box-W}.
In the followings, ``crossed'' means diagrams in which the vertices $A$ attached are flipped.
The box-diagrams without would-be NG bosons contribute to twist-2 operator.
\begin{align}
\text{Fig.}~\ref{fig:W1}
=&
-
\frac{1}{8}
\frac{i}{(4\pi)^2}
g_W^4
\frac{m_f}{m_W^2}
,
\\
\text{Fig.}~\ref{fig:W2} + \text{(crossed)}
=&
\frac{1}{8}
\frac{i}{(4\pi)^2}
g_W^4
m_f
\Biggl(
f_{B1}
+
m_A^2
\left(
2 f_{B2}
-6 f_{B3}
\right)
\Biggr)
\nonumber\\
&
+
\frac{i}{(4\pi)^2}
g_W^4
\left(
p^{\mu}q_{\mu} \slashchar{p}
-
\frac{1}{4} p^2 \slashchar{q}
\right)
\left(
f_{B2} - f_{B3}
\right)
,
\\
\text{Fig.}~\ref{fig:W3} + 
\text{Fig.}~\ref{fig:W4} + 
\text{(crossed)}
=&
-
\frac{i}{(4\pi)^2}
g_W^2
\frac{m_{H^{\pm}}^2 - m_A^2}{v^2} 
m_f
\Biggl(
f_{B1}
+
2 m_{A}^2
f_{B4}
\Biggr)
,
\end{align}
and where $p$ and $q$ are four-momenta of the dark matter and the quark,
respectively. Note that we ignore the momentum transfer between the dark
matter and the quark.
The definitions of $f_{B1}$, $f_{B2}$, and $f_{B3}$ are
given in Appendix~\ref{sec:loopfunction_box}, and their argument here
is $(m_W, m_{H^{\pm}}, m_A)$.

\section{One-loop higgs vertex corrections}\label{sec:oneloop_vertex}
We calculate one-loop corrections to the dark matter coupling to the
Higgs boson. We interested in the case that the coupling is highly
suppressed at the tree level. Hence we take $\lambda_A =0$, in our
calculation. We denote $q^2$ as the
momentum of the Higgs boson, and treat the Higgs boson as off-shell,
because what we need is the difference between $q^2=m_h^2$ case and
$q^2=0$ case. Hence we ignore terms independent from $q^2$ in the
following calculations.
The sum of the diagrams we calculate in this section 
gives $- i v \delta \Gamma_h$, where
$\delta \Gamma_h$ is defined in Eq.~(\ref{eq:eff_int}).

\subsection{$Z$ boson contribution}
\begin{figure}[tb]
\subfigure[]{
 \includegraphics[width=0.18\hsize]{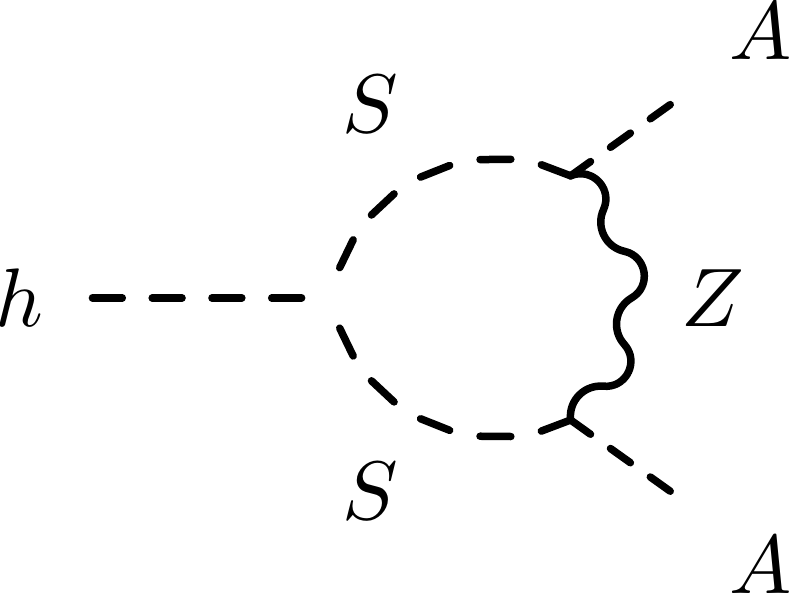} 
\label{fig:hZ1}
}  
\subfigure[]{
 \includegraphics[width=0.18\hsize]{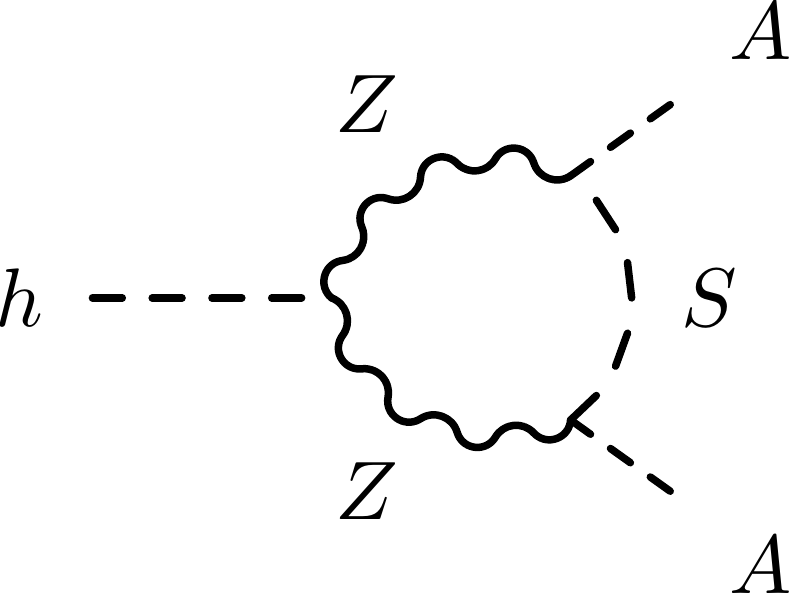} 
\label{fig:hZ2}
}  
\subfigure[]{
 \includegraphics[width=0.18\hsize]{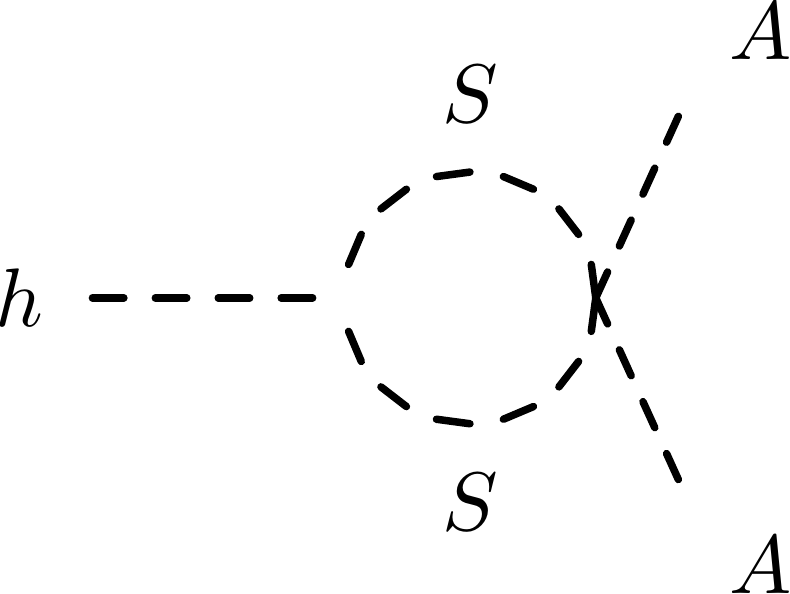} 
\label{fig:hZ3}
}  
\subfigure[]{
 \includegraphics[width=0.18\hsize]{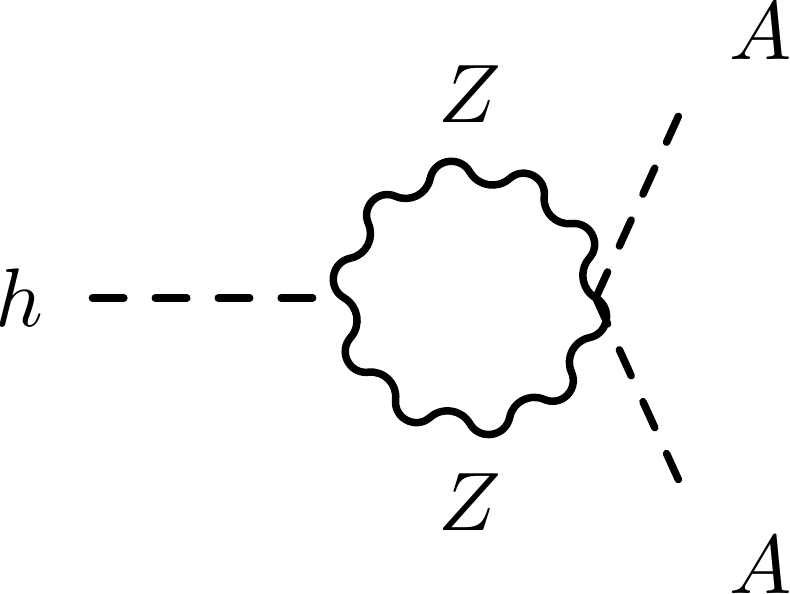} 
\label{fig:hZ4}
}  
\subfigure[]{
 \includegraphics[width=0.18\hsize]{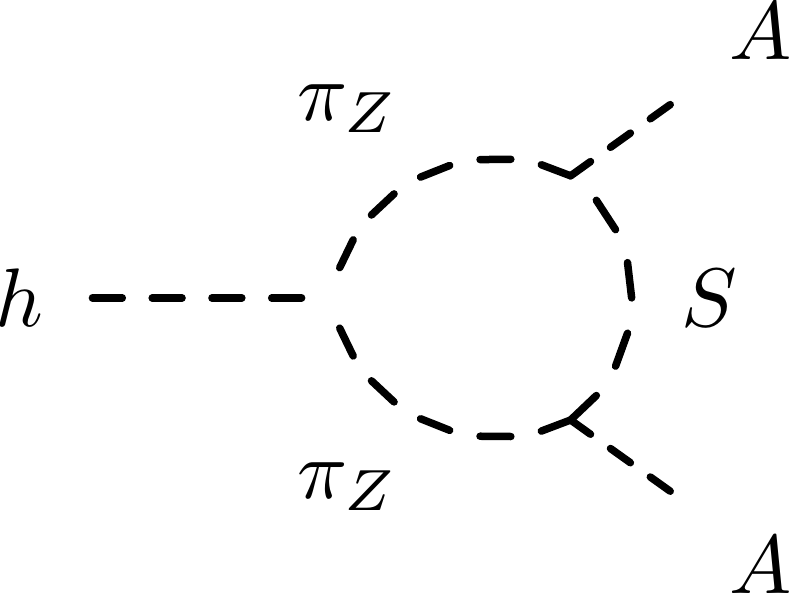} 
\label{fig:hZ5}
}  
\subfigure[]{
 \includegraphics[width=0.18\hsize]{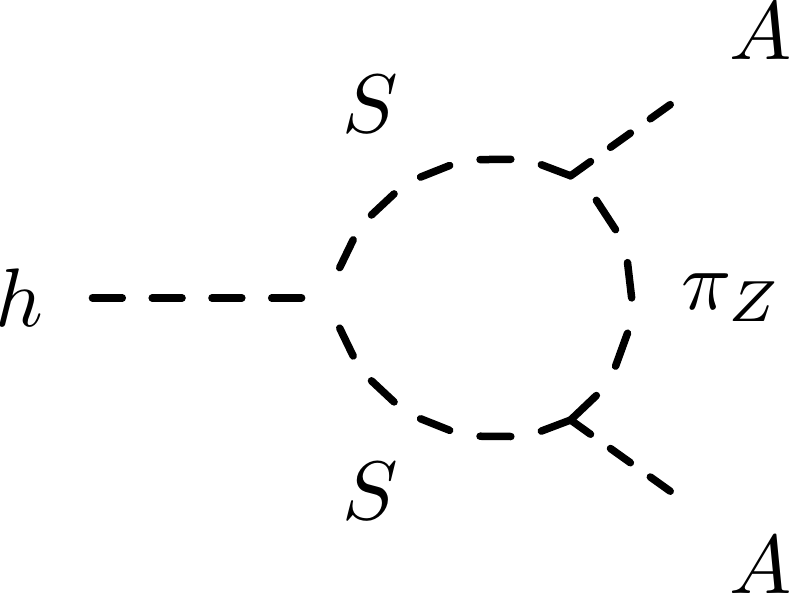} 
\label{fig:hZ6}
}  
\subfigure[]{
 \includegraphics[width=0.18\hsize]{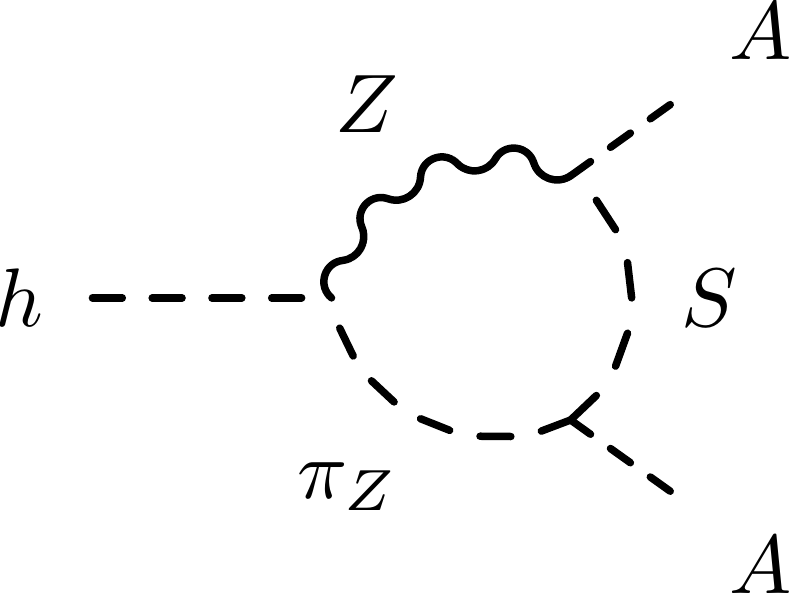} 
\label{fig:hZ7}
}  
\subfigure[]{
 \includegraphics[width=0.18\hsize]{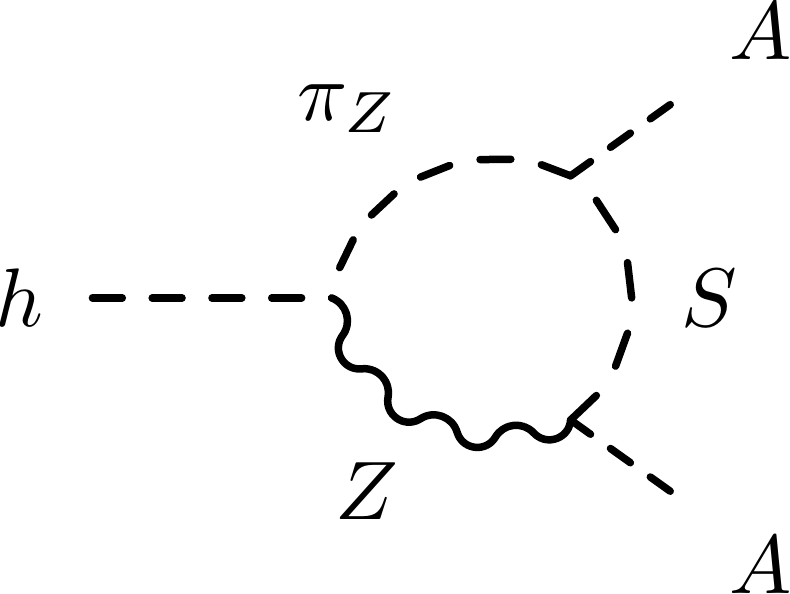} 
\label{fig:hZ8}
}  
\subfigure[]{
 \includegraphics[width=0.18\hsize]{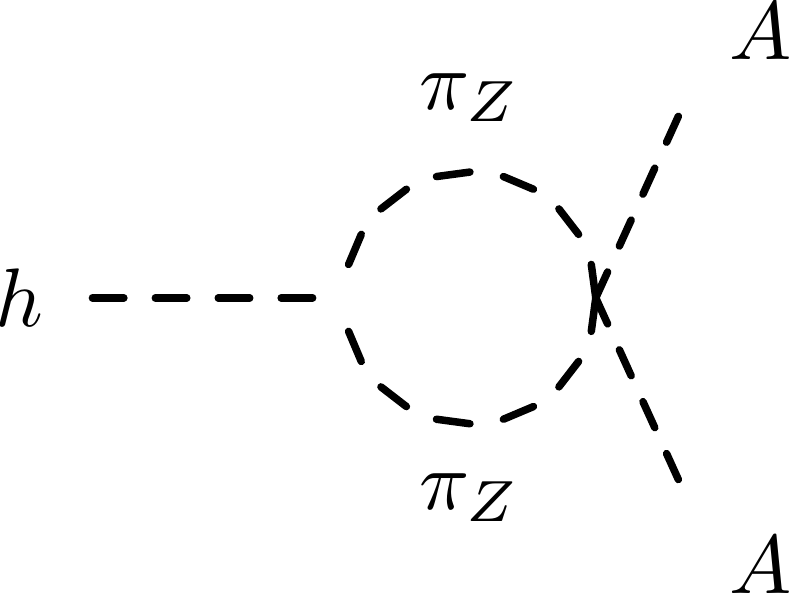} 
\label{fig:hZ9}
}  
\caption{
The diagrams for the vertex correction with neutral particles.
}
\end{figure}
Up to the $q^2$-independent terms, we find
\begin{align}
\text{Fig.}~\ref{fig:hZ1}
=&
\frac{2i}{(4\pi)^2}
\frac{m_Z^2}{v^2}
\frac{m_{S}^2 - m_{A}^2}{v}
\left(
F_1 ( m_S^2, q^2)
+
(-m_Z^2 + 2 m_S^2 + 2 m_A^2 - 2 q^2)
F_2 ( m_S^2, m_Z^2, q^2)
\right)
,
\\
\text{Fig.}~\ref{fig:hZ2}
=&
\frac{2i}{(4\pi)^2}
\frac{m_Z^2}{v^2}
\frac{m_{Z}^2}{v}
\left(
-2 F_1 ( m_Z^2, q^2)
+
(-m_Z^2 + 2 m_S^2 + 2 m_A^2 - \frac{1}{2} q^2)
F_2 ( m_Z^2, m_S^2, q^2)
\right)
,
\\
\text{Fig.}~\ref{fig:hZ3}
=&
\frac{2i}{(4\pi)^2}
\lambda_2
\frac{m_{S}^2 - m_{A}^2}{v}
F_1 ( m_S^2, q^2)
,
\\
\text{Fig.}~\ref{fig:hZ4}
=&
\frac{8i}{(4\pi)^2}
\frac{m_Z^2}{v^2}
\frac{m_{Z}^2}{v}
F_1 ( m_Z^2, q^2)
,
\\
\text{Fig.}~\ref{fig:hZ5}
=&
- \frac{i}{(4\pi)^2}
\frac{m_{h}^2}{v}
\left(
\frac{m_{S}^2 - m_{A}^2}{v}
\right)^2
F_2 ( m_Z^2, m_S^2, q^2)
,
\\
\text{Fig.}~\ref{fig:hZ6}
=&
- \frac{2i}{(4\pi)^2}
\left(
\frac{m_{S}^2 - m_{A}^2}{v}
\right)^3
F_2 ( m_S^2, m_Z^2, q^2)
,
\\
\text{Fig.}~\ref{fig:hZ7}
+
\text{Fig.}~\ref{fig:hZ8}
=&
\frac{2i}{(4\pi)^2}
\frac{m_Z^2}{v^2}
\left(
\frac{m_{S}^2 - m_{A}^2}{v}
\right)
\left(
F_1 ( m_Z^2, q^2)
-
( m_S^2 - m_A^2 + q^2)
F_2 ( m_Z^2, m_S^2, q^2)
\right)
,
\\
\text{Fig.}~\ref{fig:hZ9}
=&
\frac{i}{(4\pi)^2}
\frac{m_h^2}{v^2}
\left(
\frac{m_{S}^2 - m_{A}^2}{v}
\right)
F_1 ( m_Z^2, q^2)
,
\end{align}
where $F_1$ and $F_2$ are defined in the Appendix \ref{sec:loopFunc}.

\subsection{$W$ boson contribution}
\begin{figure}[tb]
\subfigure[]{
 \includegraphics[width=0.18\hsize]{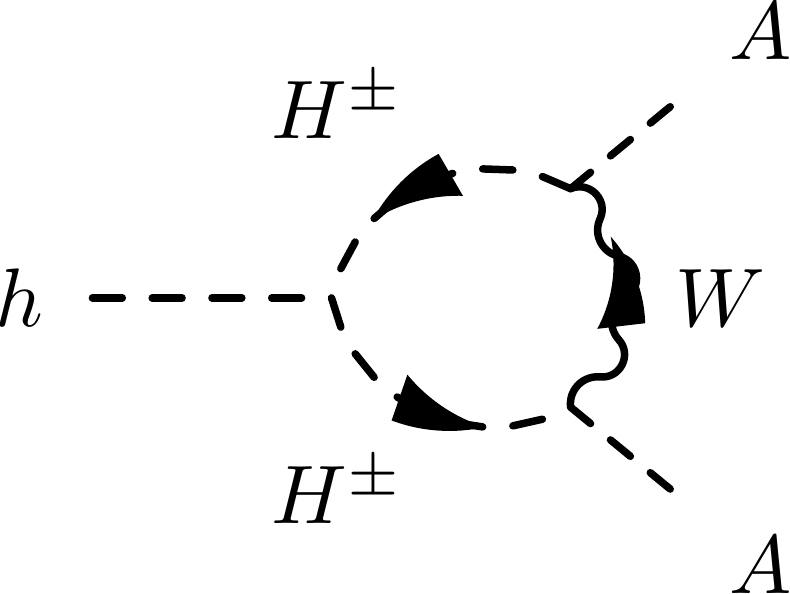} 
\label{fig:hW1}
}  
\subfigure[]{
 \includegraphics[width=0.18\hsize]{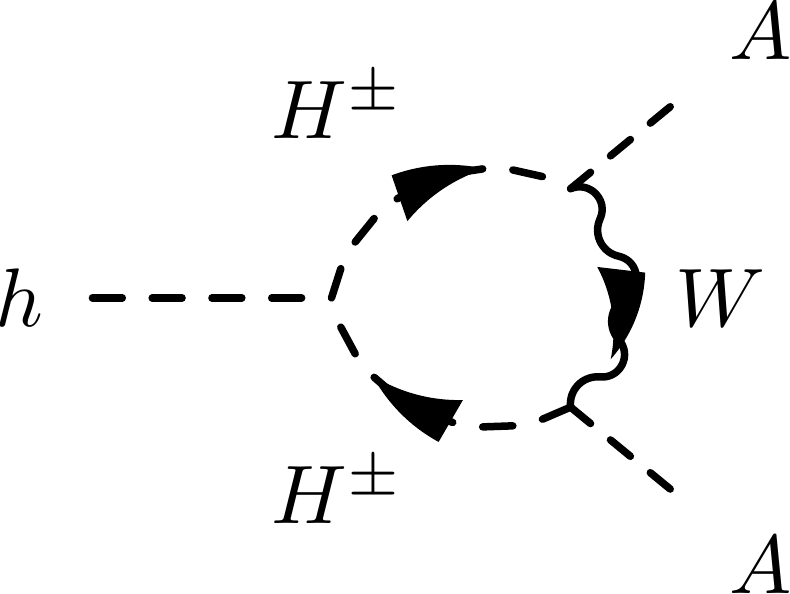} 
\label{fig:hW2}
}  
\subfigure[]{
 \includegraphics[width=0.18\hsize]{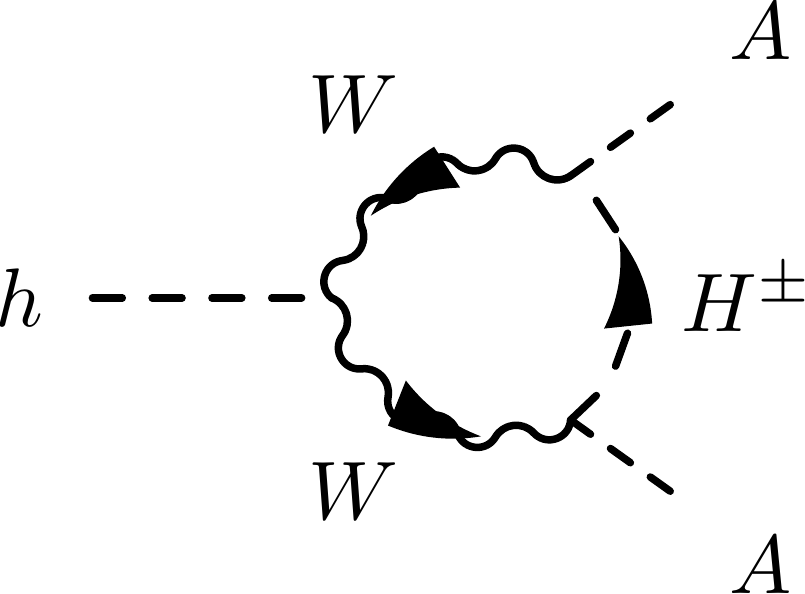} 
\label{fig:hW8}
}  
\subfigure[]{
 \includegraphics[width=0.18\hsize]{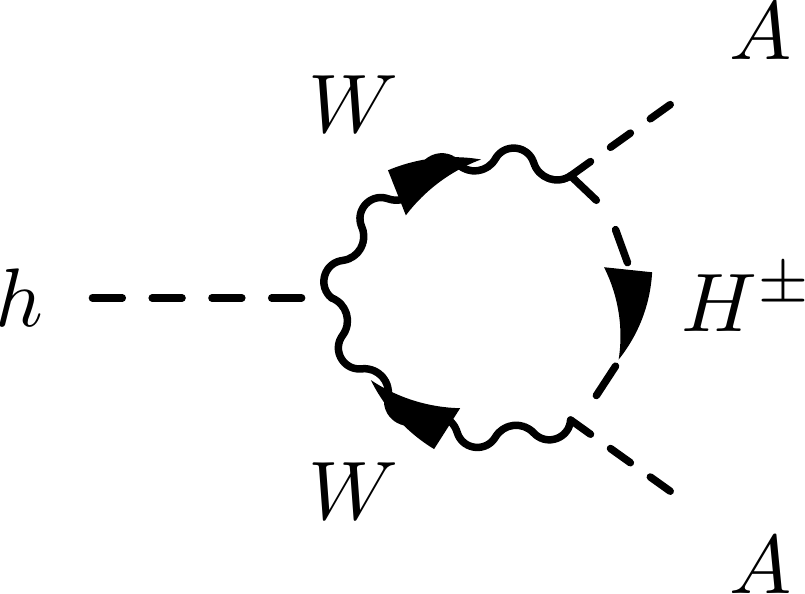} 
\label{fig:hW9}
}  
\subfigure[]{
 \includegraphics[width=0.18\hsize]{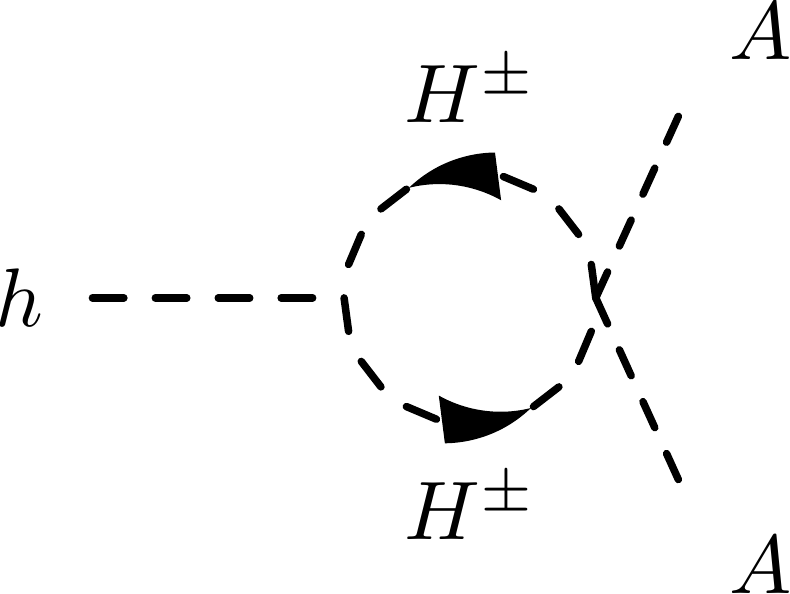} 
\label{fig:hW5}
}  
\subfigure[]{
 \includegraphics[width=0.18\hsize]{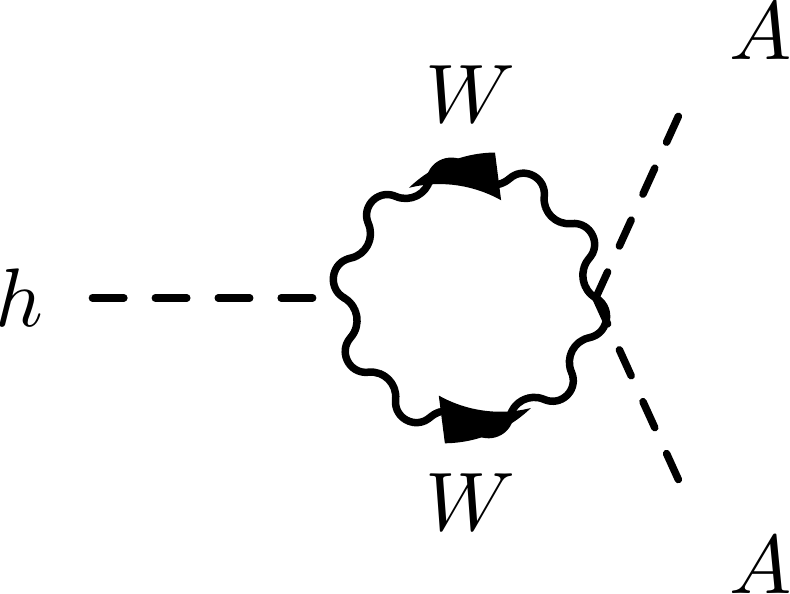} 
\label{fig:hW6}
}  
\subfigure[]{
 \includegraphics[width=0.18\hsize]{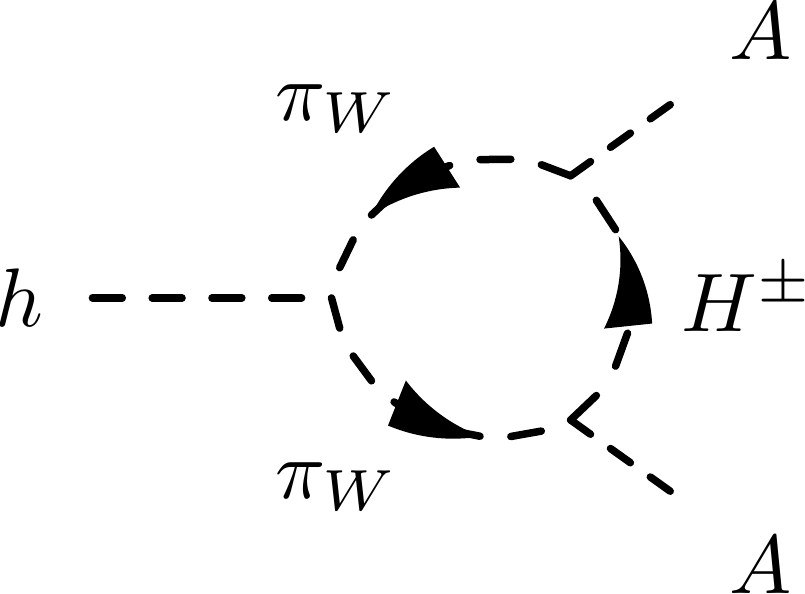} 
\label{fig:hW14}
}  
\subfigure[]{
 \includegraphics[width=0.18\hsize]{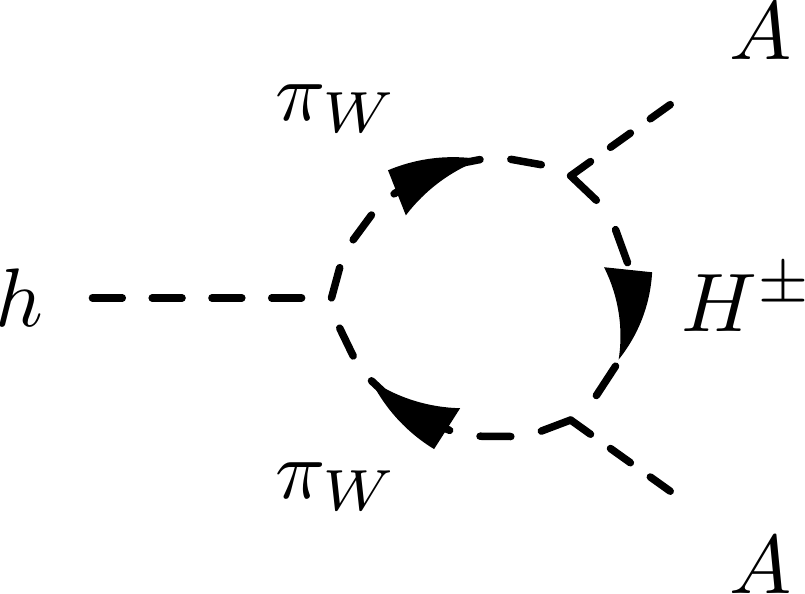} 
\label{fig:hW15}
}  
\subfigure[]{
 \includegraphics[width=0.18\hsize]{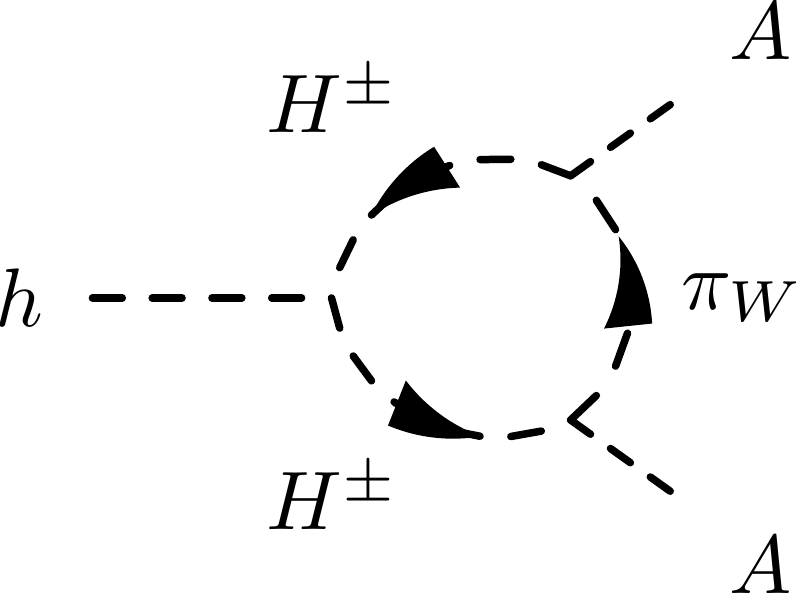} 
\label{fig:hW3}
}  
\subfigure[]{
 \includegraphics[width=0.18\hsize]{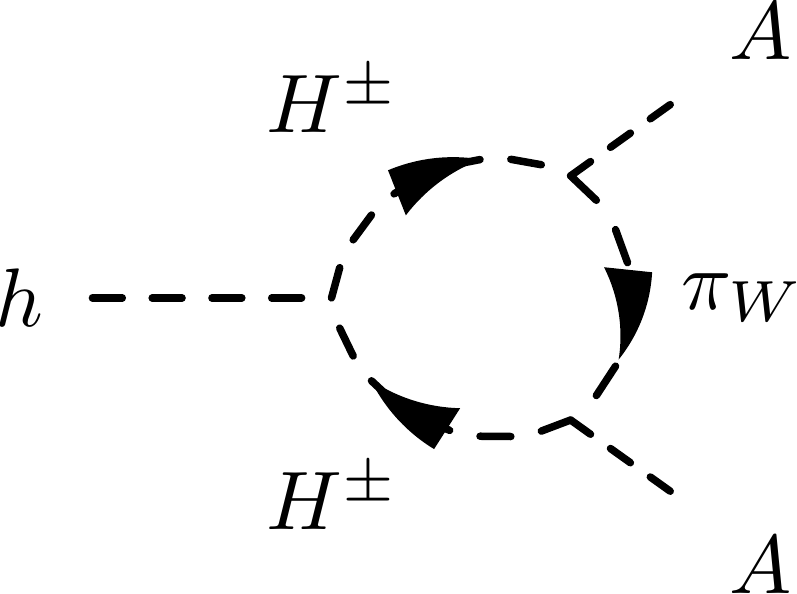} 
\label{fig:hW4}
}  
\subfigure[]{
 \includegraphics[width=0.18\hsize]{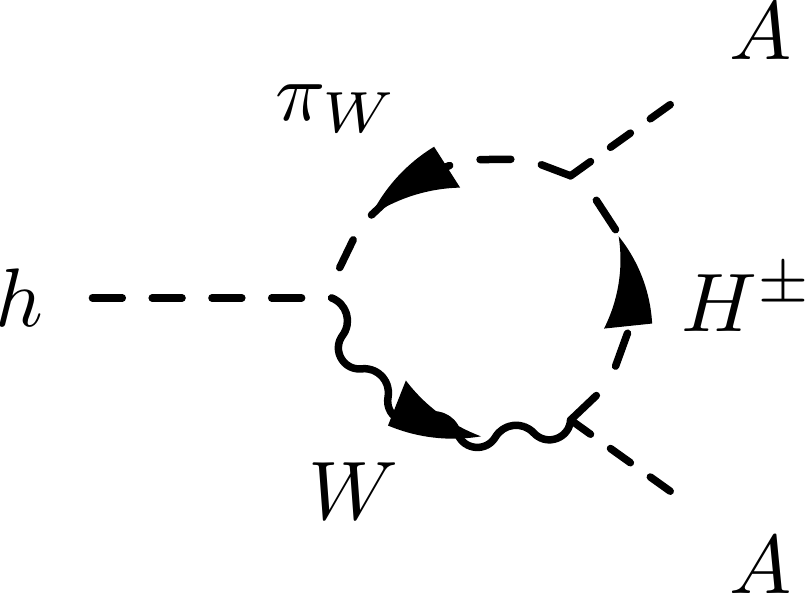} 
\label{fig:hW10}
}  
\subfigure[]{
 \includegraphics[width=0.18\hsize]{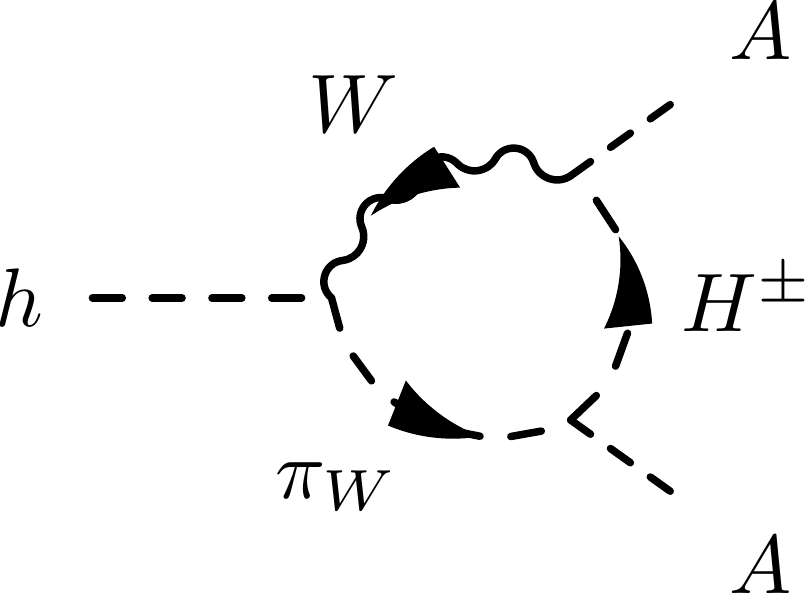} 
\label{fig:hW11}
}  
\subfigure[]{
 \includegraphics[width=0.18\hsize]{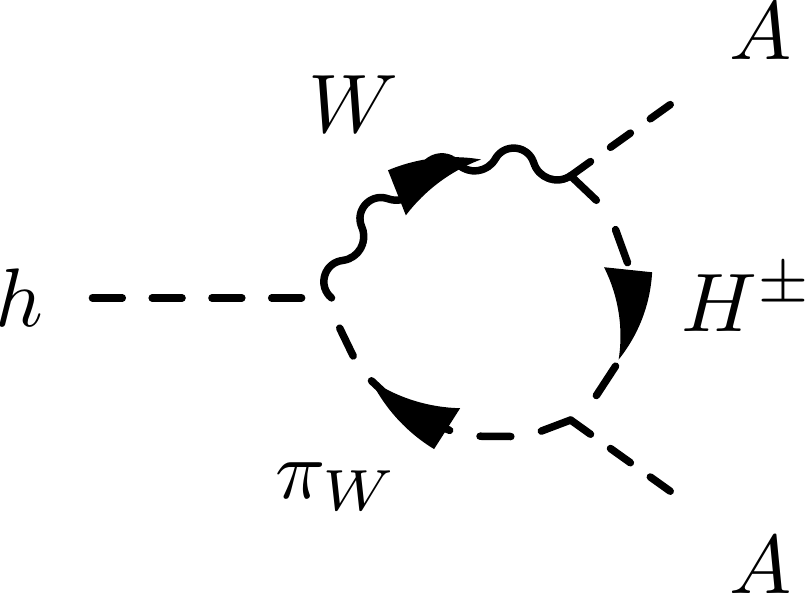} 
\label{fig:hW12}
}  
\subfigure[]{
 \includegraphics[width=0.18\hsize]{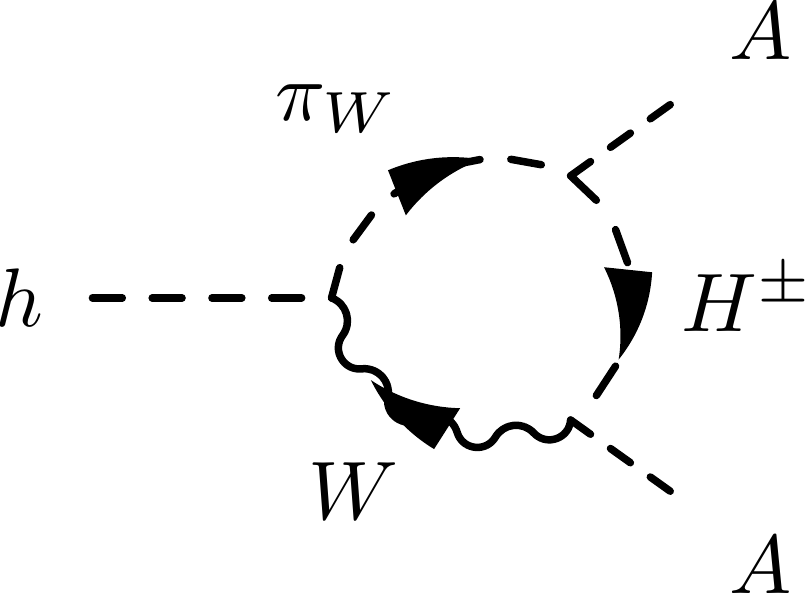} 
\label{fig:hW13}
}  
\subfigure[]{
 \includegraphics[width=0.18\hsize]{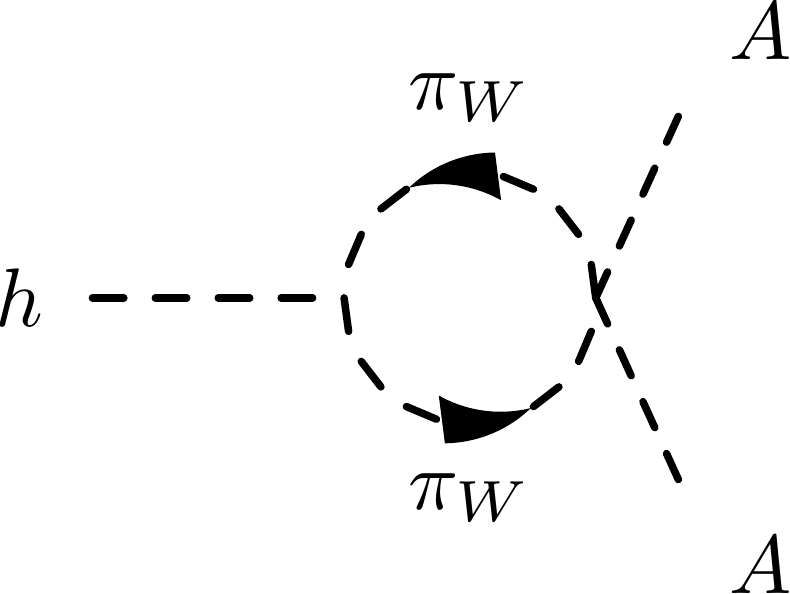} 
\label{fig:hW7}
}  
\caption{
The diagrams for the vertex correction with charged particles.
}
\end{figure}
Up to the $q^2$-independent terms, we find
\begin{align}
&\text{Fig.}~\ref{fig:hW1}
+
\text{Fig.}~\ref{fig:hW2}
\nonumber\\
=&
\frac{4i}{(4\pi)^2}
\frac{m_W^2}{v^2}
\frac{m_{H^{\pm}}^2 - m_{A}^2}{v}
\left(
F_1 ( m_{H^{\pm}}^2, q^2)
+
(-m_W^2 + 2 m_{H^{\pm}}^2 + 2 m_A^2 - 2 q^2)
F_2 ( m_{H^{\pm}}^2, m_W^2, q^2)
\right)
,
\\
&
\text{Fig.}~\ref{fig:hW8}
+
\text{Fig.}~\ref{fig:hW9}
\nonumber\\
=&
-\frac{4i}{(4\pi)^2}
\frac{m_W^4}{v^3}
\left(
2 F_1 ( m_{W}^2, q^2)
+
\left(m_W^2 - 2 m_{H^{\pm}}^2 - 2 m_A^2 + \frac{1}{2} q^2 \right)
F_2(m_W^2, m_{H^{\pm}}, q^2)
\right)
,
\\
\nonumber\\
&\text{Fig.}~\ref{fig:hW5}
\nonumber\\
=&
\frac{4i}{(4\pi)^2}
\lambda_2
\frac{m_{H^{\pm}}^2 - m_{A}^2}{v}
F_1 ( m_{H^{\pm}}^2, q^2)
,
\\
\nonumber\\
&\text{Fig.}~\ref{fig:hW6}
\nonumber\\
=&
\frac{16 i}{(4\pi)^2}
\frac{m_W^4}{v^3}
F_1 ( m_{W}^2, q^2)
,
\\
\nonumber\\
&\text{Fig.}~\ref{fig:hW14}
+
\text{Fig.}~\ref{fig:hW15}
\nonumber\\
=&
-\frac{2i}{(4\pi)^2}
\frac{m_h^2}{v}
\left(
\frac{m_{H^{\pm}}^2 - m_A^2}{v}
\right)^2
F_2(m_W^2, m_{H^{\pm}}, q^2)
,
\\
\nonumber\\
&\text{Fig.}~\ref{fig:hW3}
+
\text{Fig.}~\ref{fig:hW4}
\nonumber\\
=&
- \frac{4i}{(4\pi)^2}
\left(
\frac{m_{H^{\pm}}^2 - m_{A}^2}{v}
\right)^3
F_2 ( m_{H^{\pm}}^2, m_W^2, q^2)
,
\\
\nonumber\\
&
\text{Figs.}~\ref{fig:hW10}
+
~\ref{fig:hW11}
+
~\ref{fig:hW12}
+
~\ref{fig:hW13}
\nonumber\\
=&
\frac{4i}{(4\pi)^2}
\frac{m_W^2}{v^2}
\frac{m_{H^{\pm}}^2 - m_A^2}{v}
\left(
F_1 ( m_{W}^2, q^2)
-
\left(m_{H^{\pm}}^2 - m_A^2 + q^2 \right)
F_2(m_W^2, m_{H^{\pm}}, q^2)
\right)
,
\\
\nonumber\\
&
\text{Fig.}~\ref{fig:hW7}
\nonumber\\
=&
\frac{2i}{(4\pi)^2}
\frac{m_h^2}{v^2}
\frac{m_{H^{\pm}}^2 - m_{A}^2}{v}
F_1 ( m_{W}^2, q^2)
.
\end{align}


\section{Gluon contribution at two-loop level}
\label{sec:NN-sub-diagrams}
\begin{figure}[tb]
\subfigure[]{
 \includegraphics[width=0.20\hsize]{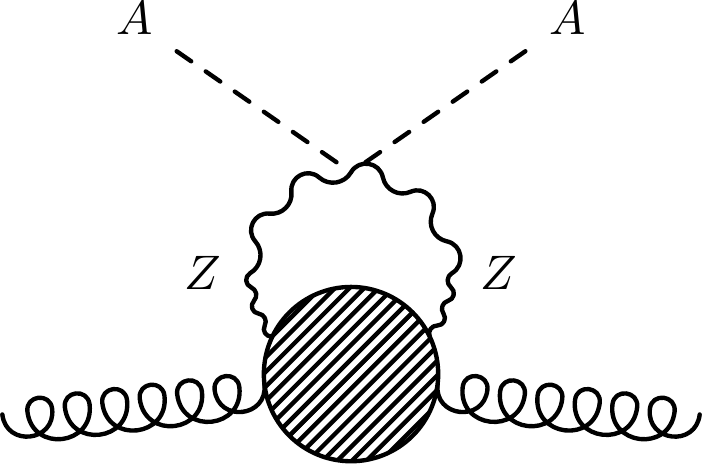} 
\label{fig:2loop_Z1}
}  
\quad
\subfigure[]{
 \includegraphics[width=0.20\hsize]{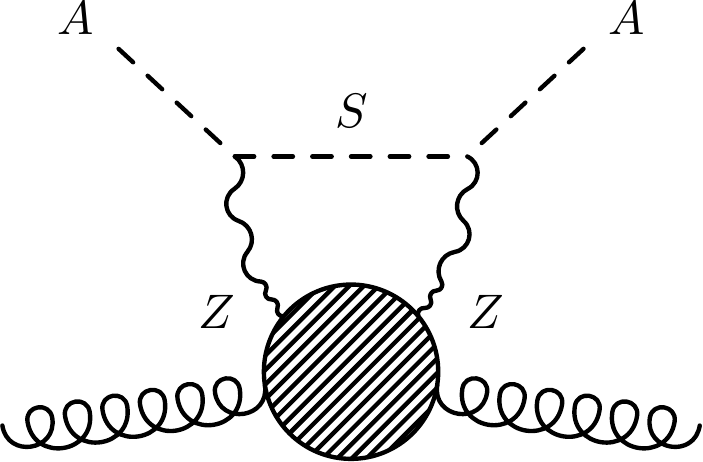} 
\label{fig:2loop_Z2}
}  
\quad
\subfigure[]{
 \includegraphics[width=0.20\hsize]{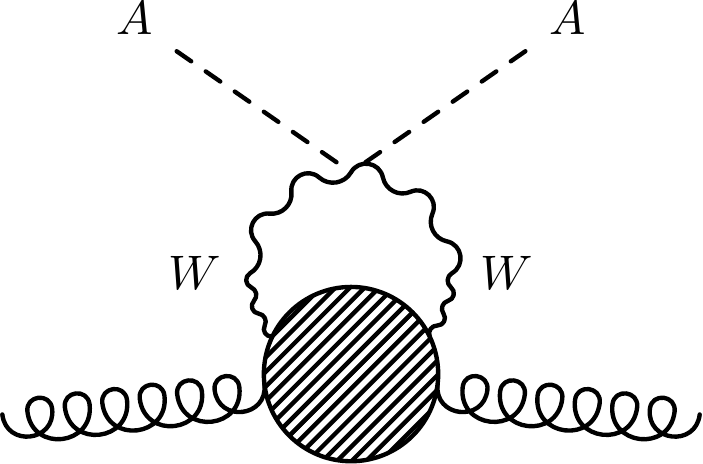} 
\label{fig:2loop_W1}
}  
\quad
\subfigure[]{
 \includegraphics[width=0.20\hsize]{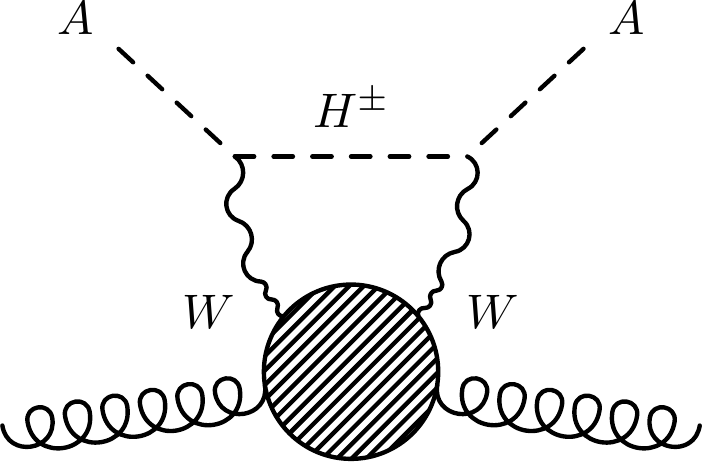} 
\label{fig:2loop_W2}
}  
\\
\subfigure[]{
 \includegraphics[width=0.20\hsize]{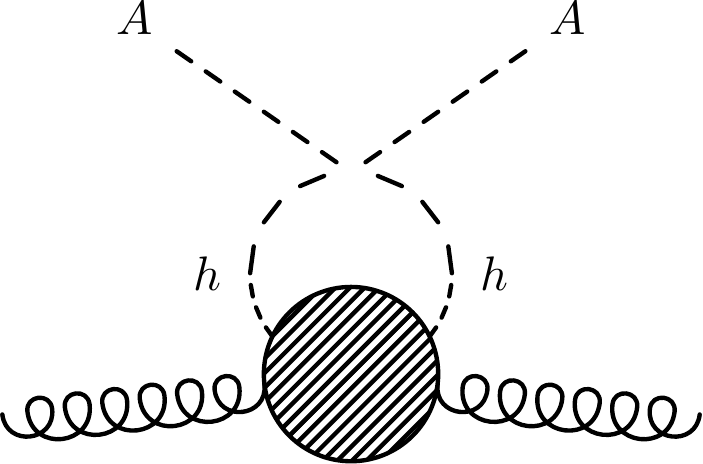} 
}  
\quad
\subfigure[]{
 \includegraphics[width=0.20\hsize]{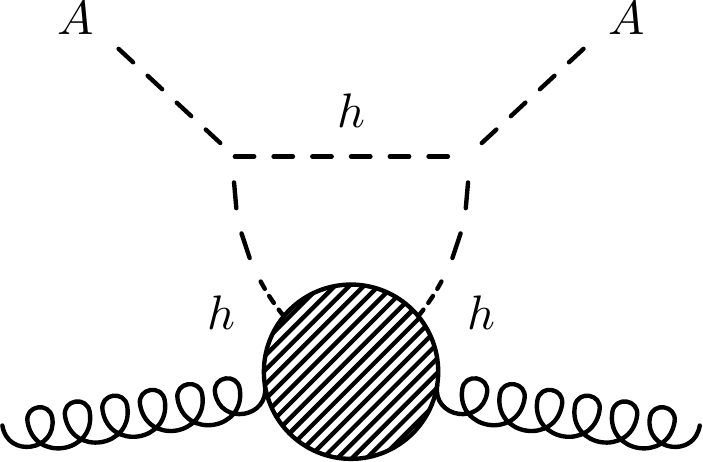} 
}  
\caption{
The diagram we calculate in this section. The shaded quark loop diagram.
We suppressed NG boson contributions. The last two diagrams are
 proportional to $\l_A$ which is much smaller than the other
 couplings, so we ignore their contributions.
}
\label{fig:2loop}
\end{figure}

The effective operator $A_0^2 G^a_{\m\n} G^{a\m\n}$ also give non-negligible contribution.
Two-loop diagrams shown in Fig.~\ref{fig:2loop} give contributions to this operator.
The shaded region contains quark loop diagram. There are also would-be Nambu-Goldstone (NG) bosons
 contributions, but we suppressed them in the Figures.
The last two diagrams in Fig.~\ref{fig:2loop} are
 proportional to $\lambda_A$ which is much smaller than the other
 couplings, so we ignore their contributions.
In this subsection, we describe an evaluation of them
by taking a method which is used for a calculation of the cross section of wino dark matter-nucleon scattering \cite{Hisano:2010fy, Hisano:2010ct, Hisano:2011cs}.
 Note that
 the operator with the gluon 
 field strength at two-loop order is the same as the operator without gluon
 field at one-loop order as we have discussed in Sec.~\ref{sec:3.1}.

\subsection{Two-point functions in the gluon background field}
\begin{figure}[tb]
\subfigure[]{
 \includegraphics[width=0.25\hsize]{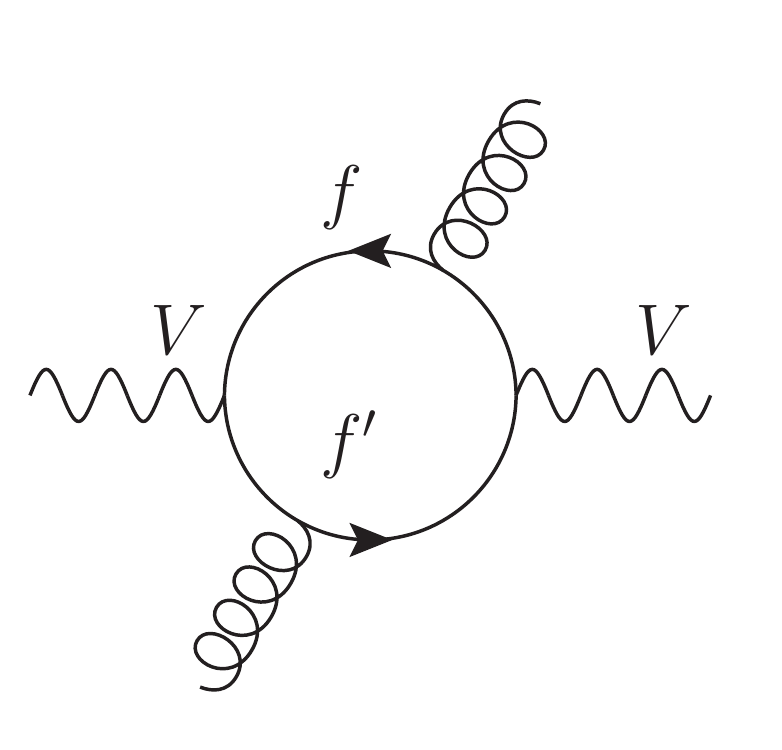} 
\label{fig:NNb1}
}  
\quad
\subfigure[]{
 \includegraphics[width=0.25\hsize]{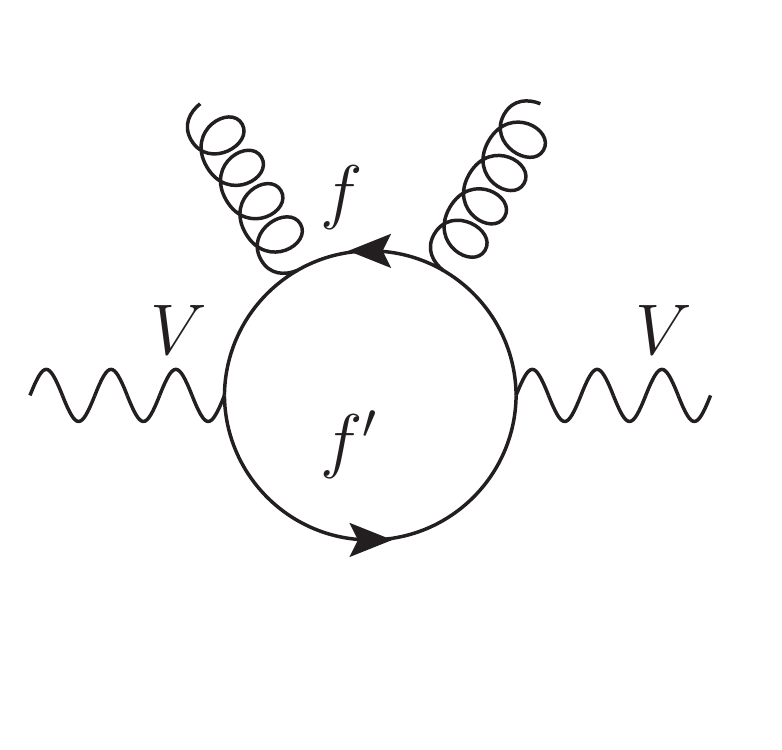} 
\label{fig:NNc1}
}  
\quad
\subfigure[]{
 \includegraphics[width=0.25\hsize]{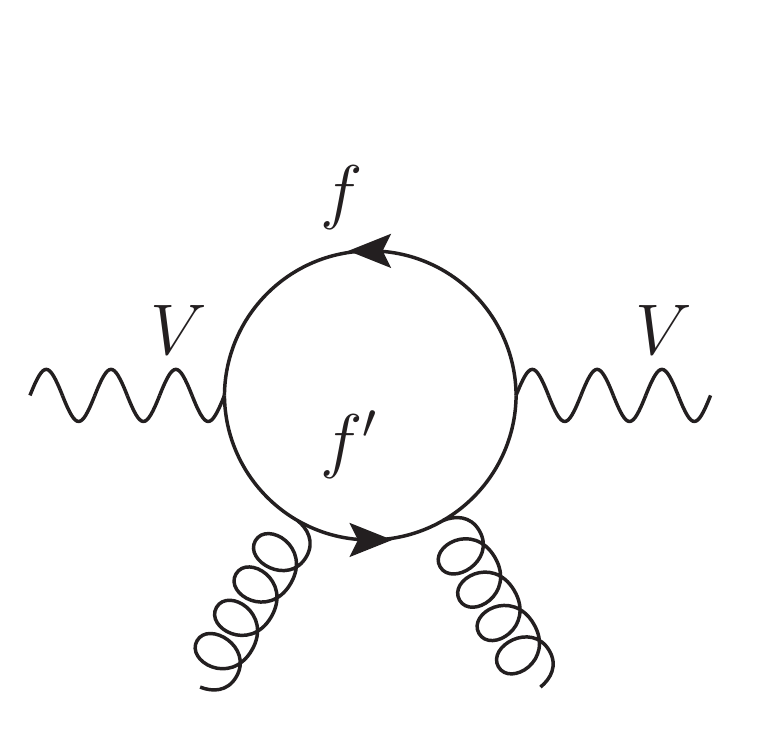} 
\label{fig:NNc2}
}  
\caption{One-loop corrections for two point function of gauge boson in gluon background field.}\label{fig:NN_VV}
~\\
\subfigure[]{
 \includegraphics[width=0.25\hsize]{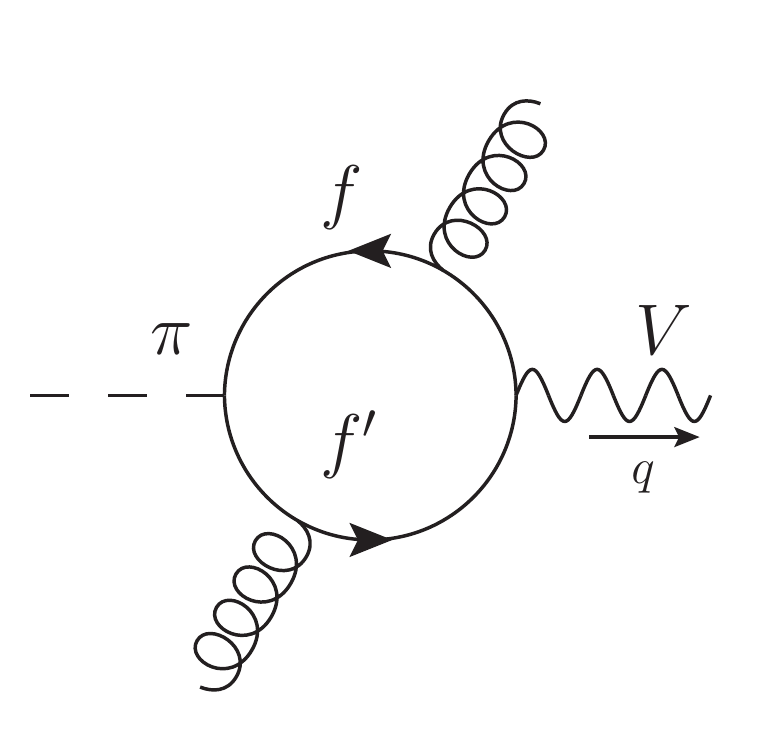} 
\label{fig:NNb1Vpi}
}  
\quad
\subfigure[]{
 \includegraphics[width=0.25\hsize]{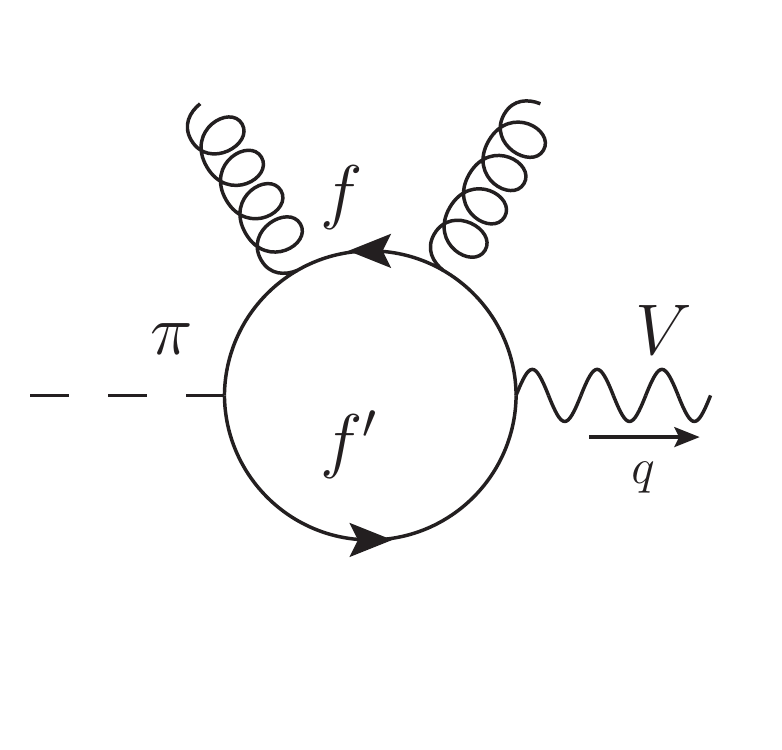} 
\label{fig:NNc1Vpi}
}  
\quad
\subfigure[]{
 \includegraphics[width=0.25\hsize]{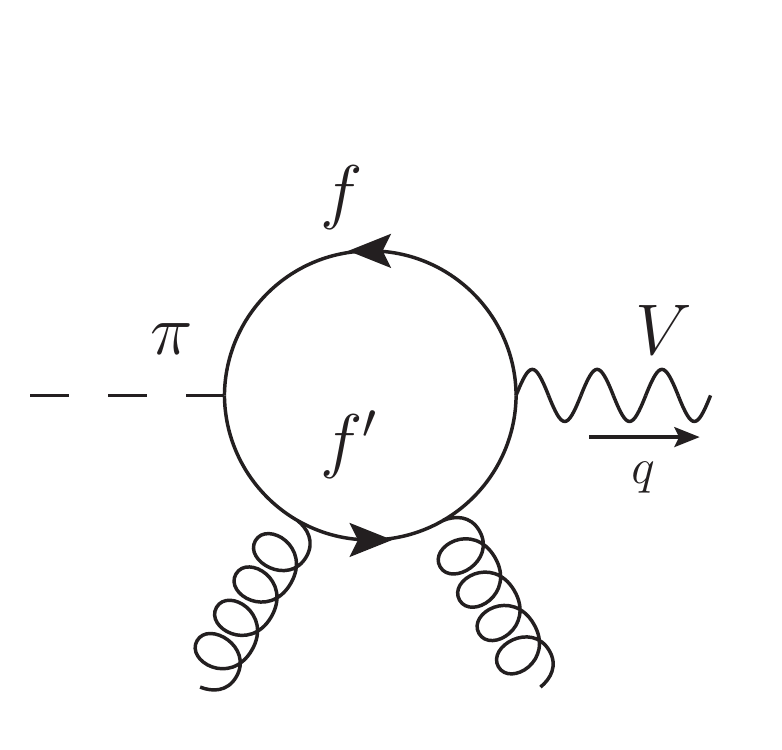} 
\label{fig:NNc2Vpi}
}  
\caption{One-loop corrections for two point function of gauge boson and pseudo-NG boson in gluon background field.}\label{fig:NN_Vpi}
~\\
\subfigure[]{
 \includegraphics[width=0.25\hsize]{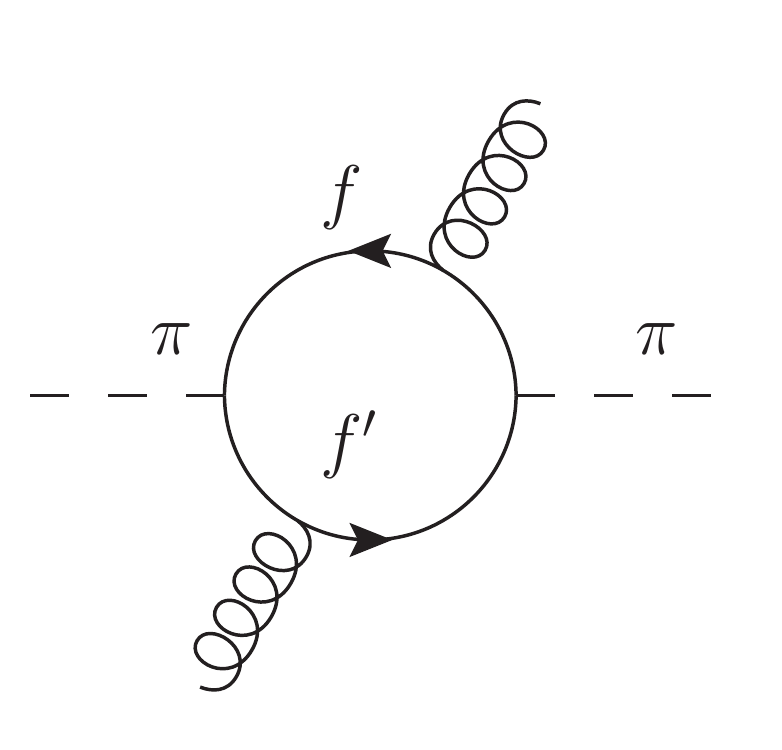} 
\label{fig:NNb1pipi}
}  
\quad
\subfigure[]{
 \includegraphics[width=0.25\hsize]{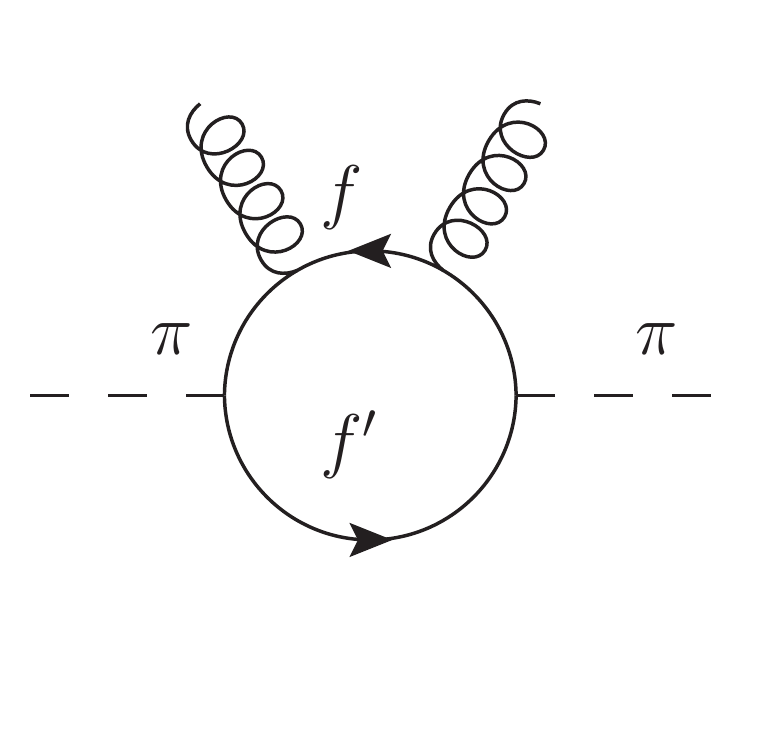} 
\label{fig:NNc1pipi}
}  
\quad
\subfigure[]{
 \includegraphics[width=0.25\hsize]{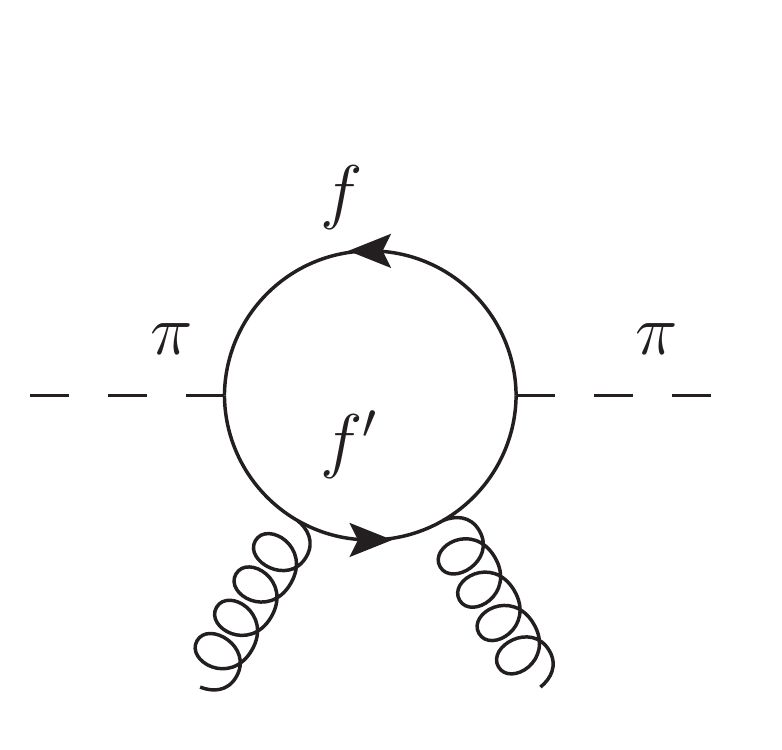} 
\label{fig:NNc2pipi}
}  
\caption{One-loop corrections for two point function of pseudo-NG boson in gluon background field.}\label{fig:NN_pipi}
\label{fig:NN}
\end{figure}
First, we evaluate quark loop sub-diagrams in the two-loop diagrams
shown in Fig.~\ref{fig:NN_VV}, \ref{fig:NN_Vpi} and \ref{fig:NN_pipi}.
For this purpose, we calculate one-loop corrections for two-point functions of gauge boson / pseudo-NG boson in the gluon background field
by taking the Fock-Schwinger gauge $x^\m A^a_\m=0$ for the gluon field, where $x^\m$ is the position four vector.
In the following of this paper, we only take into account gluon twist-0 operator and neglect higher twist operators,
{\it i.e.}, a product of gluon field strength can be substitute as,
\begin{align}
G_{\m\rho}^aG^a_{\n\s} \to \frac{1}{12}(g_{\m\n}g_{\rho\s}-g_{\m\s}g_{\n\rho}) G^a_{\m\n} G^{a\m\n}.
\end{align}
Thanks to these simplifications, 
two-point function of $W$ boson and pseudo-NG boson $\pi_W$ can be factorized as,
\begin{align}
i\Pi_{WW}^{(j)\a\b} &= -\frac{1}{6}\frac{i g_s^2}{16\pi^2} G^a_{\m\n} G^{a\m\n} \left( A_W^{(j)}(q^2)g^{\a\b} + B_W^{(j)}(q^2)q^\a q^\b \right),\\
i\Pi_{W\pi_W}^{(j)\a} &= -\frac{1}{6}\frac{i g_s^2}{16\pi^2} G^a_{\m\n} G^{a\m\n} C_W^{(j)}(q^2) q^\a,\\
i\Pi_{\pi_W\pi_W}^{(j)} &= -\frac{1}{6}\frac{i g_s^2}{16\pi^2} G^a_{\m\n} G^{a\m\n} D_W^{(j)}(q^2),
\end{align}
where $j=1,2$ and $3$ express generation of quarks which give the contribution to the two-point function.
Also, for $Z$ boson and $\pi_Z$,
\begin{align}
i\Pi_{ZZ}^{(f)\a\b} &= -\frac{1}{6}\frac{i g_s^2}{16\pi^2}G^a_{\m\n} G^{a\m\n} \left( A_Z^{(f)}(q^2)g^{\a\b} + B_Z^{(f)}(q^2)q^\a q^\b \right),\\
i\Pi_{Z\pi_Z}^{(f)\a} &= -\frac{1}{6}\frac{i g_s^2}{16\pi^2}G^a_{\m\n} G^{a\m\n} C_Z^{(f)}(q^2) q^\a,\\
i\Pi_{\pi_Z\pi_Z}^{(f)} &= -\frac{1}{6}\frac{i g_s^2}{16\pi^2}G^a_{\m\n} G^{a\m\n} D_Z^{(f)}(q^2),
\end{align}
where $f=u,d,s,c,b$ and $t$.
In $\Pi_{W\pi_W}^{(i)\a}(q^2)$ and $\Pi_{Z\pi_Z}^{(f)\a}(q^2)$, $q$ is momentum of gauge boson and its direction is out-going.

As noted in Refs.~\cite{Hisano:2010fy, Hisano:2010ct, Hisano:2011cs},
for the evaluation of the above two-point functions, we have to be careful for double-counting.
The loop integral in diagram \ref{fig:NNc1}, \ref{fig:NNc2}, \ref{fig:NNc1Vpi}, \ref{fig:NNc2Vpi}, \ref{fig:NNc1pipi} and \ref{fig:NNc2pipi} dominates
when the internal momentum is around a mass of quark emitting gluons.
In these diagrams,
if the quark emitting gluons is light quarks ({\it i.e.,} up, down or strange),
the dominant contribution comes from a region in which the internal momentum is smaller than QCD confinement scale.
In such a region, perturbative calculation cannot be reliable, and
the corresponding effect should be included in the evaluation of
$\langle N | m_q \bar q q| N \rangle$
\cite{Novikov:1983gd}. 
Therefore, the diagrams in which up, down or strange quark emitting two gluons
should be removed in the evaluation of the above $A$, $B$, $C$ and $D$ function.
On the other hand, the loop integral in diagram \ref{fig:NNb1}, \ref{fig:NNb1Vpi} and \ref{fig:NNb1pipi}
 dominates when the internal momentum is around external momentum $q$,
which is the order of $m_W$ or $m_Z$.
Therefore, this diagram always should be took into account of all of the quarks.
We assume $m_c$ and $m_b$ is larger than QCD confinement scale, but much smaller than $m_W$, $m_t$, $m_A$.
The charged gauge/pseudo-NG bosons obtain the contributions from up and down quark as,
\begin{align}
A_W^{(1)}(q^2) = \frac{g_W^2}{2} \frac{1}{q^2},\qquad
B_W^{(1)}(q^2) = -\frac{g_W^2}{2} \frac{1}{q^4}, \qquad
C_W^{(1)}(q^2) = 0, \qquad
D_W^{(1)}(q^2) = 0.
\end{align}
From charm and strange quark,
\begin{align}
A_W^{(2)}(q^2) = \frac{g_W^2}{2} \frac{1}{q^2},\qquad
B_W^{(2)}(q^2) = 0, \qquad
C_W^{(2)}(q^2) = \frac{g_W^2}{2} \frac{1}{2m_W} \frac{2}{q^2}, \qquad
D_W^{(2)}(q^2) = 0.
\end{align}
From top and bottom quark,
\begin{align}
A_W^{(3)} = \frac{g_W^2}{2}\left( \frac{1}{q^2-m_t^2} -\frac{1}{2}\frac{m_t^2}{(q^2-m_t^2)^2} \right),\qquad
B_W^{(3)}(q^2) = \frac{g_W^2}{2}\frac{1}{(q^2-m_t^2)^2}, \qquad\\
C_W^{(3)}(q^2) = \frac{g_W^2}{2}\frac{1}{2m_W} \frac{4q^2-3m_t^2}{(q^2-m_t^2)^2}, \qquad
D_W^{(3)}(q^2) = \frac{g_W^2}{2}\frac{m_t^2}{2m_W^2}\frac{5q^2-4m_t^2}{(q^2-m_t^2)^2}.
\end{align}

Neutral current couplings of quark $f$ are defined as $g_{f_L} = g_Z(T_{3f} - s_W^2 Q_f )$ and $g_{f_R} = -g_Z s_W^2 Q_f $.
The neutral gauge/pseudo-NG bosons obtain the contributions from up, down and strange quark as,
\begin{align}
A_Z^{(f)} = \frac{g_{f_L}^2 + g_{f_R}^2}{q^2},\qquad
B_Z^{(f)}(q^2) = -\frac{g_{f_L}^2+g_{f_R}^2}{q^4}, \qquad
C_Z^{(f)}(q^2) = 0, \qquad
D_Z^{(f)}(q^2) = 0.
\end{align}
From charm and bottom quark,
\begin{align}
A_Z^{(f)} = \frac{g_{f_L}^2 - 4 g_{f_L}g_{f_R} + g_{f_R}^2}{q^2},\quad
B_Z^{(f)}(q^2) = \frac{g_{f_L}^2+g_{f_R}^2}{q^4}, \quad
C_Z^{(f)}(q^2) = \frac{g_Z^2}{2m_Z} \frac{1}{q^2}, \quad
D_Z^{(f)}(q^2) = 0.
\end{align}
From top quark, 
\begin{align}
A^{(t)}_Z(q^2)
=& (g_{t_L}^2+g_{t_R}^2 )\int_0^1 dx \left( \frac{-w(1-w)}{\D(w)} + \frac{m_t^2 (2-5w+5w^2)}{[\D(w)]^2} + \frac{m_t^4 (-2+6w-6w^2)}{[\D(w)]^3} \right) \nonumber\\
& \qquad + \frac{g_Z^2}{4}\int_0^1 dx \left( \frac{m_t^2 (-1+2w-2w^2)}{[\D(w)]^2} + \frac{m_t^4 w(1-w)}{[\D(w)]^3} \right),\\
B_Z^{(t)}(q^2) =& -(g_{t_L}^2+g_{t_R}^2) \int_0^1 dw \left[ \frac{w^2(1-w)^2}{[\D(w)]^2}- \frac{2m_t^2 w(1-w)(1-3w+3w^2)}{[\D(w)]^3}   \right], \\
C_Z^{(t)}(q^2) =& -\frac{3g_Z^2}{4}\frac{m_t^2}{m_Z} \int_0^1 dw \left[ \frac{1-2w+2w^2}{2[\D(w)]^2}- \frac{m_t^2 (1-3w+3w^2)}{3[\D(w)]^3}   \right], \\
D_Z^{(t)}(q^2) =& \frac{g_Z^2}{4}\frac{m_t^2}{m_Z^2} \int_0^1 dw \left[ -\frac{3w(1-w)}{\D(w)} -\frac{3m_t^2(1+w-w^2)}{[\D(w)]^2} + \frac{4m_t^4 (1-3w+3w^2)}{[\D(w)]^3}   \right],
\end{align}
where $\D(w) \equiv m_t^2-w(1-w)q^2$.

\subsection{Effective interaction for dark matter-gluon scattering}
Next, by using the self-energy functions which have been evaluated so far,
we evaluate the $\Gamma^{G}_{\text{Box}}$ which is the coefficient of
the effective operator $A^2 G^a_{\m\n} G^{a\m\n}$ as defined in Eq.~(\ref{eq:eff_int}).
We take the Feynman-'t Hooft gauge for electroweak gauge bosons, and find
$\Gamma^{G}_{\text{Box}}$ is expressed as,
\begin{align}
-
\frac{\alpha_s}{4 \pi}
\Gamma^{G}_{\text{Box}}
 =& \sum_i f_{G,W}^{(i)} + \sum_f f_{G,Z}^{(f)}
, \\
f_{G,W}^{(i)}
=& \frac{ig_W^2}{12}\frac{g_s^2}{16\pi^2} \sum_i \int\frac{d^d \ell}{(2\pi)^d}\biggl[
\frac{3\ell^2 + 4\ell p - 4m_H^2}{[(\ell+p)^2-m_H^2][\ell^2-m_W^2]^2} A^{(i)}_W(\ell^2) \nonumber\\
& \qquad\qquad\qquad\qquad\qquad\qquad - \frac{m_H^2-m_A^2}{m_W^2} \frac{\ell^2 + 2\ell p}{[(\ell+p)^2-m_H^2][\ell^2-m_W^2]^2} \tilde B^{(i)}_W(\ell^2)
\biggr],\\
f_{G,Z}^{(f)}
=& \frac{ig_Z^2}{24}\frac{g_s^2}{16\pi^2} \sum_f \int\frac{d^d \ell}{(2\pi)^d}\biggl[
\frac{3\ell^2 + 4\ell p - 4m_S^2}{[(\ell+p)^2-m_S^2][\ell^2-m_Z^2]^2} A^{(f)}_Z(\ell^2) \nonumber\\
& \qquad\qquad\qquad\qquad\qquad\qquad - \frac{m_S^2-m_A^2}{m_Z^2} \frac{\ell^2 + 2\ell p}{[(\ell+p)^2-m_S^2][\ell^2-m_Z^2]^2} \tilde B^{(f)}_Z(\ell^2)
\biggr],
\end{align}
where
$\tilde B^{(i)}_W \equiv m_W^2 B^{(i)}_W - 2m_W C^{(i)}_W + D^{(i)}_W$ and
$\tilde B^{(f)}_Z \equiv m_Z^2 B^{(f)}_Z - 2m_Z C^{(f)}_Z + D^{(f)}_Z$.
%
$X_n$ and $Y_n$ which are defined in the Appendix \ref{sec:XY_2loop} are useful for the evaluations of two-loop diagrams.
For the convenience, 
we define $X_n^{Wq}$, $X_n^{Wt}$, $X_n^{Zq}$ and $\tilde X_n^{Zt}$ as,
\begin{align}
X_n^{Wq} &\equiv X_n(m_A^2,m_{H^\pm}^2,m_W^2,0), \\
X_n^{Wt} &\equiv X_n(m_A^2,m_{H^\pm}^2,m_W^2,m_t^2), \\
X_n^{Zq} &\equiv X_n(m_A^2,m_S^2,m_Z^2,0), \\
\tilde X_n^{Zt}(w) &\equiv X_n(m_A^2,m_S^2,m_Z^2,w^{-1}(1-w)^{-1}m_t^2).
\end{align}
$Y_n^{Wq}$, $Y_n^{Wt}$, $Y_n^{Zq}$ and $\tilde Y_n^{Zt}(w)$ are also defined in the same manner.
Finally, the contributions to $f_G$ is written as,
\begin{align}
f_{G,W}^{(1)} &= -\frac{g_W^2 g_s^2}{12(16\pi^2)^2}
\frac{g_W^2}{2}\left( 3X_0^{Wq} + 4m_A^2Y_1^{Wq} - 4m_{H^\pm}^2 X_1^{Wq} + (m_{H^\pm}^2-m_A^2)(X_1^{Wq} + 2m_A^2Y_2^{Wq}) \right), \\
%
f_{G,W}^{(2)} &= -\frac{g_W^2 g_s^2}{12(16\pi^2)^2}
\frac{g_W^2}{2}\left( 3X_0^{Wq} + 4m_A^2Y_1^{Wq} - 4m_{H^\pm}^2 X_1^{Wq} + \frac{2}{m_W^2} (m_{H^\pm}^2-m_A^2)(X_0^{Wq} + 2m_A^2Y_1^{Wq}) \right), \\
%
f_{G,W}^{(3)} & = -\frac{g_W^2 g_s^2}{12(16\pi^2)^2}
\frac{g_W^2}{2}\Biggl[ (3X_0^{Wt} + 4m_A^2 Y_1^{Wt} + (3m_t^2-4m_{H^\pm}^2) X_1^{Wt}) \nonumber\\
& \qquad\qquad\qquad\qquad\qquad - \frac{m_t^2}{2}(3X_1^{Wt} + 4m_A^2 Y_2^{Wt} + (3m_t^2-4m_{H^\pm}^2) X_2^{Wt}) \nonumber\\
& \qquad\qquad\qquad\qquad\qquad - \frac{m_{H^\pm}^2-m_A^2}{m_W^2} \left(-4 + \frac{5m_t^2}{2m_W^2}\right)(X_0^{Wt} + 2m_A^2 Y_1^{Wt} + m_t^2 X_1^{Wt}) \nonumber\\
& \qquad\qquad\qquad\qquad\qquad - (m_{H^\pm}^2-m_A^2) \left(1 - \frac{m_t^2}{m_W^2} + \frac{m_t^4}{2m_W^4}\right) (X_1^{Wt} + 2m_A^2 Y_2^{Wt} + m_t^2 X_2^{Wt}) 
\Biggr].
\end{align}
For up, down and strange quarks ($f=u,d,s$),
\begin{align}
f_{G,Z}^{(f)} & = -\frac{g_Z^2 g_s^2}{24(16\pi^2)^2}
(g_{f_L}^2+g_{f_R}^2)\biggl[ (3X_0^{Zq} + 4m_A^2Y_1^{Zq} - 4m_S^2 X_1^{Zq}) + (m_S^2-m_A^2)(X_1^{Zq} + 2m_A^2Y_2^{Zq}) \biggr].
\end{align}
For charm and bottom quarks ($f=c,b$),
\begin{align}
f_{G,Z}^{(f)} & = -\frac{g_Z^2 g_s^2}{24(16\pi^2)^2}
\biggl[ (g_{f_L}^2-4g_{f_L}g_{f_R}+g_{f_R}^2)(3X_0^{Zq} + 4 m_A^2Y_1^{Zq} - 4m_S^2 X_1^{Zq}) \nonumber\\
& \qquad\qquad\qquad - \frac{m_S^2-m_A^2}{m_Z^2} \left[ -g_Z^2 (X_0^{Zq} + 2m_A^2Y_1^{Zq})
+ \left( g_{f_L}^2 + g_{f_R}^2 \right)m_Z^2(X_1^{Zq} + 2m_A^2Y_2^{Zq}) \right]
\biggr].
\end{align}
For top-quark, 
\begin{align}
f_{G,Z}^{(t)} &= -\frac{g_Z^2 g_s^2}{24(16\pi^2)^2}
\sum_{n=1}^3 \int_0^1 dw {(-1)^n m_t^{2(n-1)} \over w^n(1-w)^n} \nonumber\\
&\qquad\qquad\qquad \times \Biggl[ 
\left(3 \tilde X_{n-1}^{Zt}(w) + 4 m_A^2 \tilde Y_n^{Zt}(w) + \left(\frac{3m_t^2}{w(1-w)}-4m_S^2\right)\tilde X_n^{Zt}(w) \right) g_{An}(w)  \nonumber\\
&\qquad\qquad\qquad\qquad - \frac{m_S^2-m_A^2}{m_Z^2} \left(\tilde X_{n-1}^{Zt}(w) + 2 m_A^2 \tilde Y_n^{Zt}(w) + \frac{m_t^2}{w(1-w)} \tilde X_n^{Zt}(w) \right) g_{Bn}(w) \Biggr].
\end{align}
Here, $g_{An}$ and $g_{Bn}$ are functions which satisfy,
\begin{align}
A^{(t)}_Z(q^2)
&= \int_0^1 dw \left( \frac{g_{A1}(w)}{\D(w)} + \frac{m_t^2 g_{A2}(w)}{[\D(w)]^2} + \frac{m_t^4 g_{A3}(w)}{[\D(w)]^3} \right), \\
\tilde B_Z^{(t)}(q^2)
&= \int_0^1 dw \left( \frac{g_{B1}(w)}{\D(w)} + \frac{m_t^2 g_{B2}(w)}{[\D(w)]^2} + \frac{m_t^4 g_{B3}(w)}{[\D(w)]^3} \right),
\end{align}
where $\D(w)=m_t^2-w(1-w)q^2$.
Explicit form of $g_{An}$ and $g_{Bn}$ are given by,
\begin{align}
g_{A1}(w) &= -(g_{t_L}^2+g_{t_R}^2) w(1-w), \\
g_{A2}(w) &= (g_{t_L}^2+g_{t_R}^2) (2-5w+5w^2) + \frac{g_Z^2}{4}(-1+2w-2w^2), \\
g_{A3}(w) &= (g_{t_L}^2+g_{t_R}^2) (-2+6w-6w^2) + \frac{g_Z^2}{4} w(1-w), \\
g_{B1}(w) &= -\frac{3g_Z^2}{4}\frac{m_t^2}{m_Z^2} w(1-w), \\
g_{B2}(w) &= -(g_{t_L}^2+g_{t_R}^2)\frac{m_Z^2}{m_t^2} w^2(1-w)^2 + \frac{g_Z^2}{4}(3-6w+6w^2) + \frac{g_Z^2}{4}\frac{m_t^2}{m_Z^2}(-3-3w+3w^2), \\
g_{B3}(w) &= (g_{t_L}^2+g_{t_R}^2)\frac{m_Z^2}{m_t^2} w(1-w)(2-6w+6w^2) + \frac{g_Z^2}{4}(-2+6w-6w^2) + \frac{g_Z^2}{4}\frac{m_t^2}{m_Z^2}(4-12w+12w^2).
\end{align}

\section{Loop functions for radiative corrections} \label{sec:loopFunc}
In this appendix, we summarize loop functions which are useful for
the evaluation of the radiative correction on the spin-independent cross section.
$B_i$, $B_i'$, $C_i$ and $D_i$ functions which appears in this appendix are
the Passarino-Veltman functions \cite{Passarino:1978jh} and the derivative
with respect to the momentum.
Our convention is same as used by {\tt LoopTools}~\cite{Hahn:1998yk}.
The explicit definitions of Passarino-Veltman functions are given as,
\begin{align}
\int\frac{d^4 \ell}{(2\pi)^d} \frac{1}{ [\ell^2-m_Z^2][(\ell+p)^2-m_S^2] }
&= {i\over 16\pi^2} B_0(p^2,m_Z^2,m_S^2), \\
\int\frac{d^4 \ell}{(2\pi)^d} \frac{\ell^\m}{ [\ell^2-m_Z^2][(\ell+p)^2-m_S^2] }
&= {i\over 16\pi^2} p^\m B_1(p^2,m_Z^2,m_S^2), \\
\int\frac{d^4 \ell}{(2\pi)^d} \frac{1}{ [\ell^2-m_Z^2]^2[(\ell+p)^2-m_S^2] }
&= {i\over 16\pi^2} C_0(0,p^2,p^2,m_Z^2,m_Z^2,m_S^2), \\
\int\frac{d^4 \ell}{(2\pi)^d} \frac{\ell^\m}{ [\ell^2-m_Z^2]^2[(\ell+p)^2-m_S^2]}
&= {i\over 16\pi^2} p^\m C_2(0,p^2,p^2,m_Z^2,m_Z^2,m_S^2), \\
\int\frac{d^d \ell}{(2\pi)^d} \frac{1}{ [\ell^2-m_Z^2]^3[(\ell+p)^2-m_S^2] }
&= {i\over 16\pi^2} D_0(0,0,p^2,p^2,0,p^2,m_Z^2,m_Z^2,m_Z^2,m_S^2), \\
\int\frac{d^d \ell}{(2\pi)^d} \frac{\ell^\m}{ [\ell^2-m_Z^3]^3[(\ell+p)^2-m_S^2] }
&= {i\over 16\pi^2} p^\m D_3(0,0,p^2,p^2,0,p^2,m_Z^2,m_Z^2,m_Z^2,m_S^2).
\end{align}

\subsection{One-loop vertex}
The functions $F_1$ and $F_2$ which are used in the Appendix \ref{sec:oneloop_vertex} are defined as,
\begin{align}
 F_1(m^2, q^2)
=&
B_0 ( q^2, m^2, m^2)
,\label{eq:loopfunc_F1}\\
 F_2(m_1^2, m_2^2, q^2)
=&
- C_0 (q^2, m_A^2, m_A^2, m_1^2, m_1^2, m_2^2)
.\label{eq:loopfunc_F2}
\end{align}

\subsection{One-loop box diagrams} \label{sec:loopfunction_box}
The functions $f_{B1}, f_{B2}, f_{B3}$ and $f_{B4}$ which are used in the Appendix \ref{sec:oneloop_box} are defined as,
\begin{align}
f_{B1}(m_1,m_2,m_A)
\equiv&
\int_0^1 dx
\frac{x}{m_1^2 x + m_2^2 (1-x) - m_A^2 x (1-x)}
\nonumber\\
=&
- \frac{\partial}{\partial m_1^2} B_0(m_A^2, m_1^2, m_2^2)
,
\\
f_{B2}(m_1,m_2,m_A)
\equiv&
 \int_{xyz}
\frac{y(1-z)}{\left(m_1^2 y + m_2^2 z - m_{A}^2 z(1-z)\right)^2}
\nonumber\\
=&
\frac{1}{m_1^2} 
\frac{\partial}{\partial m_1^2} B_0 (m_A^2, m_1^2, m_2^2)
+
\frac{1}{m_1^2} 
B_0' (m_A^2, m_1^2, m_2^2)
\nonumber\\
&
+
\frac{1}{m_1^4}
\left(
B_1(m_A^2, m_2^2, m_1^2) 
-
B_1(m_A^2, m_2^2, 0)
\right)
,
\\
f_{B3}(m_1,m_2,m_A)
\equiv&
\int_{xyz}
\frac{y^2}{\left(m_1^2 y + m_2^2 z - m_{A}^2 z(1-z)\right)^2}
\nonumber\\
=&
\frac{1}{m_1^2} 
\frac{\partial}{\partial m_1^2} B_0 (m_A^2, m_1^2, m_2^2)
+
\frac{1}{m_1^2} 
B_0' (m_A^2, m_1^2, m_2^2)
\nonumber\\
&
+
\frac{1}{m_1^4}
-
2 \frac{m_2^2 - m_A^2}{m_1^6}
\left(
B_1(m_A^2, m_1^2, m_2^2) 
-
B_1(m_A^2, 0, m_2^2)
\right)
\nonumber\\
&
+
2 \frac{m_A^2}{m_1^6}
\left(
B_{11}(m_A^2, m_1^2, m_2^2) 
-
B_{11}(m_A^2, 0, m_2^2)
\right)
,
\\
f_{B4}(m_1,m_2,m_A)
\equiv&
 \int_{xyz}
\frac{yz}{\left(m_1^2 y + m_2^2 z - m_{A}^2 z(1-z)\right)^2}
\nonumber\\
=&
-
\frac{1}{m_1^2}
B_0' (m_A^2, m_1^2, m_2^2)
+
\frac{1}{m_1^4}
\left(
B_1(m_A^2, m_1^2, m_2^2)
-
B_1(m_A^2, 0, m_2^2)
\right)
.
\end{align}
Here, $\int_{xyz}$ is defined as,
\begin{align}
 \int_{xyz}
f(x,y,z)
\equiv
 \int_{x+y+z=1}
f(x,y,z)
\equiv
 \int_{0}^{1} dz
 \int_{0}^{1-z} dy
f(1-y-z,y,z)
.
\end{align}

%
%
%


\subsection{Loop functions for dark matter-gluon scattering} \label{sec:XY_2loop}
Here, we summarize some loop functions which are useful for
the evaluation of the coefficient of effective interaction between dark matter and gluon.

\subsubsection{Definitions of $X$, $Y$ functions}
We define the following two types of loop functions:
\begin{align}
\int\frac{d^4 \ell}{(2\pi)^4} \frac{1}{[(\ell+p)^2-m_S^2][\ell^2-m_Z^2]^2[\ell^2-m_t^2]^n} &=
{i\over 16\pi^2} X_n(p^2,m_S^2,m_Z^2,m_t^2), \\
\int\frac{d^4 \ell}{(2\pi)^4} \frac{\ell^\m}{[(\ell+p)^2-m_S^2][\ell^2-m_Z^2]^2[\ell^2-m_t^2]^n} &=
{i \over 16\pi^2} p^\m Y_n(p^2,m_S^2,m_Z^2,m_t^2).
\end{align}

\subsubsection{$X$, $Y$ in $B$, $C$, $D$-function }
$X$ and $Y$ functions which are defined in the previous subsection are rewritten by Passarino-Veltman functions \cite{Passarino:1978jh}:
\begin{align}
X_0(m_A^2,m_S^2,m_Z^2,m_t^2) &= C_0^{(Z)}, \\
X_1(m_A^2,m_S^2,m_Z^2,m_t^2) &= -\frac{C_0^{(Z)}}{m_t^2-m_Z^2}+\frac{B_0^{(t)}-B_0^{(Z)}}{(m_t^2-m_Z^2)^2},\\
X_2(m_A^2,m_S^2,m_Z^2,m_t^2) &= \frac{C_0^{(t)} + C_0^{(Z)}}{(m_t^2-m_Z^2)^2}+\frac{-2B_0^{(t)}+2B_0^{(Z)}}{(m_t^2-m_Z^2)^3},\\
X_3(m_A^2,m_S^2,m_Z^2,m_t^2) &= \frac{D_0^{(t)}}{(m_t^2-m_Z^2)^2} + \frac{-2C_0^{(t)} - C_0^{(Z)}}{(m_t^2-m_Z^2)^3} + \frac{3B_0^{(t)}-3B_0^{(Z)}}{(m_t^2-m_Z^2)^4},\\
%
Y_1(m_A^2,m_S^2,m_Z^2,m_t^2) &= -\frac{C_2^{(Z)}}{m_t^2-m_Z^2}+\frac{B_1^{(t)}-B_1^{(Z)}}{(m_t^2-m_Z^2)^2},\\
Y_2(m_A^2,m_S^2,m_Z^2,m_t^2) &= \frac{C_2^{(t)} + C_2^{(Z)}}{(m_t^2-m_Z^2)^2}+\frac{-2B_1^{(t)}+2B_1^{(Z)}}{(m_t^2-m_Z^2)^3},\\
Y_3(m_A^2,m_S^2,m_Z^2,m_t^2) &= \frac{D_3^{(t)}}{(m_t^2-m_Z^2)^2} + \frac{-2C_2^{(t)} - C_2^{(Z)}}{(m_t^2-m_Z^2)^3} + \frac{3B_1^{(t)}-3B_1^{(Z)}}{(m_t^2-m_Z^2)^4},
\end{align}
where 
$B_{i}^{(X)}$, $C_{i}^{(X)}$ and $D_{i}^{(X)}$ are
\begin{align}
 B_{i}^{(X)} &\equiv B_{i}(m_A^2,m_X^2,m_S^2) ,\\
 C_{i}^{(X)} &\equiv C_{i}(0,m_A^2,m_A^2,m_X^2,m_X^2,m_S^2) ,\\
 D_{i}^{(X)} &\equiv D_{i}(0,0,m_A^2,m_A^2,0,m_A^2,m_X^2,m_X^2,m_X^2,m_S^2) .
\end{align}

\subsubsection{$C$, $D$ in $B_0$ and $\q B_0/\q q^2$}
All the external lines should satisfy the on-shell condition when we use {\tt LoopTools}.
For this technical reason, {\tt LoopTools-2.12} cannot evaluate
$C_{0/2}^{(Z/t)}$ and $D_{0/3}^{(t)}$ directly.
In this case we need to convert this function to other functions.
In this subsection, we express $C_{0/2}^{(Z/t)}$ and $D_{0/3}^{(t)}$ as combinations of $B_0$ and $\q B_0/\q q^2$.


\begin{align}
C_0^{(Z)}
=& \frac{\q}{\q m_Z^2} B_0(m_A^2,m_Z^2,m_S^2) \nonumber\\
=& \frac{1}{m_Z^4+m_S^4+m_A^4-2m_Z^2m_S^2-2m_A^2m_Z^2-2m_A^2m_S^2} \nonumber\\
& \qquad\times \left[
(m_S^2-m_Z^2+m_A^2)(-B_0(m_A^2,m_Z^2,m_S^2) + B_0(0,m_Z^2,m_Z^2) +2)
- 2m_S^2 \log \frac{m_S^2}{m_Z^2}
\right].
\end{align}

\begin{align}
C_2^{(Z)}
&= \frac{\q}{\q m_Z^2}B_1(m_A^2,m_Z^2,m_S^2) 
= \frac{\q}{\q m_A^2}B_0(m_A^2,m_Z^2,m_S^2).
\end{align}

\begin{align}
D_0^{(Z)}
=& \frac{1}{2} \frac{\q^2}{\q (m_Z^2)^2} B_0(m_A^2,m_Z^2,m_S^2) \nonumber\\
=& \frac{2m_S^2 m_A^2}{(m_Z^4+m_S^4+m_A^4-2m_Z^2m_S^2-2m_A^2m_Z^2-2m_A^2m_S^2)^2}  \nonumber\\
& \qquad\times \biggl[ -B_0(m_A^2,m_Z^2,m_S^2) + B_0(0,m_Z^2,m_Z^2) + \frac{m_Z^2-m_S^2-m_A^2}{2m_A^2}\log\frac{m_S^2}{m_Z^2} \nonumber\\
& \qquad\qquad - \frac{m_A^6 - 3(m_Z^2+m_S^2)m_A^4 + 3(m_Z^2-m_S^2)^2 m_A^2 - (m_Z^2+m_S^2)(m_Z^2-m_S^2)^2 }{4m_Z^2m_S^2m_A^2 } \biggr].
\end{align}

\begin{align}
D_3^{(Z)}
&= \frac{1}{2}\frac{\q^2}{\q (m_Z^2)^2}B_1(m_A^2,m_Z^2,m_S^2) \nonumber\\
&= \frac{1}{2}\frac{\q}{\q m_A^2}\frac{\q}{\q m_Z^2}B_0(m_A^2,m_Z^2,m_S^2) \nonumber\\
&= \frac{m_Z^2+m_S^2-m_A^2}{(m_Z^4+m_S^4+m_A^4-2m_Z^2m_S^2-2m_A^2m_Z^2-2m_A^2m_S^2)^2} \nonumber\\
& \qquad\qquad\times \left(
(m_S^2-m_Z^2+m_A^2)(-B_0(m_A^2,m_Z^2,m_S^2) + B_0(0,m_Z^2,m_Z^2) +2)
- 2m_S^2 \log \frac{m_S^2}{m_Z^2}
\right) \nonumber\\
& \qquad + \frac{1}{2}\frac{1}{m_Z^4+m_S^4+m_A^4-2m_Z^2m_S^2-2m_A^2m_Z^2-2m_A^2m_S^2} \nonumber\\
& \qquad\qquad\times \left(
-B_0(m_A^2,m_Z^2,m_S^2) + B_0(0,m_Z^2,m_Z^2) +2
- (m_S^2-m_Z^2+m_A^2)\frac{\q}{\q m_A^2} B_0(m_A^2,m_Z^2,m_S^2 )
\right).
\end{align}


\begin{thebibliography}{99}

\bibitem{1207.7214} 
  G.~Aad {\it et al.}  [ATLAS Collaboration],
  Phys.\ Lett.\ B {\bf 716}, 1 (2012)
  [arXiv:1207.7214 [hep-ex]].


\bibitem{1207.7235} 
  S.~Chatrchyan {\it et al.}  [CMS Collaboration],
  Phys.\ Lett.\ B {\bf 716}, 30 (2012)
  [arXiv:1207.7235 [hep-ex]].

\bibitem{Bertone:2004pz} 
For a review, see, \textit{e.g.},
  G.~Bertone, D.~Hooper and J.~Silk,
  Phys.\ Rept.\  {\bf 405}, 279 (2005)
  [hep-ph/0404175].


\bibitem{PHRVA.D18.2574} 
  N.~G.~Deshpande and E.~Ma,
  Phys.\ Rev.\ D {\bf 18}, 2574 (1978).

\bibitem{hep-ph/0603188} 
  R.~Barbieri, L.~J.~Hall and V.~S.~Rychkov,
  Phys.\ Rev.\ D {\bf 74}, 015007 (2006)
  [hep-ph/0603188].


\bibitem{PRLTA.39.165} 
  B.~W.~Lee and S.~Weinberg,
  Phys.\ Rev.\ Lett.\  {\bf 39}, 165 (1977).

\bibitem{NUPHA.B310.693} 
  M.~Srednicki, R.~Watkins and K.~A.~Olive,
  Nucl.\ Phys.\ B {\bf 310}, 693 (1988).

\bibitem{NUPHA.B360.145} 
  P.~Gondolo and G.~Gelmini,
  Nucl.\ Phys.\ B {\bf 360}, 145 (1991).


\bibitem{Chacko:2005un} 
  Z.~Chacko, H.~S.~Goh and R.~Harnik,
  JHEP {\bf 0601}, 108 (2006)
  [hep-ph/0512088].

\bibitem{Goh:2007dh} 
  H.~S.~Goh and C.~A.~Krenke,
  Phys.\ Rev.\ D {\bf 76}, 115018 (2007)
  [arXiv:0707.3650 [hep-ph]].


\bibitem{0712.1234} 
  E.~M.~Dolle and S.~Su,
  Phys.\ Rev.\ D {\bf 77}, 075013 (2008)
  [arXiv:0712.1234 [hep-ph]].

\bibitem{1105.5403} 
  J.~Mrazek, A.~Pomarol, R.~Rattazzi, M.~Redi, J.~Serra and A.~Wulzer,
  Nucl.\ Phys.\ B {\bf 853}, 1 (2011)
  [arXiv:1105.5403 [hep-ph]].


\bibitem{Ma:2006km} 
  E.~Ma,
  Phys.\ Rev.\ D {\bf 73}, 077301 (2006)
  [hep-ph/0601225].

\bibitem{1302.3936} 
  M.~Aoki, J.~Kubo and H.~Takano,
  Phys.\ Rev.\ D {\bf 87}, no. 11, 116001 (2013)
  [arXiv:1302.3936 [hep-ph]].

\bibitem{1303.7356} 
  Y.~Kajiyama, H.~Okada and T.~Toma,
  Phys.\ Rev.\ D {\bf 88}, no. 1, 015029 (2013)
  [arXiv:1303.7356].




\bibitem{0808.1729} 
  E.~Ma,
  Phys.\ Lett.\ B {\bf 671}, 366 (2009)
  [arXiv:0808.1729 [hep-ph]].


\bibitem{1304.1603} 
  E.~Ma,
  Phys.\ Lett.\ B {\bf 723}, 161 (2013)
  [arXiv:1304.1603 [hep-ph]].


\bibitem{1007.0871} 
  M.~Hirsch, S.~Morisi, E.~Peinado and J.~W.~F.~Valle,
  Phys.\ Rev.\ D {\bf 82}, 116003 (2010)
  [arXiv:1007.0871 [hep-ph]].

\bibitem{1205.3442} 
  L.~Lavoura, S.~Morisi and J.~W.~F.~Valle,
  JHEP {\bf 1302}, 118 (2013)
  [arXiv:1205.3442 [hep-ph]].



\bibitem{1110.5334} 
  T.~A.~Chowdhury, M.~Nemevsek, G.~Senjanovic and Y.~Zhang,
  JCAP {\bf 1202}, 029 (2012)
  [arXiv:1110.5334 [hep-ph]].


\bibitem{1204.4722} 
  D.~Borah and J.~M.~Cline,
  Phys.\ Rev.\ D {\bf 86}, 055001 (2012)
  [arXiv:1204.4722 [hep-ph]].


\bibitem{1207.0084} 
  G.~Gil, P.~Chankowski and M.~Krawczyk,
  Phys.\ Lett.\ B {\bf 717}, 396 (2012)
  [arXiv:1207.0084 [hep-ph]].

\bibitem{1302.2614} 
  J.~M.~Cline and K.~Kainulainen,
  Phys.\ Rev.\ D {\bf 87}, no. 7, 071701 (2013)
  [arXiv:1302.2614 [hep-ph]].

\bibitem{1304.2055} 
  A.~Ahriche and S.~Nasri,
  JCAP {\bf 1307}, 035 (2013)
  [arXiv:1304.2055].

\bibitem{0707.0633} 
  T.~Hambye and M.~H.~G.~Tytgat,
  Phys.\ Lett.\ B {\bf 659}, 651 (2008)
  [arXiv:0707.0633 [hep-ph]].

\bibitem{1202.0288} 
  J.~O.~Gong, H.~M.~Lee and S.~K.~Kang,
  JHEP {\bf 1204}, 128 (2012)
  [arXiv:1202.0288 [hep-ph]].











\bibitem{PHRVA.D76.095011} 
  Q.~H.~Cao, E.~Ma and G.~Rajasekaran,
  Phys.\ Rev.\ D {\bf 76}, 095011 (2007)
  [arXiv:0708.2939 [hep-ph]].

\bibitem{PHRVA.D81.035003} 
  E.~Dolle, X.~Miao, S.~Su and B.~Thomas,
  Phys.\ Rev.\ D {\bf 81}, 035003 (2010)
  [arXiv:0909.3094 [hep-ph]].

\bibitem{PHRVA.D82.035009} 
  X.~Miao, S.~Su and B.~Thomas,
  Phys.\ Rev.\ D {\bf 82}, 035009 (2010)
  [arXiv:1005.0090 [hep-ph]].

\bibitem{1206.6316} 
  M.~Gustafsson, S.~Rydbeck, L.~Lopez-Honorez and E.~Lundstrom,
  Phys.\ Rev.\ D {\bf 86}, 075019 (2012)
  [arXiv:1206.6316 [hep-ph]].

\bibitem{1303.7102} 
  M.~Krawczyk, D.~Sokolowska and B.~Swiezewska,
  J.\ Phys.\ Conf.\ Ser.\  {\bf 447}, 012050 (2013)
  [arXiv:1303.7102 [hep-ph]].





\bibitem{1303.6191} 
  M.~Aoki, S.~Kanemura and H.~Yokoya,
  Phys.\ Lett.\ B {\bf 725}, 302 (2013)
  [arXiv:1303.6191 [hep-ph]].

\bibitem{1401.6698} 
  A.~Arhrib, R.~Benbrik and T.~C.~Yuan,
  Eur.\ Phys.\ J.\ C {\bf 74}, 2892 (2014)
  [arXiv:1401.6698 [hep-ph]].










\bibitem{PHRVA.D85.095021} 
  A.~Arhrib, R.~Benbrik and N.~Gaur,
  Phys.\ Rev.\ D {\bf 85}, 095021 (2012)
  [arXiv:1201.2644 [hep-ph]].

\bibitem{1212.4100} 
  B.~Swiezewska and M.~Krawczyk,
  Phys.\ Rev.\ D {\bf 88}, no. 3, 035019 (2013)
  [arXiv:1212.4100 [hep-ph]].

\bibitem{1305.6266} 
  M.~Krawczyk, D.~Sokolowska, P.~Swaczyna and B.~Swiezewska,
  JHEP {\bf 1309}, 055 (2013)
  [arXiv:1305.6266 [hep-ph]].










\bibitem{astro-ph/0703512} 
  M.~Gustafsson, E.~Lundstrom, L.~Bergstrom and J.~Edsjo,
  Phys.\ Rev.\ Lett.\  {\bf 99}, 041301 (2007)
  [astro-ph/0703512 [ASTRO-PH]].

\bibitem{0811.1798} 
  P.~Agrawal, E.~M.~Dolle and C.~A.~Krenke,
  Phys.\ Rev.\ D {\bf 79}, 015015 (2009)
  [arXiv:0811.1798 [hep-ph]].

\bibitem{0901.1750} 
  S.~Andreas, M.~H.~G.~Tytgat and Q.~Swillens,
  JCAP {\bf 0904}, 004 (2009)
  [arXiv:0901.1750 [hep-ph]].

\bibitem{1306.4681} 
  C.~Garcia-Cely and A.~Ibarra,
  JCAP {\bf 1309}, 025 (2013)
  [arXiv:1306.4681 [hep-ph]].




\bibitem{hep-ph/0607067} 
  D.~Majumdar and A.~Ghosal,
  Mod.\ Phys.\ Lett.\ A {\bf 23}, 2011 (2008)
  [hep-ph/0607067].

\bibitem{JCAP.0702.028} 
  L.~Lopez Honorez, E.~Nezri, J.~F.~Oliver and M.~H.~G.~Tytgat,
  JCAP {\bf 0702}, 028 (2007)
  [hep-ph/0612275].




\bibitem{Silveira:1985rk} 
  V.~Silveira and A.~Zee,
  Phys.\ Lett.\ B {\bf 161}, 136 (1985).

\bibitem{McDonald:1993ex} 
  J.~McDonald,
  Phys.\ Rev.\ D {\bf 50}, 3637 (1994)
  [hep-ph/0702143 [HEP-PH]].

\bibitem{Burgess:2000yq} 
  C.~P.~Burgess, M.~Pospelov and T.~ter Veldhuis,
  Nucl.\ Phys.\ B {\bf 619}, 709 (2001)
  [hep-ph/0011335].
 


\bibitem{PHRVA.D87.075025} 
  M.~Klasen, C.~E.~Yaguna and J.~D.~Ruiz-Alvarez,
  Phys.\ Rev.\ D {\bf 87}, 075025 (2013)
  [arXiv:1302.1657 [hep-ph]].

\bibitem{hep-ph/0703056} 
  A.~Pierce and J.~Thaler,
  JHEP {\bf 0708}, 026 (2007)
  [hep-ph/0703056 [HEP-PH]].


\bibitem{PHRVA.D79.035013} 
  E.~Lundstrom, M.~Gustafsson and J.~Edsjo,
  Phys.\ Rev.\ D {\bf 79}, 035013 (2009)
  [arXiv:0810.3924 [hep-ph]].



\bibitem{PHRVA.D80.055012} 
  E.~M.~Dolle and S.~Su,
  Phys.\ Rev.\ D {\bf 80}, 055012 (2009)
  [arXiv:0906.1609 [hep-ph]].

\bibitem{1107.1991} 
  D.~Sokolowska,
  arXiv:1107.1991 [hep-ph].

\bibitem{1303.3010} 
  A.~Goudelis, B.~Herrmann and O.~St{\aa}l, 
  JHEP {\bf 1309}, 106 (2013)
  [arXiv:1303.3010 [hep-ph]].

\bibitem{JCAP.1406.030} 
  A.~Arhrib, Y.~L.~S.~Tsai, Q.~Yuan and T.~C.~Yuan,
  JCAP {\bf 1406}, 030 (2014)
  [arXiv:1310.0358 [hep-ph]].

\bibitem{Abe:2014gua} 
  T.~Abe, R.~Kitano and R.~Sato,
  arXiv:1411.1335 [hep-ph].



\bibitem{1003.3125} 
  L.~Lopez Honorez and C.~E.~Yaguna,
  JHEP {\bf 1009}, 046 (2010)
  [arXiv:1003.3125 [hep-ph]].

\bibitem{Ade:2013zuv} 
  P.~A.~R.~Ade {\it et al.}  [Planck Collaboration],
  Astron.\ Astrophys.\  {\bf 571}, A16 (2014)
  [arXiv:1303.5076 [astro-ph.CO]].



\bibitem{JCAP.1101.002} 
  L.~Lopez Honorez and C.~E.~Yaguna,
  JCAP {\bf 1101}, 002 (2011)
  [arXiv:1011.1411 [hep-ph]].


\bibitem{Hisano:2010ct} 
  J.~Hisano, K.~Ishiwata and N.~Nagata,
  Phys.\ Rev.\ D {\bf 82}, 115007 (2010)
  [arXiv:1007.2601 [hep-ph]].


\bibitem{Shifman:1978zn} 
  M.~A.~Shifman, A.~I.~Vainshtein and V.~I.~Zakharov,
  Phys.\ Lett.\ B {\bf 78}, 443 (1978).

\bibitem{Peskin:1995ev} 
  M.~E.~Peskin and D.~V.~Schroeder,
  Reading, USA: Addison-Wesley (1995) 842 p

\bibitem{Pumplin:2002vw} 
  J.~Pumplin, D.~R.~Stump, J.~Huston, H.~L.~Lai, P.~M.~Nadolsky and W.~K.~Tung,
  JHEP {\bf 0207}, 012 (2002)
  [hep-ph/0201195].

\bibitem{Hisano:2010fy} 
  J.~Hisano, K.~Ishiwata and N.~Nagata,
  Phys.\ Lett.\ B {\bf 690}, 311 (2010)
  [arXiv:1004.4090 [hep-ph]].





\bibitem{Belanger:2013oya} 
  G.~Belanger, F.~Boudjema, A.~Pukhov and A.~Semenov,
  Comput.\ Phys.\ Commun.\  {\bf 185}, 960 (2014)
  [arXiv:1305.0237 [hep-ph]].



\bibitem{1310.8214} 
  D.~S.~Akerib {\it et al.}  [LUX Collaboration],
  Phys.\ Rev.\ Lett.\  {\bf 112}, 091303 (2014)
  [arXiv:1310.8214 [astro-ph.CO]].



%
\bibitem{Aprile:2012zx} 
  E.~Aprile [XENON1T Collaboration],
  arXiv:1206.6288 [astro-ph.IM].

\bibitem{Feng:2014uja} 
  J.~L.~Feng, S.~Ritz, J.~J.~Beatty, J.~Buckley, D.~F.~Cowen, P.~Cushman, S.~Dodelson and C.~Galbiati {\it et al.},
  arXiv:1401.6085 [hep-ex].

\bibitem{Billard:2013qya} 
  J.~Billard, L.~Strigari and E.~Figueroa-Feliciano,
  Phys.\ Rev.\ D {\bf 89}, 023524 (2014)
  [arXiv:1307.5458 [hep-ph]].

%
%
%
%
%
%
%
%
%
%
%




%



\bibitem{Hisano:2011cs} 
  J.~Hisano, K.~Ishiwata, N.~Nagata and T.~Takesako,
  JHEP {\bf 1107}, 005 (2011)
  [arXiv:1104.0228 [hep-ph]].


\bibitem{Novikov:1983gd} 
  V.~A.~Novikov, M.~A.~Shifman, A.~I.~Vainshtein and V.~I.~Zakharov,
  Fortsch.\ Phys.\  {\bf 32}, 585 (1984).





\bibitem{Passarino:1978jh} 
  G.~Passarino and M.~J.~G.~Veltman,
  Nucl.\ Phys.\ B {\bf 160}, 151 (1979).

\bibitem{Hahn:1998yk} 
  T.~Hahn and M.~Perez-Victoria,
  Comput.\ Phys.\ Commun.\  {\bf 118}, 153 (1999)
  [hep-ph/9807565].



\end{thebibliography}
\end{document}